\documentclass[10pt]{article}
\usepackage{amsmath,amssymb}
\usepackage{amsthm}

\batchmode

\newtheorem{thm}{Theorem}

\newtheorem{cor}[thm]{Corollary}
\newtheorem{lem}[thm]{Lemma}
\newtheorem{defn}{Definition}

\newcommand{\be}{\begin{equation}}
\newcommand{\ee}{\end{equation}}

\newcommand{\pr}{\noindent{\bf Proof}. }
\newcommand{\re}{\noindent{\bf Remark}. }
\newcommand{\res}{\noindent{\bf Remarks}. }

\newcommand{\pa}{\partial}
\newcommand{\one}{\cO(1)}

\newcommand{\const}{\textrm{const}}

\newcommand{\supp}{ \mathrm{ supp  }}
\newcommand{\hs}{ \hspace{1cm}}

\newcommand{\bmu}{ \bar  \mu}
\newcommand{\un}{\underline}
\newcommand{\loc}{ \textrm{loc}}
\newcommand{\Vol}{\textrm{Vol}}
\newcommand{\tk}{\bbT^{-k}_{\sM +\sN -k}}

\newcommand{\sZ}{\mathsf{Z}}
\newcommand{\sM}{\mathsf{M}}
\newcommand{\sN}{\mathsf{N}}
\newcommand{\sB}{\mathsf{B}}

\newcommand{\al}{\alpha}
\newcommand{\De}{\Delta}
\newcommand{\de}{\delta}
\newcommand{\ga}{\gamma}

\newcommand{\ka}{\kappa}

\newcommand{\la}{\lambda}
\newcommand{\Om}{\Omega}
\newcommand{\om}{\omega}

\newcommand{\ep}{\epsilon}

\newcommand{\vep}{\varepsilon}

\newcommand{\cB}{{\cal B}}
\newcommand{\cC}{{\cal C}}
\newcommand{\cD}{{\cal D}}

\newcommand{\cO}{{\cal O}}

\newcommand{\cH}{{\cal H}}
\newcommand{\cS}{{\cal S}}

\newcommand{\cR}{{\cal R}}

\newcommand{\cK}{{\cal K}}

\newcommand{\cM}{{\cal M}}
\newcommand{\cN}{{\cal N}}
\newcommand{\cW}{{\cal W}}

\newcommand{\cL}{{\cal L}}

\newcommand{\cZ}{{\cal Z}}

\newcommand{\bbR}{{\mathbb{R}}}
\newcommand{\bbZ}{{\mathbb{Z}}}
\newcommand{\bbC}{{\mathbb{C}}}

\newcommand{\bbT}{{\mathbb{T}}}

\newcommand{\sD}{\mathsf{D}}

\begin{document}

\title{The Renormalization Group According to Balaban\\ I. Small Fields }
\author{ 
J. Dimock
\thanks{dimock@buffalo.edu}\\
Dept. of Mathematics \\
SUNY at Buffalo \\
Buffalo, NY 14260 }
\maketitle

\begin{abstract}
This is   an expository account of Balaban's approach to the renormalization group.  The method is 
illustrated with a treatment of  the the ultraviolet problem  for     the scalar $\phi^4$ 
model  on  toroidal lattice  in dimension  $d=3$.    This yields   another proof of the stability bound.
In this first paper we  analyze the small field contribution to the partition function.
\end{abstract}

\section{Introduction}

\subsection{overview}
Balaban  has developed a very powerful  renormalization group method    for    analyzing
  lattice  quantum  field theories.      The characteristic feature is that after each renormalization 
group transformation  a  split is introduced into regions  where  the fields  are large and 
regions where the fields are small.  Then one sums over  all possible splittings to cover  the
entire function space.   In  small field regions one can perform  a detailed analysis  
of  the effective actions,  for example carrying  out  perturbative  renormalization procedures.
The  large field region  is treated  crudely   but  makes  a small  contribution due to 
the fundamental stability  of the interaction.

Using this  approach Balaban  has  been able  to  treat   some  important  problems in quantum 
field theory.    These include   an analyis of the ultraviolet  problem  for scalar  QED  in $d=3$, \cite{Bal82a}  -  \cite{Bal83b},
the ultraviolet problem for  Yang  Mills  in  $d=3,4\ $  \cite{Bal84a} -  \cite{Bal89b},    and  the infrared  problem  for   an  N- component
scalar  field  in $d \geq  3$    with a  potential which has  a deep   minimum  on  the surface of  a  sphere, \cite{Bal95} -  \cite{Bal98c}     
  (known as the 
the''N-vector model''  or   ''linear  $\sigma$-model'').

These are  all very  difficult  problems   and  for each problem the analysis extends  over many  papers.
 As  a result  others have been slow to adopt  the approach  (exceptions  are   \cite{Imb86}, \cite{Kin86}, \cite{BIJ88},  \cite{BOS89},  \cite{BOS91},  \cite{Dim04}).
But    the basic  strategy  is   fairly straightforward     and it seems 
worthwhile  to  expose  it  in a simpler model.  That is the purpose of this  paper.

The model  we choose  is  the  scalar  $\phi^4$  model  in dimension    $d=3$   and in  a finite volume.
This  is  a special case  of  the scalar  QED  model  treated by  Balaban,  \cite{Bal82a}- \cite{Bal83c}.  
However   we  also incorporate many of the improvements   which  can be  found in   \cite{Bal95} -  \cite{Bal98c}.    The  
  treatment  is     also    more  efficient,   avoiding the analysis of  large orders  of 
perturbation theory.   Indeed we  avoid perturbation theory entirely and use a dynamical systems approach to renormalization.     

The   analysis stretches over  three papers.   In this first  paper  we  study a modified model in which  the fields  are all small
(i.e bounded).
The second paper develops an  expansion in small and large field regions  in which the results of the first paper are the
leading terms.    The third paper   establishes the convergence of the expansion.
The treatment is mostly self-contained,   however we sometimes  refer to the original papers for   technical results.

\subsection{the model}  The basic torus is   $\bbT_{\sM}  =   (\bbR  / L^{\sM} \bbZ )^3$
where  $L$ is a fixed large positive number  and $\sM \geq 0$ is a fixed nonnegative
integer.  It has volume  $\Vol ( \bbT_{\sM})  =  L^{3\sM}$.  In this torus we consider  lattices  with spacing  $L^{-\sN}$
defined by  
 \begin{equation}
\bbT_{\sM}^{-\sN}  =   ( L^{-\sN}\bbZ  / L^{\sM} \bbZ )^3  
\end{equation}
also  with volume $ \Vol (\bbT_{\sM}^{-\sN})    =  \Vol( \bbT_{\sM}) =  L^{3\sM}$.
If  $\sN < \sN'$   then   $\bbT_{\sM}^{-\sN}  \subset  \bbT_{\sM}^{-\sN'}  \subset \bbT_{\sM}$

Let   $\phi:  \bbT_{\sM}^{-\sN}  \to  \bbR$ be a scalar  field on the lattice.
The lattice version of the  $\phi^4$ model  is defined by the density 
\begin{equation}  \label{def1}
 \rho^N( \phi)  
=   \exp \left( - \frac12   < \phi,   (-\De  +  \bar \mu)  \phi>  +  V^{\sN}(\phi)   \right) 
=   \exp \left( - \frac12 \|  \pa \phi  \|^2  +    \bar \mu  \|  \phi  \|^2  +  V^{\sN}(\phi)   \right) 
\end{equation}
Here the  inner product  is defined with a weighted sum written as an integral:
\begin{equation}  \label{three0}
<u,v>   =   \int  u(x)v(x)  dx   \equiv   L^{-3\sN}    \sum_{x \in  \bbT_{\sM}^{-\sN}  }  u(x)v(x)   
\end{equation}
If     $\{e_{\mu}\} =  \{e_1,e_2,e_3\}$ are oriented unit basis vectors  the      
   derivative 
in the direction $e_{\mu}$   is   
\begin{equation}  \label{lattice1}
(\pa_{\mu}  \phi   )(x)   = ( \phi (x +L^{-\sN} e_{\mu}) -  \phi (x))/ L^{-\sN}
\end{equation}
and the Laplacian  is   $\De   = - \pa^*  \pa$.
The parameter  $\bar \mu  >0  $ is a  fixed     mass-squared .
The  potential has the form
\begin{equation}
V^{\sN}(\phi)  =  \vep^{\sN}     \Vol (\bbT_{\sM} )     +    \frac12     \mu^{\sN }      \int  \phi^2 (x)  dx   
  +  \frac{1}{4}   \la       \int  \phi^4 (x)  dx  
\end{equation}
Here   $\la >0$  is a  fixed   coupling constant.  The   parameters  $\vep^{\sN},  \mu^{\sN}$   are  energy  and   mass-squared counterterms
which  we  allow  to   depend  on the lattice spacing  $L^{-\sN}$.   This is renormalization.   The coupling constant 
$\la$  requires no renormalization in this model. 
 
We use  our  renormalization group    method to study     the  partition   function   
\begin{equation}
\sZ_{\sM,\sN}=\int  \rho^{\sN}(  \phi )  \ d\phi^{\sM, \sN}  \hs   \hs     d  \phi^{\sM, \sN} = 
  \  \prod_{x  \in    \bbT_{\sM}^{-\sN}  }  d  \phi(x)   \
\end{equation}
Actually it    is convenient to  study the  relative partition function  
$\sZ_{\sM,\sN}/\sZ_{\sM,\sN}(0)$  where   $\sZ_{\sM,\sN}(0)$  is the free field  partition 
 with  $V^{\sN}  =0$.

A  result of the analysis is  the following stability bound,  whose proof comes in the final   paper.

\begin{thm}    \label{major}
   For  any  $\la,  \bar \mu>0$    there is  a choice  of  renormalization  counterterms  
     $E^{\sN},  \mu^{\sN}$   and a constant  $c$   such that  
  such  that  
\begin{equation}
\exp \Big(  -  c \rm{  Vol }( \bbT_{\sM} )  \Big)    \leq     \frac{ \sZ_{\sM,\sN} }{  \sZ_{\sM,\sN}(0)}    \leq   \exp \Big(    c \rm{  Vol }( \bbT_{\sM} )  \Big)      
\end{equation}
for  all    $\sM,\sN$.
\end{thm}

This  result  is  not  new.   The upper bound was    first   obtained   by  Glimm and Jaffe    \cite{GlJa73}.    Their  results were extended by     Feldman and Osterwalder    \cite{FeOs76}  who 
established various   infinite volume limits   for weak coupling.
We have already mentioned  the   analysis of  Balaban     \cite{Bal82a}  -  \cite{Bal83c}  which includes this result.
  An alternative     renormalization group treatment for 
this problem was given by  Brydges,  Dimock, and Hurd  \cite{BDH95} where further references can be
found.

\subsection{the scaled model}
Before proceeding we scale the problem up to a unit lattice
with large volume.  This changes our ultraviolet problem to  an
infrared problem and puts us in  the natural home for the renormalization 
group.
The new   lattice  is  $\bbT_{\sM+ \sN}^{0}$ with volume  $ L^{3(\sM+ \sN)}$
and unit lattice spacing.  For fields 
$\Phi:   \bbT_{\sM+ \sN}^{0}  \to  \bbR$       we 
define
\begin{equation}
 \rho^{\sN}_0( \Phi) =       \rho^{\sN}( \Phi_{L^{-\sN}})  \end{equation}
 where $  \Phi_{L^{-\sN}}:   \bbT^{-\sN}_{\sM }  \to  \bbR$
is defined by 
\begin{equation}
  \Phi_{L^{-N}}(x)  = L^{N/2}\Phi(L^{N}x) 
\end{equation}
Making the change of variables  $\phi   =    \Phi_{L^{-N}}$     in the partition function we    have  
\begin{equation}
\sZ_{\sM,\sN}=\int  \rho^{\sN}_0(  \Phi )  \ d\Phi^{\sM, \sN}  \hs   \hs     d  \Phi^{\sM, \sN} = 
  \  \prod_{x  \in    \bbT_{\sM+ \sN}^{0}  }  d  \Phi(x)   \
\end{equation}
There  should actually be a factor  $ ( L^{N/2})^{ |\bbT_{\sM+ \sN}^{0} |/2}$
here,  but  since it makes no contribution to the relative partition function we have dropped it.

This scaling  preserves the Laplacian term and we have 
\begin{equation}  \label{def20}
 \rho^{\sN}_0(\Phi)  
=\exp \left( - \frac12  <\Phi, (- \De +   \bar  \mu^{\sN}_0)\Phi> - V^{\sN}_0(\Phi)  \right) 
\end{equation}
where  the  inner  product is now on the unit lattice and
\begin{equation}
V^{\sN}_0(\Phi)  =  \vep^{\sN}_0     | \bbT^0_{\sM+\sN}  |   +  \frac12   \mu^{\sN}_0     \sum_x \Phi^2 (x)    + \frac14   \la^{\sN}_0  \sum_x  \Phi^4 (x)  
\end{equation} 
The   fixed  coupling constants have  scaled to   
\begin{equation}  
\bar \mu^{\sN}_0  = L^{-2{\sN}}\bar  \mu   \ \ \ \ \   \la^{\sN}_0 = L^{-\sN} \la
\end{equation}
and the counterterms  have   scaled to  
\begin{equation}  
\vep^{\sN}_0  =  L^{-3\sN}   E^{\sN}
   \ \ \ \ \ \mu^{\sN}_0  = L^{-2\sN}\mu^N
\end{equation}

The  subscripts "zero"  indicate that we are
at the starting point of our renormalization group iteration.   In the following we generally omit the superscript
$\sN$.  Thus   $ \rho^N_0,  V^N_0$  are denoted     $ \rho_0,  V_0$,     and   $ \la^N_0,   \bar \mu^N_0,    \mu^N_0,
\vep^N_0$    are  denoted    $ \la_0,   \bar \mu_0,    \mu_0,  \vep_0$.

\section{the RG transformation}

\subsection{block  averaging}
The renormalization  group  (RG)  is a series of transformations which average
out the short distance features of the model, leaving only the   
the long distance properties in which we are interested (now that 
we have scaled the model).

First we define averaging operators.  On the   lattice   $L^{-k}\bbZ^3$,  or any  associated toroidal lattice,   the averaging operator
$Q$   takes functions  $f$ on  $L^{-k}\bbZ^3$ to functions  $Qf$    on  $L^{-k+1}\bbZ^3$ by
\begin{equation}
(Q f)(y)   =  L^{-3}  \sum_{x \in B(y)} f(x)  
\end{equation}
Here  $B(y)$  is  cubes  of  $L^3$ sites   ($L$ on a side) in   $L^{-k}\bbZ^3$   centered on  $y \in   L^{-k+1}\bbZ^3$. 
It can be written  
\begin{equation} 
 B(y) =  \{ x \in  L^{-k}\bbZ^3:  |x-y|  \leq L^{-k+1}/2\}
\end{equation}
The distance  is  $|x-y| = \sup_{\mu} |x_{\mu} -y_{\mu}|$  and we assume   $L$ is odd.
The  transpose operator  $ Q^T$  with respect to the inner product  (\ref{three0})     takes   functions
on  $L^{-k+1}\bbZ^3$ to functions  on  $L^{-k}\bbZ^3$.  It is computed to be  
\begin{equation}
\begin{split}
(Q ^Tf)(x) &   = f(y)  \ \ \    \textrm{ if   } \ \ \      x \in B(y) \\
\end{split}
\end{equation}
Then   $QQ^T=I$ while  
 $Q^T Q$   is a  projection operator onto  the range of $Q^T$ which is functions constant on the cubes.
\bigskip

 Now  starting with the density   $ \rho_{0}$  on functions  $\Phi_0: \bbT^0_{\sM + \sN} \to  \bbR$  we define a
transformed density $\tilde \rho_1$    on  functions $\Phi_1: \bbT^1_{\sM+ \sN} \to \bbR $ by 
\begin{equation}
\begin{split}  \label{first0}
 \tilde  \rho_{1} ( \Phi_1)  = & \cN^{-1}_{aL ,  \bbT^1_{\sM + \sN}}  \int \exp \left(-  \frac{a}{2L^2} 
\|\Phi_1-Q\Phi_0\|^2 \right)    \rho_{0} ( \Phi_0) \ \ d\Phi_0 \\
 = & \cN^{-1}_{aL ,  \bbT^1_{\sM + \sN}} \int \exp \left(- \frac12    aL
|\Phi_1-Q\Phi_0|^2 \right)    \rho_{0} ( \Phi_0) \ \ d\Phi_0 \\
\end{split}
\end{equation}
Here in the first  expression norms are taken with the natural metric for the lattice  so 
  $\| \Phi_1\|^2 =
L^3  \sum_{x  \in  \bbT^1_{\sM+ \sN}}   |\Phi_1(x)|^2 $.
In the second expression we use an unweighted sum  
 $| \Phi_1|^2 =
  \sum_{x  \in  \bbT^1_{\sM+ \sN}}   |\Phi_1(x)|^2 $.
The positive constant  $a$ is arbitrary  and the normalization constant  
is 
\footnote{In general  if  $\Om$ is a set and  $\Phi: \Om \to \bbR$ we define
\[     \cN_{a, \Om}    =    \int \exp \left(- \frac12    a|\Phi|^2 \right)  d \Phi   
=   \left(  \frac{2 \pi} {a}  \right)^{         | \Om| /2 } \] }
\begin{equation}
 \cN_{aL ,  \bbT^1_{\sM + \sN}} 
=    \int \exp \left(- \frac12    aL|\Phi_1|^2 \right)  d \Phi_1   
=   \left(  \frac{2 \pi} {aL}  \right)^{         | \bbT^1_{\sM + \sN}| /2 }
\end{equation}
the  constant is chosen so that   
\begin{equation}
\int \tilde   \rho_{1} ( \Phi_1)\  \ d\Phi_1 =
\int   \rho_{0} ( \Phi_0)    \ d\Phi_0
\end{equation}

Now  one  scales  back to the unit lattice.     
A  function       $\Phi_1: \bbT^0_{\sM+ \sN-1} \to  \bbR$   scales   up  to     $\Phi_{1,L}:  \bbT^1_{\sM+ \sN} \to \bbR  $ defined    by  
\begin{equation}
\Phi_{1,L}(x) =    L^{-1/2} \Phi_1(x/L)
\end{equation}
We  define
  \begin{equation}
   \rho_{1} ( \Phi_1)  =  \tilde  \rho_{1} ( \Phi_{1,L}) L^{  - | \bbT^0_{\sM +\sN -1} |/2}  =\tilde  \rho_{1} ( \Phi_{1,L}) L^{  - | \bbT^1_{\sM +\sN } |/2}  
\end{equation}  
 This preserves the integral
   \begin{equation}
  \int       \rho_{1} ( \Phi_1)  d \Phi_1
=  \int   \rho_{0} ( \Phi_0)    \ d\Phi_0
\end{equation}

  We compute,  taking account that   $ \cN^{-1}_{aL ,  \bbT^1_{\sM + \sN}}  L^{  - | \bbT^1_{\sM +\sN } |/2}  = \cN^{-1}_{a ,  \bbT^1_{\sM + \sN}} $,
  \begin{equation}
\begin{split}  \label{first}
 \rho_{1} ( \Phi_1) 
  = &  \cN^{-1}_{a ,  \bbT^1_{\sM + \sN}} \int \exp \left(-  \frac{a}{2L^2} 
\|\Phi_{1,L} -Q\Phi_0\|^2 \right)    \rho_{0} ( \Phi_0) \ \ d\Phi_0 \\
 = & \cN^{-1}_{a ,  \bbT^1_{\sM + \sN}} \int \exp \left(-   \frac{a}{2L^2} 
\|\Phi_{1,L}-Q\phi_L\|^2     \right)    \rho_{0} (\phi_L) \ \ d\phi^{(1)} \\
 = & \cN^{-1}_{a ,  \bbT^1_{\sM + \sN}}\int \exp \left(-   \frac{a}{2} 
\|\Phi_{1}-Q\phi\|^2     \right)    \rho_{0} (\phi_L) \ \ d\phi ^{(1)}\\
\end{split}
\end{equation}
In  the second step we have made the change of variables   by  $\Phi_0    =  \phi_{L}$ where   
   $\phi:  \bbT^{-1}_{\sM+ \sN-1} \to   \bbR$.   Then   
\begin{equation}
  d \Phi_0  = L^{- |\bbT^{-1}_{\sM+ \sN-1}| /2}  d \phi   \equiv   d  \phi^{(1)} 
 \end{equation}
In the last step we  use  that  $Q$ is scale invariant:    $Q  \phi_L = ( Q \phi)_L$.

We  repeat this step a number of times.  After  $k$ steps  we will have a density $\rho_k( \Phi_k)$ defined  on
functions $ \Phi_k:   \bbT^0_{\sM+ \sN-k} \to  \bbR$.    The next step is 
to   define a density on  functions   $ \Phi_{k+1}:   \bbT^1_{\sM+ \sN-k} \to  \bbR$ by
\begin{equation}
\begin{split}  \label{kth}
 \tilde  \rho_{k+1} ( \Phi_{k+1})  = & \cN^{-1}_{aL ,  \bbT^1_{\sM + \sN-k}}  \int \exp \left(-  \frac{a}{2L^2} 
\|\Phi_{k+1}-Q\Phi_k\|^2 \right)    \rho_{k} ( \Phi_k) \ \ d\Phi_k \\
 = & \cN^{-1}_{aL ,  \bbT^1_{\sM + \sN-k}} \int \exp \left(- \frac12    aL
|\Phi_{k+1}-Q\Phi_k|^2 \right)    \rho_{k} ( \Phi_k) \ \ d\Phi_k \\
\end{split}
\end{equation}
Then  one  scales  back to the unit lattice.    If     $ \Phi_{k+1}:   \bbT^0_{\sM+ \sN-k-1} \to  \bbR$ then 
 $ \Phi_{k+1,L}:   \bbT^1_{\sM+ \sN-k} \to  \bbR$  and we  define
  \begin{equation}   \label{scaleddensity}
 \rho_{k+1} ( \Phi_{k+1})  =  \tilde  \rho_{k+1} ( \Phi_{k+1,L}) L^{ -  | \bbT^1_{\sM +\sN -k} |/2}   
 \end{equation}
Then     we still have the normalization
 \begin{equation}  \label{preserve}
  \int       \rho_{k+1} ( \Phi_{k+1})  d \Phi_{k+1}
   =  \int       \rho_{k} ( \Phi_k)  d \Phi_k
=  \int   \rho_{0} ( \Phi_0)    \ d\Phi_0
\end{equation}

The various averaging operators can be composed into a single averaging 
operation over  large cubes.    Let     $Q_k = Q^k$  be averaging  operator over cubes   $B_k(y)$   with  $L^{3k}$ sites  ($L^k$ on a side).    The operator  $Q_k$ maps functions on    $\bbT^{-k}_{\sM+ \sN-k} $ to functions on   $\bbT^0_{\sM+ \sN-k} $  and  is given by 
\begin{equation}
(Q_kf)(y)    =   L^{-3k}  \sum_{x  \in B_k(y)}  f(x)   =  \int_{|x-y|< 1/2}   f(x) dx
\end{equation}

\begin{lem}
$\rho_k(\Phi_k)$ can be written  
  \begin{equation}
\begin{split}  \label{second}
 \rho_{k} ( \Phi_k)  
 =  \cN^{-1}_{a_k, \bbT^0_{\sM+ \sN-k}  }\ \int \exp \left(-  \frac{a_k}{2} 
\|\Phi_{k} -Q_k \phi \|^2 \right)    \rho_{0} ( \phi_{L^k}) \ \ d^{(k)}\phi \\
\end{split}
\end{equation}
where     
\begin{equation}
 \phi:    \bbT^{-k}_{\sM+ \sN-k}  \to   \bbR  \hs     d  \phi^{(k)} 
 = L^{- k|\bbT^{-k}_{\sM+ \sN-k}| /2}  d \phi   
 \end{equation}
  and 
\begin{equation}
a_k  =   a \frac{1-L^{-2}}{1- L^{-2k}}
\end{equation}
\end{lem}
\bigskip

\re  The following proof is maybe not the shortest,  but it develops  some  machinery we want to use later.
\bigskip

\pr    The proof is by induction.   Assuming it is true for $k$  we compute
\begin{equation}    \label{queen}
 \tilde  \rho_{k+1} ( \Phi_{k+1}) 
= \const
\int \exp \left(- \frac12   \frac{a}{L^2}
\|\Phi_{k+1}-Q\Phi_k\|^2 -  \frac{a_k}{2} 
\|\Phi_{k} -Q_k \phi \|^2 \right)    \rho_{0} ( \phi_{L^k}) \ \ d^{(k)}\phi\ \ d\Phi_k 
\end{equation}
The   expression inside the exponential has a minimum  in $\Phi_k$ when 
\begin{equation}
\Big(a_k  +  \frac{a}{L^2}    Q^TQ  \Big)\Phi_k   =    a_k Q_{k} \phi    + \frac{a}{L^2} Q^T  \Phi_{k+1}   
\end{equation}
This has the solution $\Phi_k  =  \Psi_k$  where
\begin{equation}  \label{kingmaker0}
\begin{split}
  \Psi_k(\Phi_{k+1}, \phi) =  &a_k^{-1} \Big( I- \frac{aL^{-2}}{a_k+ aL^{-2}}Q^T  Q \Big)
  \Big( a_kQ_{k}  \phi +  \frac{a}{L^2}  Q^T  \Phi_{k+1}  \Big)
\\
 =  & Q_{k}  \phi  
- \frac{aL^{-2}}{a_k+ aL^{-2}}Q^T  Q_{k+1} \phi  +  \frac{aL^{-2}}{a_k+ aL^{-2}}Q^T  \Phi_{k+1}   \\
\end{split}
\end{equation}  
We   compute using  $QQ^T=1$
\begin{equation}
\begin{split}
Q \Psi_k =   &
   Q_{k+1}  \phi  
- \frac{aL^{-2}}{a_k+ aL^{-2}}   Q_{k+1} \phi +  \frac{aL^{-2}}{a_k+ aL^{-2}}  \Phi_{k+1}   \\
= &
 \frac{a_k}{a_k+ aL^{-2}}   Q_{k+1} \phi +  \frac{aL^{-2}}{a_k+ aL^{-2}}  \Phi_{k+1}   \\
\end{split}
\end{equation}  
Thus     
\begin{equation}  \label{potpie}
\Phi_{k+1}-Q    \Psi_k  =\frac{a_k}{a_k+ aL^{-2}}(  \Phi_{k+1}-  Q_{k+1} \phi )
\end{equation}
and so  we have
\begin{equation}    \label{tee}
\begin{split}
  \frac{a}{2L^2} \|\Phi_{k+1}-Q \Psi_k\|^2
=&  \frac{a}{2L^2} \left( \frac{a_k}{a_k + a L^{-2}} \right)^2  \| \Phi_{k+1}  - Q_{k+1}  \phi \|^2  \\
=&    \frac{a_{k+1}}{2L^2}  \frac{a_k}{a_k + a L^{-2}}   \| \Phi_{k+1}  - Q_{k+1}  \phi \|^2  \\
\end{split}
\end{equation}
Here we   use the identity 
\begin{equation}   \label{ak}
a_{k+1}   = \frac{a_ka}{a_k+ aL^{-2}}
\end{equation}

On the other hand  from  (\ref{kingmaker0})   and  $\|Q^T \Phi \|^2  =  \|  \Phi\|^2$
\begin{equation}  \label{tea}
\begin{split}
\frac12    a_k \|\Psi_k -   Q_{k}  \phi  \|^2 = & \frac12    a_k    \left( \frac{aL^{-2}}{a_k+ aL^{-2}}\right)^2
  \|  \Phi_{k+1}   -   Q_{k+1} \phi \|^2  \\
  = & \frac{a_{k+1}}{2L^2}      \frac{aL^{-2}}{a_k+ aL^{-2}}
  \|  \Phi_{k+1}   -   Q_{k+1} \phi  \|^2  \\
 \end{split} 
\end{equation}  

Combining (\ref{tee}) and (\ref{tea})  gives the value at the minimum as
\begin{equation}  \label{fifty}
 \frac{a}{2L^2} \|\Phi_{k+1}-Q \Psi_k\|^2 +  \frac12    a_k \|\Psi_k -   Q_{k}  \phi \|^2
=  \frac{a_{k+1}}{2L^2}      \|  \Phi_{k+1}   -   Q_{k+1} \phi  \|^2  
\end{equation}

Now  in the integral  (\ref{queen})  expand around the minimizer.  We write  $\Phi_k = \Psi_k +Z$ and integrate over
$Z$.  The term with no $Z$'s is   (\ref{fifty}). The linear terms in $Z$ vanish  and the terms quadratic  in $Z$  when integrated over $Z$ yield a constant.
Thus we have  
\begin{equation}    \label{queen2}
 \tilde  \rho_{k+1} ( \Phi_{k+1}) 
= \const
\int \exp \left(-   \frac{a_{k+1}}{2L^2}      \|  \Phi_{k+1}   -   Q_{k+1} \phi  \|^2         \right)    \rho_{0} ( \phi_{L^k}) \ \ d^{(k)}\phi    
\end{equation}
Replacing  $\Phi_{k+1}$  by   $\Phi_{k+1,L}$  now with  $\Phi_{k+1}:\bbT^0_{\sM + \sN -k-1} \to \bbR$,   and  replacing   $\phi$ by  $\phi_L$  with  now  $ \phi:    \bbT^{-k-1}_{\sM+ \sN-k-1}  \to   \bbR$
we  find 
\begin{equation}    \label{queen3}
   \rho_{k+1} ( \Phi_{k+1}) 
= \const
\int \exp \left(-   \frac{a_{k+1}}{2}      \|  \Phi_{k+1}   -   Q_{k+1} \phi  \|^2         \right)    \rho_{0} ( \phi_{L^{k+1}}) \ \ d^{(k+1)}\phi    
\end{equation}
But the constant must   be  $  \cN^{-1}_{a_{k+1}, \bbT^0_{\sM+ \sN-k-1}  }$ in order to preserve the identity (\ref{preserve}).
This completes the proof.

\subsection{free flow  }  \label{freeflow}
Now suppose  we  only  keep  the  quadratic part of  $\rho_0$  so  that  
\begin{equation}  \label{def2}
  \rho_0 ( \Phi_0)
=\exp \left( - \frac12  <\Phi_0, (- \De +  \bmu_0)\Phi_0> \right)
 \end{equation}
 Inserting  this  in  (\ref{second}) yields 
   \begin{equation}
\begin{split}  \label{third}
 \rho_{k} ( \Phi_k)  
 =  \cN^{-1}_{a_k, \bbT^0_{\sM+ \sN-k}  }\ \int \exp (-S_k(\Phi_k, \phi) )   \ \ d^{(k)}\phi \\
\end{split}
\end{equation}
where  
\begin{equation}   \label{norton}
S_k(\Phi_k, \phi)  =       \frac{a_k}{2} 
\|\Phi_{k} -Q_k \phi \|^2  +     \frac12  <\phi, (- \De +  \bmu_k)\phi>
  \end{equation}
wtih     
\begin{equation}
 \bmu_k=     L^{2k}  \bmu_0  =   L^{-2(N-k)}  \bmu 
\end{equation}

To  compute  this we   look for the minimizer of   $S_k(\Phi_k, \phi)$ in  $\phi$.
The  minimizer    satisfies the equation
\begin{equation}
( -\De  +\bmu_k +  a_k\ Q_k^TQ_k) \phi   =  a_k Q_k^T \Phi_k
\end{equation}
The solution involves  the inverse
\begin{equation}
G_k =    ( -\De  +\bmu_k +  a_k\ Q_k^TQ_k)^{-1} 
\end{equation}
and has the form     $\phi  = \phi_k( \Phi_k)$
defined by   
\begin{equation}
\phi_k(\Phi_k)  =a_k  G_k Q_k^T \Phi_k   
\end{equation}
 We  shift the integration in  (\ref{third})  so  it is  centered on the minimum.
We  take   $\phi  =  \phi_k  +   \cZ$    where       $\cZ:  \bbT^{-k}_{\sM + \sN -k} \to   \bbR$
 is the new integration variable.      The cross terms vanish and   so  
  \begin{equation} 
 S_k(\Phi_k,    \phi_k  +   \cZ)  =  S_k(\Phi_k,    \phi_k  ) +   \frac12  <\cZ, (- \De + \bmu_k  +  a_k Q_k^T Q_k)\cZ>  
 \end{equation}   Then  
 (\ref{third}) becomes 
 \begin{equation}  \label{only}
 \rho_{k} ( \Phi_k)  
 =  Z_k\    \exp (-S_k(\Phi_k, \phi_k) ) 
\end{equation}
where  
\begin{equation}
\begin{split}  
 Z_k  
 =  \cN^{-1}_{a_k, \bbT^0_{\sM+ \sN-k}  }\ \int \exp\Big (- \frac12  <\cZ, (- \De + \bmu^N_k  +  a_k Q_k^T Q_k)\cZ>    \Big)   \ \ d^{(k)}\cZ \\
\end{split}
\end{equation}

Here is another representation of   $S_k(\Phi_k, \phi_k)$.  With 
  $\phi_k   =  a_k G_k Q_k^T  \Phi_k$   we have  
\begin{equation}  \label{spiffy}
\begin{split}
 S_k(\Phi_k, \phi_k )   =& \frac{a_k}{2} \| \Phi_k \|^2   - a_k  <\phi_k ,  Q_k^T  \Phi_k>  
  +   \frac12 \Big<\phi_k, (   - \De     + \bmu_k  + a_k Q_k^T Q_k )\phi_k \Big>  \\
  =&   \frac{a_k}{2}  \| \Phi_k \|^2   - a^2_k <\Phi_k ,Q_k G_k Q_k^T  \Phi_k>  
  + \frac{a_k^2}{2}   \Big<G_k Q_k^T\Phi_k, \Big(   - \De     + \bmu_k  +  Q_k^T Q_k \Big)G_k Q_k^T\Phi_k \Big>  \\
   =& \frac{a_k}{2} \| \Phi_k \|^2    -  \frac{a_k^2}{2}    <\Phi_k ,Q_k G_k Q_k^T  \Phi_k> \\
   \equiv   &   \frac12  <\Phi_k, \De_k   \Phi_k>  
\\
\end{split}
\end{equation}
where  $\De_k=  a_k -  a_k^2Q_kG_kQ_k^T$.

\subsection{single step free flow }

 Now  we   rederive  the  identity   $\rho_{k} ( \Phi_k)  
 =  Z_k\    \exp (-S_k(\Phi_k, \phi_k) ) $  for the free case by an   inductive  procedure.    We   assume  that   the identity holds  for  for  $\rho_k$   and  
show it holds for  $\rho_{k+1}$.     This approach will generate some useful identities and provide 
guidance    for  the treatment  of  the  general case with the potential added.

Starting with the identity for $k$   we  have
\begin{equation}   \label{manx}
\begin{split}
 \tilde  \rho_{k+1} ( \Phi_{k+1})    = & \cN^{-1}_{aL ,  \bbT^1_{\sM + \sN-k}} Z_k \int \exp \left(-  \frac{a}{2L^2} 
\|\Phi_{k+1}-Q\Phi_k\|^2 - S_k(\Phi_k, \phi_k) \right)  \ \ d\Phi_k \\
  = & \cN^{-1}_{aL ,  \bbT^1_{\sM + \sN-k}} Z_k \int \exp \Big(-  J( \Phi_{k+1},  \Phi_k,  \phi_k  )   \Big)  \ \ d\Phi_k \\
\end{split}
\end{equation}
where   
\begin{equation}
J( \Phi_{k+1},  \Phi_k,  \phi  )   \equiv    \frac{a}{2L^2} \|\Phi_{k+1}-Q\Phi_k\|^2 +   
\frac12    a_k
\|\Phi_k-Q_k  \phi\|^2     +  \frac12     \| \pa \phi\|^2  + \frac12   \bmu_k  \|  \phi\|^2
\end{equation}
Here   $\Phi_{k+1}, \Phi_k,  \phi$     are fields on    $\bbT^{1}_{\sM+ \sN-k},\bbT^{0}_{\sM+ \sN-k},\bbT^{-k}_{\sM+\sN-k}$   respectively.

To  compute the integral we want to minimize    $J( \Phi_{k+1},  \Phi_k,  \phi_k ) $ 
in  $\Phi_k$.   But   $J( \Phi_{k+1},  \Phi_k,  \phi_k ) $   is the minimum  value   of    $J( \Phi_{k+1},  \Phi_k,  \phi  ) $  in  $\phi$. 
 This  suggests we study the minimizer  of   $J( \Phi_{k+1},  \Phi_k,  \phi  ) $  in  $\Phi_k ,  \phi $  simultaneously.
 \footnote{  If   $A,B$  are two sets  and  $f: A \times B \to  \bbR$  then  
 \[    \inf_{y \in B}  ( \inf_{x \in A}  f(x,y)  )=  \inf_{x \in A, y \in B} f(x,y) 
 =   \inf_{x \in A} (   \inf_{y \in B}   f(x,y) ) \]    If  the minimizers are unique then they must come at the same point.} 
 
For   the next lemma we  need the operator  
\begin{equation}
 G_{k+1}^0   =   \Big( - \De  + \bmu_k  +   a_{k+1}L^{-2}Q^T_{k+1}  Q_{k+1} \Big)^{-1} 
\end{equation}
defined on functions  on  $\bbT^{-k}_{\sM + \sN-k}$.    This scales to $G_{k+1}$ as we will see.

\begin{lem} {  \  }  \label{tungsten}
\begin{enumerate}
\item Given  $\Phi_{k+1}$
the  unique  minimum  of  $J( \Phi_{k+1},  \Phi_k, \phi)$    comes  at     $(\Phi_k, \phi)  =  (\Psi_k,  \phi^0_{k+1})$  where    
\begin{equation}  \label{hunger}
 \phi^0_{k+1} =     \phi^0_{k+1} ( \Phi_{k+1}) =     L^{-2} a_{k+1}  G_{k+1}^0 Q_{k+1}^T  \Phi_{k+1}  
 \end{equation}
and    where   
\begin{equation}   \label{kingmaker}
\Psi_k  =     \Psi_k (\Phi_{k+1}, \phi^0_{k+1})   =   
    Q_{k} \phi^0_{k+1} 
- \frac{aL^{-2}}{a_k+ aL^{-2}}Q^T   Q_{k+1} \phi^0_{k+1} +  \frac{aL^{-2}}{a_k+ aL^{-2}}Q^T  \Phi_{k+1}   
\end{equation}  
\item   The minimizer in  $\phi$  can also be  written  $\phi_k( \Psi_k)$  so we have the identity
between functions of   $\Phi_{k+1}$
\begin{equation}  \label{someday}
 \phi^0_{k+1} =\phi_k(   \Psi_k)
 \end{equation}
\item    The   value  of   $J( \Phi_{k+1},  \Phi_k, \phi)$  at the minimizers  is     
\begin{equation}  S^0_{k+1}(\Phi_{k+1},\phi^0_{k+1})=
\frac12   \frac{ a_{k+1}}{L^2}
\|\Phi_{k+1}-Q_{k+1}  \phi^0_{k+1}\|^2     +  \frac12     \| \pa \phi^0_{k+1}\|^2  + \frac12   \bmu_{k}  \|  \phi^0_{k+1}\|^2
\end{equation}
\end{enumerate}
\end{lem}
\bigskip

\pr  
  The  variational equations  for the minimizer in    $\Phi_k, \phi$ are 
\begin{equation}  \label{cherry0}
 \begin{split}
 \Big(a_k  +  \frac{a}{L^2}    Q^TQ  \big)\Phi_k   =  &  a_k Q_{k} \phi    + \frac{a}{L^2} Q^T  \Phi_{k+1}   \\
( - \De  + \bmu_k  + a_k  Q_{k} ^T Q_{k} ) \phi 
=&  a_k  Q_{k} ^T \Phi_k     \\
\end{split}
\end{equation}
both of which we have seen before.
The  first  is solved by   $\Phi_k  =   \Psi_k=  \Psi_k(\Phi_{k+1}, \phi)$. 
Substituting this into the second   and solving  for    $\phi$  we  find  that
the minimizer  comes at     $\phi = a_k G_kQ_k^T \Psi_k  \equiv  \phi_k(\Psi_k)$.

We  further analyze the second equation at   $\Phi_k  =   \Psi_k$. 
Using      (\ref{ak})   the right side is evaluated as
\begin{equation}  \label{stinky}
a_kQ_k^T    \Psi_k  = a_k Q^T_k Q_{k}  \phi  
- a_{k+1}L^{-2}Q^T_{k+1} Q_{k+1} \phi  + a_{k+1}L^{-2}Q_{k+1}^T  \Phi_{k+1} 
\end{equation}
 The  second    equation then becomes
 \begin{equation}
( - \De  + \bmu_k  +   a_{k+1}L^{-2}Q^T_{k+1}  Q_{k+1} )  \phi  
=   a_{k+1}L^{-2}  Q_{k+1}^T \Phi_{k+1} 
\end{equation}
which  has the solution    $ \phi=    \phi^0_{k+1}  =     L^{-2} a_{k+1}  G_k^0 Q_{k+1}^T  \Phi_{k+1}  $.
This establishes  (\ref{hunger}),  (\ref{kingmaker}),  and (\ref{someday}).

The   value at the minimum  is  
\begin{equation}
\begin{split}
J( \Phi_{k+1},  \Psi_k,  \phi^0_{k+1}  )   =&    \frac{a}{2L^2} \|\Phi_{k+1}-Q    \Psi_k\|^2 +   
\frac12    a_k
\| \Psi_k-Q_k  \phi^0_{k+1}\|^2     +  \frac12     \| \pa \phi^0_{k+1}\|^2  + \frac12   \bmu_k  \|  \phi^0_{k+1}\|^2\\
   = &  \frac{a_{k+1}}{2L^2}   \| \Phi_{k+1}  - Q_{k+1}  \phi^0_{k+1}  \|^2    + \frac12   \| \pa \phi^0_{k+1}\|^2
     + \frac12   \bmu_{k}  \|  \phi^0_{k+1}\|^2\\
\end{split}
\end{equation}
Here we used  (\ref{fifty}).
This completes the proof.

\begin{lem}   \label{tiny}
 Let   $\rho_{k} ( \Phi_k)  
 =  Z_k\    \exp (-S_k(\Phi_k, \phi_k) ) $.  Then  
  $\tilde  \rho_{k+1} $  is given by   
 \begin{equation}    \label{understand}
  \tilde  \rho_{k+1} ( \Phi_{k+1})    =   Z_k \ \cN^{-1}_{aL ,  \bbT^1_{\sN + \sM-k}}   (2 \pi)^{| \bbT^0_{\sM +\sN-k}|/2}( \det  C_k  )^{1/2} \exp\Big(  -   S^0_{k+1}(\Phi_{k+1},  \phi^0_{k+1}) \Big)
 \end{equation}
 where   
\begin{equation}
C_k   =  \Big(\De_{k} +  \frac{a}{L^2} Q^TQ  \Big)^{-1}
\end{equation}
\end{lem}
\bigskip

 \pr  
 We  calculate $\tilde  \rho_{k+1}$   given in  (\ref{manx})  by expanding around the minimum  is  $\Phi_k$.
We   we  write   $\Phi_k = \Psi_k +Z$
and integrate over  $Z:     \bbT^0_{\sM + \sN-k} \to  \bbR$ instead of  $\Phi_k$.    We  have   from  (\ref{someday})
\begin{equation}  \label{undo}
\phi_k(\Psi_k +Z) =   \phi_{k+1}^0  +   \cZ_k 
\end{equation}
where  $\cZ_k:   \bbT^{-k}_{\sM + \sN -k}  \to  \bbR$ is defined by  
\begin{equation}
\cZ_k  =  \phi_k(Z)  =   a_{k}  G_k Q_{k}^T Z  
\end{equation}
Thus we  are expanding     $ J( \Phi_{k+1},  \Phi_k,  \phi_k  )$   around the minimum in the last two variables.   We claim  that   
\begin{equation}    \label{eighty}
\begin{split}
J( \Phi_{k+1},    \Psi_k+Z,   \phi_{k+1}^0  +   \cZ_k )   =  &   S^0_{k+1}(\Phi_{k+1},  \phi^0_{k+1})  +        \frac{a}{2L^2}\|QZ\|^2  + S_k(Z, \cZ_k)
 \\
 =  &   S^0_{k+1}(\Phi_{k+1},  \phi^0_{k+1})  +       
    \frac12 \Big<Z,  \Big(\De_{k} +  \frac{a}{L^2} Q^TQ  \Big)  Z  \Big>  \\
    \end{split}
\end{equation}
Indeed we have already seen that  $  S^0_{k+1}(\Phi_{k+1}, \phi^0_{k+1})$  is the minimum value.  The linear 
terms in $Z, \cZ_k$   must  vanish.  The quadratic  terms   in  $Z, \cZ_k$  are as indicated.   The  second form follows from
(\ref{spiffy}). 

 Inserting this last expression into  (\ref{manx})  yields  
\begin{equation}   \label{manx2}
\begin{split}
 \tilde  \rho_{k+1} ( \Phi_{k+1})    = & \cN^{-1}_{aL ,  \bbT^1_{\sN + \sM-k}} Z_k
 \exp(  -    S^0_{k+1}(\Phi_{k+1},  \phi^0_{k+1}))  \\
&  \int     \exp \left(   - \frac12 \Big<Z, (\De_{k} +  \frac{a}{L^2} Q^TQ )  Z  \Big> \right)  dZ \\
\end{split}
\end{equation}
We evaluate the last integral as  $ (2 \pi)^{| \bbT^0_{\sM +\sN-k}|/2}( \det  C_k  )^{1/2}$   which gives the result.

 \begin{lem} (scaling)   With  $\tilde \rho_{k+1}$ given by  (\ref{understand}),   the scaled density  $\rho_{k+1}$   as  defined by  (\ref{scaleddensity})  is  for   $\Phi_{k+1}:  \bbT^{0}_{\sM +\sN -k-1} \to  \bbR$      \begin{equation}
\rho_{k+1}( \Phi_{k+1} ) =    Z_{k+1}    \exp \Big(  -   S_{k+1}(\Phi_{k+1},  \phi_{k+1}) \Big)
\end{equation}
Furthermore   
 \begin{equation}
 \phi^0_{k+1}( \Phi_{k+1,L})  =  [\phi_{k+1}(\Phi_{k+1})]_L
\end{equation}  
 and
 \begin{equation}   \label{Ziterate}
Z_{k+1}  =   Z_k \ \cN^{-1}_{a ,  \bbT^1_{\sM + \sN-k}}   (2 \pi)^{| \bbT^0_{\sM +\sN-k}|/2}( \det  C_k  )^{1/2}
\end{equation}
\end{lem}
\bigskip

 \pr   We  scale by   $f_L(x)  =  L^{-1/2}f(x/L)$.  The averaging operator $Q$ is scale invariant and
 $ \bmu_k =  L^{-2}  \bmu_{k+1}$ so we  compute
 \begin{equation}
\Big(   - \De  + \bmu_k  +   a_{k+1}L^{-2}Q^T_{k+1}  Q_{k+1} \Big) f_L
=    L^{-2}    \Big[ \big(   - \De  + \bmu_{k+1}  + a_{k+1} Q^T_{k+1}  Q_{k+1} \big) f  \Big]_L
 \end{equation}
 It follows that   the inverses  satisfy $ G_{k+1}^0  f_L   =   L^2 [ G_{k+1}   f  ]_L $
 and so 
 \begin{equation}
\begin{split}
 \phi^0_{k+1} ( \Phi_{k+1,L})  = &  L^{-2}  a_{k+1}  G^0_{k+1} Q_{k+1}^T  \Phi_{k+1,L}  
   =   [ a_{k+1}  G_{k+1} Q_{k+1}^T  \Phi_{k+1} ]_L
    =  [\phi_{k+1}(\Phi_{k+1})]_L  \\
\end{split}
\end{equation}

 Now in       $\rho_{k+1}(\Phi_{k+1})$  we  have    
  \begin{equation}  \label{71}
\begin{split}
 S^0_{k+1}(\Phi_{k+1,L}, \phi_{k+1,L})= &
\frac12   \frac{ a_{k+1}}{L^2}
\|\Phi_{k+1,L}-Q_{k+1}  \phi_{k+1,L}\|^2     +  \frac12     \| \pa \phi_{k+1,L}\|^2  + \frac12   \bmu_{k}  \|  \phi_{k+1,L}\|^2\\
  =&
\frac12   a_{k+1}
\|\Phi_{k+1}-Q_{k+1}  \phi_{k+1}\|^2     +  \frac12     \| \pa \phi_{k+1}\|^2  + \frac12   \bmu_{k+1}  \|  \phi_{k+1}\|^2\\
=&   S_{k+1}(  \Phi_{k+1}, \phi_{k+1}  )   \\
\end{split}
\end{equation}
Thus     
   \begin{equation}
\rho_{k+1}( \Phi_{k+1} )=    Z_k \ \cN^{-1}_{a ,  \bbT^1_{\sN + \sM-k}}   (2 \pi)^{| \bbT^0_{\sM +\sN-k}|/2}( \det  C_k  )^{1/2}
   \exp \Big(  -   S_{k+1}(\Phi_{k+1},  \phi_{k+1}) \Big)
\end{equation}
and the constant is identified  as   $Z_{k+1}$.

\subsection{random walk expansion}     \label{randomwalk}

We  develop   a random  walk expansion for the   the Green's function $G_k  = ( -\De + \bmu_k  +  a_k  Q_k^T  Q_k)^{-1}$ on     $\bbT^{-k}_{\sM +\sN-k}$.   This will gives  us  estimates  on $G_k$   and also  provide the basis of 
localized approximations to $G_k$.

The random walk expansion is based on localized inverses which we now define.
Let  $M=  L^{m} $ for some positive integer $m$,  
and    let  $\square_z$  be  a large    cube  in  $\bbT^{-k}_{\sM + \sN -k}$   of linear size   $M$  
   centered on  points  $ z \in \bbT^{m}_{\sM + \sN -k}$.  (Warning:  $M$ is not the same as the volume parameter  $\sM$.) The  centers  are
a distance   $M$  apart  so the  $\square_z$  partition the lattice.      Also  let  $\tilde  \square_z$  be the union of all   $M$-cubes 
touching $\square_z$.      The  $\tilde  \square_z$  are   
overlapping    cubes  in  $\bbT^{-k}_{\sM + \sN -k}$   of linear size   $3M$    still   
   centered on  points  $ z \in \bbT^{m}_{\sM + \sN -k}$.          For  $\tilde   \square  = \tilde   \square_z$
let     $\De_{\tilde  \square} $   be  the  Laplacian   on   $\tilde   \square $ with Neumann boundary 
conditions.  This means that   in  
$
 < \phi,   - \De_{\tilde  \square}  \phi>  =  \sum_{ \mu} \int  |  \pa_{\mu}   \phi(x)  |^2 dx  
$
only  terms  with both  $x,  x+ L^{-k} e_{\mu}  \in \tilde     \square$  contribute.
Now    restrict  the operator    $ -\De + \bmu_k  +  a_k  Q_k^T  Q_k$    to $\tilde    \square$  with the Neumann conditions  and  take the inverse 
defining  
\begin{equation}
G_k( \tilde   \square)   =    \big[ -\De  + \bmu_k  +  a_k  Q_k^T  Q_k \big]_{\tilde   \square}^{-1}
\end{equation}

We  give  some estimates on these operators.  Here and throughout the paper we employ the convention that  $C$  stands
for a constant  depending on $L$,  but  no other parameters.  It may change  from line to  line.  Also  $\one$ stands for a constant
independent of all parameters.

   Let    $\De_y = B_k(y)$  be  unit  cubes in  $\bbT^{-k}_{\sM + \sN -k}$  centered on  points
  $y  \in   \bbT^0_{\sM + \sN -k}$.  These partition the lattice  and  any  large  cube  $\square$ or  $\tilde  \square$.
Let  $   y,y'  \in  \tilde   \square$   and  let     $x \in   \De_y \subset  \tilde \square$  and    $\supp f  \subset  \De_{y'} \subset  \tilde \square$.  Then  for some constants  $C$   and   $\ga_0 = \cO(L^{-2})$
\begin{equation}  \label{sycamore}
\begin{split}
|( G_k(\tilde   \square)f )(x) |  \leq &  C    e^{  -   \ga_0  d(y,y') } \|f\|_{\infty}\\
|( \pa G_k(\tilde   \square)f)(x)  |  \leq &  C      e^{  -   \ga_0  d(y,y') } \|f\|_{\infty}\\
\end{split}
 \end{equation}
 We  also estimate the Holder derivative of  $\pa G_k(\tilde   \square)$  of   order  $1/2 < \al < 1$.
 For      $x,x'  \in \De(y)$  and   $\supp f  \subset  \De(y')$
 \begin{equation}   \label{maple}
| (\de_{\al }\pa G_k(\tilde   \square)f)(x,x')|  \leq   C    e^{  -   \ga_0  d(y,y') } \|f\|_{\infty}  
 \end{equation}
Here     $\de_{\al}$ is defined  for  $d(x,x')  \leq   1$  by  
  \begin{equation}
 (\de_{\al }f)(x,x')  =
\frac{ f(x)  - f(x') }
{d(x,x')^{\al}  }
  \end{equation}
  and  $C$ does depend on $\al$ in  (\ref{maple}).
 We  give proofs of these estimates in   Appendix  \ref{D}.    References are     \cite{Bal83b},  \cite{Bal84b},  \cite{Bal96b}.

A  random   walk or path   is a sequence of   points  in   the lattice  $\bbT^{m}_{\sM + \sN -k}$ with spacing $M = L^{m}$ written
\begin{equation}
\om=    (  \om_0,  \om_1,  \dots,   \om_n  )
\end{equation}
such  that  $\om_j,  \om_{j+1}$  are neighbors in the sense  that  for each component 
  $|\om_{j,\mu} - \om_{j+1, \mu}| \leq   M$.    Thus  $\om_j$ has  $3^d = 9$  neighbors counting itself.
The  number of steps in the walk is  $|\om| =n$.

\begin{lem}   \label{jupiter}
The Green's function  $G_k$ has  a random  walk expansion of the form  
\begin{equation}  \label{stinger0}
G_k  =  \sum_{\om} G_{k, \om}
\end{equation}
where the sum is over  all paths  $\om$. 
 If  $M$ is sufficiently  large   the series for  $G_k,  \pa G_k,  \de_{\al} \pa G_k$   all   converge  and  give      for   $x,x'   \in  \De_y$   and      $\supp f  \subset  \De_{y'}$
  \begin{equation}  \label{gk}
\begin{split}
|( G_{k}f) (x) |  \leq &  C e^{ -\frac 12 \ga_0  d(y,y')  } \|f\|_{\infty}\\
|( \pa  G_{k}f )(x)   |   \leq &   C e^{ - \frac 12  \ga_0  d(y,y')  } \|f\|_{\infty}\\
| (\de_{\al }\pa G_{k}f)(x,x')|  \leq  &   C  e^{  - \frac 12\ga_0  d(y,y')  } \|f\|_{\infty}\\
\end{split}
 \end{equation}
 \end{lem}
\bigskip

\pr  
We  construct a partition of unity.     Let   $h \in  \cC_0^{\infty}(-2/3,2/3)$  satisfy $h \geq 0$  and    $h=1$  on  $(-1/3,1/3)$
and  
\begin{equation}
1=  \sum_{n  \in \bbZ}    h^2(x -n)  
\end{equation}
 Then   for   $z  \in  \bbT^{m}_{N+M -k}$   define $h_z$  on   $ \bbT^{-k}_{N+M -k}$  by  
 \begin{equation}
 h_z(x)   =   \prod_{\mu  =1}^d  h \left(  \frac  {x_{\mu} -  z_{\mu}}{M}  \right)
\end{equation}
Then   $h_z$  has   support   in   $\tilde   \square_z$,  and in fact in the smaller set   $\{x:  |x_{\mu} -  z_{\mu}| \leq  \frac23  M \}$.
  Hence      $h_z h_{z'} =0$ unless  $z,z'$   are neighbors.
We  have  
\begin{equation}
1 =  \sum_z    h_z^2(x) 
\end{equation}
  Furthermore
\begin{equation}  \label{hbound}
| \pa  h  |  \leq  \cO(1)M^{-1}     \hs   | \pa \pa  h |  \leq    \cO(1) M^{-2}
\end{equation}

Define   a parametrix  $G^*_k$  by  
\begin{equation}
G^*_k  =  \sum_{z}   h_z  G_k(\tilde  \square_z) h_z
\end{equation}
Then we have 
\begin{equation}
(- \De  + \bmu_k+  a_k Q_k^T Q_k)  G^*_k
=     I   -     \sum_{z}   K_z  G(\tilde  \square_z)  h_z
\equiv  I -K
\end{equation}
where  
\begin{equation}
 K_z = - \left[ ( -\De +  a_k Q_{k}^T     Q_{k}), h_z  \right]    
\end{equation}
The  solution is now
\begin{equation}
G_k  = G^*_k(I-K)^{-1}  = G^*_k \sum_{n=0}^{\infty}  K^n
\end{equation}
provided the series converges.
This can also be written as the random walk expansion
\begin{equation}  \label{brand}
\begin{split}
G_k =
&
 \sum_{n=0}^{\infty}  \sum_{\om_0, \om_1,...,\om_n}
\Big( h_{\om_0} G_k(\tilde  \square_{\om_0})   h_{\om_0}\Big)
\Big(K_{\om_1}    G_k(\tilde  \square_{\om_1})   h_{\om_1}  \Big )
\cdots     \Big (K_{\om_n}    G_k(\tilde  \square_{\om_n})    h_{\om_n}\Big)
\\
\equiv & \sum_{\om}   G_{k, \om}  \\
\end{split}
\end{equation}

Now  we  claim that    for  $x  \in   \De_y  \subset  \tilde  \square_z$ 
 \begin{equation}
|(K_z f)(x) |   \leq     \cO(1) M^{-1}  \Big (  \|1_{\De_y}f\|_{\infty}   +  \|1_{\De_y}  \pa  f  \|_{\infty} \Big )
\end{equation}
Indeed  the    term  $ \left[  -\De, h_z\right]$  is local  and involves derivatives of  $h$
  so   we  get  the  factor   $M^{-1}$  
from   (\ref{hbound}).    The   term  $\left[   Q_{k}^T     Q_{k}, h_z\right]$    is also  local and  also    
can be  expressed in term  of derivatives  of  $h_z$  since it  can be written
\begin{equation}
\Big(  \left[   Q_{k}^T     Q_{k}, h_z\right] f  \Big)  (x)   =    \int_{x' \in  \De_y} ( h_z(x') - h_z(x))f(x') dx'
\end{equation}
Combining   the bound on $K_z$     with the basic  bound  (\ref{sycamore})    on  $G_k(\tilde  \square)$ 
yields  for $x  \in   \De_y \subset  \tilde  \square_z$  and   $\supp f   \subset  \De_{y'} \subset  \tilde  \square_z$: 
 \begin{equation}
 \begin{split}
|(K_z G_k(\tilde  \square_z)f)(x) |  
\leq    &  C M^{-1}  \Big (  \|1_{\De_y} G_k(\tilde  \square_z)f\|_{\infty}   +  \|1_{\De_y}  \pa   G_k(\tilde  \square_z)f  \|_{\infty}  \Big)  \\
 \leq  &   C M^{-1}   \exp  (  - \ga_0  d(y,y')  ) \|f\|_{\infty} \\
 \end{split}
\end{equation}
(Only interior derivatives of   $G_k(\tilde  \square_z)$  appear  since $\supp\  h_z$ is well-inside  $\tilde \square$.)
We  use   this bound repeatedly  on  $G_{k, \om}$   with  $|\om| = n$.     We have for  
 $x  \in   \De_y$  and   $\supp f   \subset  \De_{y'}$   with  $y_0 = y,  y_{n+1} = y'$
 \begin{equation}
\begin{split}
&|( G_{k, \om}f )(x)| \\
 =  & \left|\Big(  ( h_{\om_0} G_k(\tilde  \square_{\om_0})    h_{\om_0}\Big) 
\Big(K_{\om_1}    G_k(\tilde  \square_{\om_1})   h_{\om_1}  \Big )
\cdots     \Big (K_{\om_n}    G_k(\tilde  \square_{\om_n})    h_{\om_n}\Big)  f \Big)(x)  \right|\\   
 \leq    &   \sum_{y_1,\dots,  y_n}  \left|   \Big(  ( h_{\om_0} G_k(\tilde  \square_{\om_0})  h_{\om_0}\Big)1_{\De_{y_1}}  
\Big(K_{\om_1}    G_k(\tilde  \square_{\om_1})   h_{\om_1}  \Big ) 1_{\De_{y_2}} 
\cdots   1_{\De_{y_n}}   \Big (K_{\om_n}    G_k(\tilde  \square_{\om_n})    h_{\om_n}\Big)    f \Big)(x) \right|\\
 \leq  &   C(C M)^{-n}   \sum_{y_1,\dots,  y_n}  \prod_{j=0}^n    e^{ - \ga_0 d(y_j,y_{j+1}) } \|f\|_{\infty}\\
\leq   &  C  (C M)^{-n}       e^{  - \frac12\ga_0  d(y,y') } \|f\|_{\infty}\\
\end{split}
\end{equation}
\bigskip

For convergence of the random walk expansion    we have     for  $M$  sufficiently large
 \begin{equation}  \label{sonic}
\begin{split}
|( G_{k}f) (x) |   \leq &    
\sum_{\om} |( G_{k, \om}f ) (x) | \\
\leq &  \sum_{n=0}^{\infty}  \sum_{\om:  |\om|  =n} C  \Big(C M^{-1}\Big)^{n}       e^{  - \frac12\ga_0  d(y,y') } \|f\|_{\infty} \\
\leq &  \sum_{n=0}^{\infty}  C \Big(C M^{-1}\Big)^{n}   (3d)^n      e^{  - \frac12\ga_0  d(y,y') } \|f\|_{\infty} \\
 \leq&   C     e^{  - \frac12\ga_0  d(y,y') } \|f\|_{\infty} \\
\end{split}
 \end{equation}
    This establishes  the bound  on   $G_{k}$  and the bounds
on  $\pa  G_{k}$  and  $\de_{\al}\pa G_{k}$   are   similar.  This completes the proof.
\bigskip

The   bounds of the lemma  yield (more elementary) global estimates:

\begin{cor}
For any   $f:  \bbT^{-k}_{\sM + \sN -k} \to  \bbR$  
  \begin{equation}  \label{gk2}
\begin{split}
| G_{k}f |,\  |\pa  G_{k}f   |,\  | \de_{\al }\pa G_{k}f| \ \leq &\  C \|f\|_{\infty}\\
\end{split}
 \end{equation}
\end{cor}
\bigskip

\pr   For  $x  \in \De_y$
\begin{equation}
| (G_{k}f )(x)|  \leq      \sum_{y'}    | (G_{k}1_{ \De_{y'}}f )(x)|   \leq       C \sum_{y'}   e^{  -\frac 12 \ga_0  d(y,y')  } \|f\|_{\infty}
\leq     C  \| f\|_{\infty}
\end{equation}
The others are similar.

\subsection{decoupling}  \label{decoupling}
We  also  we   need a version  of  $G_k$  in which the communication between sites   is systematically   weakened.

For each  $M$-cube  $\square$     introduce a variable   $ s_{  \square }$ with  $ 0 \leq s_{  \square}  \leq 1$. 
Then  define  for   $\om  =  (  \om_0, \om_1, \dots,  \om_n)$   
\begin{equation}
s_{\om}   =      \prod_{    \square  \subset   X_{\om}}
  s_{  \square}   \hs    X_{\om}   =   \bigcup_{j=1}^n  \tilde   \square_{\om_j}  
\end{equation}
Note that   $\tilde      \square_{\om_0}$ is omitted  from  $X_{\om}$.
Hence if $\om$ is only a single point  $\om_0$  (i.e. $|\om|=0$) then  $X_{\om}$  is   
empty and in this case  we set   $s_{\om} =1$.

Now we define for  $  s  =  \{ s_{\square} \}$
\begin{equation}  \label{rws}
\begin{split}
G_{k}(s) = &   \sum_{\om} s_{\om}  G_{k,\om}\\
\end{split}
\end{equation}
Then we have  
\begin{equation}
\begin{split}
G_k(1)   =    \sum_{\om} G_{k,\om}  =  G_k\\
G_k(0)   =    \sum_{\om:|\om|=1} G_{k,\om}  =  G^*_k\\
\end{split}
\end{equation}
Thus  $G_k(s)$  interpolates  between an  operator   for which   all sites  are coupled and
an operator for which keeps thing localized in  each  cube  $\tilde   \square$.

Note that   $G_k(s)$    can  be  defined   and bounded    for   $s_{\square}$   complex   and in     a  much large domain.  
We  can    take  for example  $|s_{\square}|  \leq   M^{1/2}$.   Then  in    (\ref{sonic}) instead of    $(C M^{-1})^n$
we  have   $(C|s_{\square}| M^{-1})^n  \leq       (C M^{-1/2})^n$. The   random walk  expansion still converges
if  $M$ is sufficiently  large.
The bounds  (\ref{gk})   and  (\ref{gk2})  still hold for       $G_k(s)$  with    $|s_{\square}|  \leq   M^{1/2}$.

\section{Localized functionals of the field}

\subsection{overview}  \label{overview}

Our main goal is to  follow the flow of the renormalization group  with the potential  included. 
The  detailed analysis is in the next section.   Here  we do some preliminary work.

After    $k$  steps     with   certain  small field assumptions  we will find that the density   can be    written  in  a the following form .
\begin{equation}  
 \rho_k( \Phi_k)  
= \textrm{const}  \exp \left( - S_k(\Phi_k, \phi_k)  + \vep_k  \Vol( \bbT_{\sM + \sN -k}) -\frac12  \mu_k  \int \phi_k^2   - \frac14    \la_k  \int  \phi_k^4   + E_k(\phi_k)    \right) 
\end{equation}
Here  $ \phi_k = a_k  G_k  Q^T_k  \Phi_k$  as before    and  $S_k( \Phi_k,  \phi_k)$  is the free action as  defined in  (\ref{norton}).
In the next terms  we  track the growth of      energy density $\vep_k$   and the     mass-squared   $\mu_k$.  We   could do this
as  well  with the coupling constant,  but for this model it is sufficient to  just  let  it scale and define
\begin{equation}
\la_k   =     L^k  \la_0  =   L^{-(N-k)}  \la
\end{equation}
We   will  only  be interested  in  $N,k, \la$  such that    $\la_k$  is small.

The term $E_k$  is real-valued and   contains all     non-leading  and  non-local corrections to the simple local form we have isolated.
However  it has  some local structure which we now explain. 
Again  consider    cubes  $\square$ with  side of length $M=L^{m}$    centered  on points  of 
the  lattice   $ \bbT^{m}_{\sM + \sN  -k}$  which    partition the lattice  $\bbT^{-k}_{\sM + \sN -k}$.  An  $M$-\textit{polymer}  $X$  is  a    connected unions  of such cubes.    Here \textit{connected}  means that  for any   two cubes   $\square,  \square'$   in $X$
 there is a sequence   $\square= \square_0,  \square_1,  \square_2, \dots,  \square_m = \square'$
 such that   $\square_j \subset  X$  and   $\square_j$ and $\square_{j+1}$   have  a   $d-1 =2$  dimensional face in common.

 Now we  define 
 \begin{equation}      
\cD_k =   \textrm{ all } \textrm{ $M$-polymers  }   X   \textrm{  in  }   \bbT^{-k}_{\sM + \sN -k}
\end{equation}
We  will assume       the local structure
\begin{equation}
E_k(\phi_k)  =   \sum_{X  \in \cD_k}   E_k(X,  \phi_k)
\end{equation}
where  $E_k(X,  \phi_k) $   only depend on the restriction of  $\phi_k$  to $X$.

\subsection{small fields}

Our    small field assumption is chosen  so that when it is  violated  either    the  term   
$\exp \left( - S_k(\Phi_k, \phi_k)   \right)$   or  the term   $\exp \left( -  \frac14    \la_k  \int  \phi_k^4    \right)$
 in the density is tiny. 
This is arranged as  follows.
Let    
\begin{equation}
p_k  =   p(\la_k)   =  (-   \log   \la_k  )^{p}   = ((N-k) \log L     - \log   \la  )^p
 \end{equation}
 for some positive integer  $p$.
 We   assume  always   $\la_k<1$  so    the quantity we are 
 exponentiating is positive.

 \begin{defn}   $\cS_k$  is all      functions   $\Phi_k:   \bbT^0_{M+N-k}  \to  \bbR$   such that  with  $  \phi_k = a_k  G_k  Q^T_k  \Phi_k $
on     $\bbT^{-k}_{N+M-k}$
\begin{equation}  \label{sk}
\begin{split}
 | \Phi_k  -  Q_{k}    \phi_k|  \leq &  p_k \\
 |\pa   \phi_{k}|  \leq & p_k    \\   
 |  \phi_{k}|  \leq &  \la_k^{-1/4} p_k  \\
\end{split}
\end{equation}
\end{defn}
Since  $\la_k$ is assumed small,  $p_k$ is large,  and   these    "small field" conditions actually allow rather large fields.
If  one of these  conditions   fails    then  we   gain a tiny   factor    $\cO(e^{-p_k})$.
\bigskip

In  fact    $E_k(X,  \phi_k) $   will  be the  restriction of more  general  complex-valued   functions       $E_k(X,  \phi) $   
defined  for  complex fields    $\phi:    \bbT^{-k}_{\sM + \sN -k}  \to  \bbC$.  We  want to 
choose   weaker  restrictions  on  $\phi$    so that  if  $\phi= \phi_k  \in  \cS_k$  then the new conditions are
satisfied.    We   would also like   bounds   on  $\phi,  \pa \phi,  \de_{\al} \pa \phi$  to be  all about the same size,
however it is convenient to  allow a little deviation.   This motivates the following definition:

\begin{defn}  Let  $\ep$  be a fixed small positive number.
$\cR_k $   is  all functions  $  \phi:   \bbT^{-k}_{N+M-k}    \to  \bbC$ such  that: 
\begin{equation}
\begin{split}
&  |\phi|  <  \la_k^{-1/4-3 \ep}  \\
& | \pa \phi|  <    \la_k^{-1/4- 2\ep}  \\
 & |\de_{\al}\pa \phi|    <    \la_k^{-1/4- \ep}  \\
\end{split}
\end{equation}
\end{defn}
This does the job for we have:

\begin{lem}  \label{strong}  Let    $\Phi_k  \in  \cS_k$.  Then  
\begin{enumerate}
\item   
$|\Phi_k |   \leq     2 p_k   \la_k^{-1/4}$   and      $|\pa_{\mu}  \Phi_k  | \leq   3p_k$.
\item   For   $\la_k$  sufficiently small    $ \phi_k= a_k  G_k  Q^T_k  \Phi_k  \in   \cR_k$.
\end{enumerate}
\end{lem}
\bigskip

\pr   For the first point  we have  
   \begin{equation}  
   |\Phi_k |  \leq   | \Phi_k  -  Q_{k}    \phi_{k}| 
   +      |   Q_k    \phi_{k}|  \leq    p_k    +     p_k\la_k^{-1/4}  \leq     2  p_k\la_k^{-1/4} 
\end{equation}
and also  
\begin{equation} 
\begin{split}
|(\pa_{\mu}  \Phi_k )(x) |  =    &  | ( \Phi_k )(x+ e_{\mu})- ( \Phi_k )(x)|  \\
\leq    & |Q_{k}    \phi_{k}(x+ e_{\mu} ) 
-    Q_{k}    \phi_{k}(x )|  + 2 p_k  \\
\leq  &    \| \pa   \phi_{k}\|_{\infty}
+ 2p_k   \leq   3p_k \\
\end{split}
\end{equation}
For the second  point  if  $\la_k$ is small
we  have      have  $p_k \leq  \la_k^{-\ep}$  since  
\begin{equation}
p_k  =   ( - \log \la_k )^p  \leq  p! \left(\frac{2}{\ep}\right)^p  e^{ \frac 12  \ep  (  - \log \la_k )}  
 =  p! \left(\frac{2}{\ep}\right)^p   \la_k^{-\ep/2}  \leq    \la_k^{-\ep} 
\end{equation} 
The    bounds  on  $\phi_k, \pa \phi_k$ follow  directly.
   Furthermore   by  (\ref{gk2})   and   $\| Q_k^T  \Phi_k\|_{\infty}   \leq   \| \Phi_k\|_{\infty}$   and    $p_k \leq  \cO(1) \la_k^{-\ep/2}$
\begin{equation}
\begin{split}
| \de_{\al} \pa  \phi_k| =   |a_k   \de_{\al} \pa  G_k  Q_k^T  \Phi_k|   \leq    C  \| \Phi_k\|_{\infty}
\leq   Cp_k  \la_k^{-1/4}   \leq      \la_k^{-1/4- \ep}  
\end{split}
\end{equation}   
  This completes the proof.

\subsection{norms}

The functions  $E(X, \phi)$  form a  complex  vector space.   We  add a few more conditions and define
a subspace:  
\begin{defn}
$\cK_k $ =  all $E:  \cD_k  \times  \cR_k \to  \bbC$  such  that  
\begin{equation}  \label{conditions}
\begin{split}
&  (a.)\ E(X, \phi) \textrm{ only  depends on }   \phi  \textrm{  in }  X  \in \cD_k \\
&  (b.)\   E(X, \phi)  \textrm{   is analytic and bounded in }  \phi \in \cR_k.   \\
& (c.)\     E(X, \phi)    \textrm{ is even in  }  \phi  \\
&  (d.)\   E(X, \phi)  \textrm{ is   invariant under  lattice symmetries  (translations, rotations by   $\pi/2$, reflections). } \\
\end{split}
\end{equation}
\end{defn}
In  fact we  are mainly interested in the real   subspace
\begin{equation}
\textrm{Re} ( \cK_k )   =  \{ E \in  \cK_k :  \overline{ E( \phi) }
 =   E(\overline \phi )  \}
\end{equation}
  Elements  $E(X, \phi)$ of this space 
are real  for real fields  $\phi$.

We      introduce   a  norm  on  these   spaces.    For  the $\phi$   dependence we  define for each $X \in \cD_k$:
\begin{equation} 
 \|  E(X)  \| _k   =   \sup_{\phi  \in  \cR_k}   \| E(X,  \phi)  \|
\end{equation}
We also need to describe   how   $E(X)$     decays  in $X$.  For any $X \in \cD_k$  define  $d_M(X)$  by:  
\begin{equation}   
  M d_M(X)  =  \textrm{   the length of the shortest     tree  in $X$  joining   the $ M$-cubes    in     $X$}  .
\end{equation}
Here the tree is  in  the  continuum  torus   $\bbT_{\sM + \sN -k}  = (\bbR/ (L^{\sM + \sN -k})\bbZ)^3 $.
We   expect   $E(X)$ to decay exponentially  in   $d_M(X)$  so we  define  our norm  by 
 \begin{equation}
\| E  \|_{k, \ka}  =\sup_{X}   \|  E(X)  \| _k   e^{  \ka  d_M(X)  }   
\end{equation}
The norm   depends  on  a parameter  $\ka > 0 $. 
With  any of these norms   the space  $\cK_k$  is complete  and  hence a  complex Banach space.
The space  $\textrm{Re} ( \cK_k )  $ is a real Banach space.

We   elaborate  a bit    on   the  these  definitions.   First  define  
\begin{equation}
 |X|_M   =    \Vol( X )/M^3 =  \textrm{   number of $M$-cubes in   $X$}.
\end{equation}
Then we have  the inequalities     \cite{Bal98b}
\begin{equation}  \label{ninety}
d_M(X)   \leq   |X|_M   \leq   3^d  (  1  +  d_M(X)   )
\end{equation}
here  with  $d=3$.

Secondly   it is a variation of a  standard bound   
 (see  appendix   \ref{basicsection})  that  there
are constants  $\ka_0\geq 1,  K_0$  depending only on the dimension, so that for any  $M$-cube $\square$
\begin{equation}  \label{summing0}
 \sum_{X \supset  \square}   e^{ - \ka_0  d_M(X)  }        \leq      K_0
\end{equation}
We  assume that  $\ka \geq  \ka_0$  (stronger conditions on $\ka$ later)  and then   for  $\phi \in \cR_k$
\begin{equation}  \label{summing}
   \sum_{X \supset  \square} | E(X, \phi) |   \leq    \sum_{X \supset  \square} \| E(X)  \| _k  
\leq  \| E \|_{k, \ka}   \sum_{X \supset  \square}   e^{ - \ka  d_M(X)  }   
 \leq    K_0    \| E \|_{k, \ka}  
\end{equation}

\subsection{scaling  and reblocking}

We   want to know how these   localized functionals  scale.   First  some definitions.

An   $LM$-polymer  $Y$ in $\bbT^{-k}_{\sM + \sN-k}$  is a connected  union  of   $LM= L^{m+1}$  cubes 
centered on the points   of    $\bbT^{m+1}_{\sM + \sN-k}$.   The set of  all  $LM$   polymers is 
denoted  $\cD^0_{k+1}$.   For  such  $Y$   we have that $L^{-1}Y$  is an $M$-polymer in   $\bbT^{-k-1}_{\sM + \sN-k-1}$
so  $L^{-1} \cD^0_{k+1} =  \cD_{k+1}$.     We  let  $\cK^0_{k+1}$  be  the space of all $F:  \cD^0_{k+1} \times  \cR_k \to  \bbC$
satisfying conditions like (\ref{conditions}).

Now  for    $F  \in \cK^0_{k+1}$    define the scaled down 
functional  
$F_{L^{-1}}  \in  \cK_{k+1}$  
by  \begin{equation}
      F_{L^{-1}}(X,  \phi ) =    F(LX,  \phi_L  )
\end{equation}
 This is well-defined since   $X  \in   \cD_{k+1}$   implies    $LX  \in  \cD_{k+1}^0$  and     $\phi  \in  \cR_{k+1}$
implies      $\phi_{L}  \in     \cR_k$ as the following lemma shows.

\begin{lem}  (scaling)  \label{scaling}
\begin{enumerate}   
\item  If      $\phi  \in  \cR_{k+1}$
then       $\phi_{L}  \in L^{-3/4- 3\ep} \cR_k $
\item  $
   \|  F_{L^{-1}}(X )\|_{k+1}   \leq     \| F(LX )  \|_k
$
\end{enumerate}
\end{lem}
\bigskip

\pr  The first  item follows from  $\la_{k+1}  =  L \la_k$  and 
 \begin{equation}  \label{bonnie}
\begin{split}
|\phi_L(x)|  =&  L^{-1/2}|\phi(x/L)|   \leq     L^{-1/2} \la_{k+1}^{-1/4- 3\ep} 
 =  [  L^{-3/4 - 3\ep}]   \la_{k}^{-1/4- 3\ep} \\
|\pa  \phi_L(x)|  =&  L^{-3/2}|\pa  \phi(x/L)|   \leq     L^{-3/2} \la_{k+1}^{-1/4- 2\ep} 
=  [  L^{-7/4 -2 \ep}]  \la_{k}^{-1/4-2 \ep}  \\
   | (\de_{\al}\pa  \phi)_L(x)|
  =&  L^{-3/2- \al} |\de_{\al}\pa  \phi(x/L)|  \leq     L^{-3/2-\al }\la_{k+1}^{-1/4- \ep} 
   =  [  L^{-7/4 -\al  -\ep}]  \la_{k}^{-1/4- \ep}  \\
\end{split}
\end{equation}
The  second is immediate.  This completes the proof.
\bigskip

To   prepare for scaling  we  need a reblocking operation.
If   $X \in \cD_k$   let    $ \bar  X \in \cD_{k+1}^0$  be the union of  all  $LM$  cubes  intersecting  $X$.
Given  $E  \in  \cK_k$       we   define    functionals  $ \cB E  \in \cK_{k+1}^0$    by  
\begin{equation}
  ( \cB E) (Y,  \phi)   =  \sum_{X \in \cD_k : \bar  X  =     Y }   E(X, \phi)
  \end{equation}  
  Then we  have
  \begin{equation}
  \sum_{X  \in \cD_k}  E(X,  \phi)     =  \sum_{Y  \in \cD^0_{k+1} }    ( \cB E) (Y, \phi)  
  \end{equation}

  \begin{lem}   (reblocking)  \label{citizen}  For  $\ka'  =  L(\ka - \ka_0-1)$
\begin{equation}  
 \|\cB E \|_{k, \ka'}   \leq     9 K_0  L^3    \|E\|_{k, \ka}  
\end{equation}
where the norm on the left is  defined with $d_{LM}$.
\end{lem}
\bigskip

\pr   If  $\bar  X   =   Y$  then  a  minimal  tree on   the  $M$  blocks  in $X$  is also  a tree on  the  $LM$ blocks 
in  $Y$   and so       $ M d_M(X)   \geq  LM d_{LM}(Y)$  or  
 $d_M(X)   \geq  L d_{M}(Y)$. 
Therefore
 \begin{equation}
\begin{split}  \label{study0}
 \| \cB E(Y)\|_{k}   \leq  &      \sum_{\bar  X  =     Y }    \|E(X ) \|_k 
 \leq     \|E\|_{k, \ka}    \sum_{\bar  X  =     Y }   e^{ - \ka  d_M(X)  }    \\
  \leq   &  \|E\|_{k, \ka}    e^{ -   (   \ka - \ka_0 )  L d_{M}(Y)  }    \sum_{\bar  X  =     Y  }   e^{ -   \ka_0  d_M(X)  }    \\
\end{split}
\end{equation} 
If   $\bar X =Y$ there must be an $M$-cube $\square$  so  $\square \subset  X  \subset Y$.   Using this  and   (\ref{summing0})  yields
 \begin{equation}   \label{slum}
    \sum_{X:\bar  X  =     Y  }   e^{ -    \ka_0  d_M(X)  }    
 \leq    
  \sum_{\square  \subset   Y}   \sum_{X \supset  \square }   e^{ -  \ka_0  d_M(X)  }   
  \leq   |Y|_M   \sum_{X \supset  \square }   e^{ -      \ka_0  d_M(X)  }   
  \leq     K_0  L^3
   |Y|_{LM}           
\end{equation} 
But      by  (\ref{ninety})  
$  |Y|_{LM}   \leq   9 (  1  +  d_{LM}(Y))  \leq    9 e^{   d_{LM}(Y)}$
so  we  have   
\begin{equation}    \label{study00}
 \|   \cB E(Y)\|_{k}   \leq  9 K_0  L^3 \|E\|_{k, \ka}      e^{ - ( L(\ka - \ka_0) -1)  d_{LM}(Y)  }   \leq  9 K_0  L^3 \|E\|_{k, \ka}      e^{ -\ka' d_{LM}(Y)  }  
\end{equation}   
This gives the result.
\bigskip

\re   If we combine them we have a map  $E  \to   (\cB E)_{L^{-1}}$  from  $\cK_k$   to  $\cK_{k+1}$.  
Since $d_{LM}(LX) =  d_M(X)$   we  have:  
\begin{equation}   \label{shotgun0}
\begin{split}
 \|(\cB E)_{L^{-1}} \|_{k+1, \ka'} =& \sup_{X \in \cD_{k+1}}    \leq   \|(\cB E)_{L^{-1}}(X) \|_{k+1} e^{\ka'  d_M(X)}
 \leq  \sup_{X \in \cD_{k+1}}    \|(\cB E)(LX) \|_{k} e^{\ka'  d_{LM}(LX)  }  \\
 = &   \sup_{Y \in \cD^0_{k+1}}    \|(\cB E)(Y) \|_{k} e^{\ka'  d_{LM}(Y)  }
=   \|\cB E \|_{k, \ka'}    \leq         9 K_0  L^3    \|E\|_{k, \ka}  \\
\end{split} 
\end{equation}
 Assuming  $\ka \leq  \ka'$ (a condition that  $\ka$ be large)   we have  
  \be    \label{shotgun}
   \|(\cB E)_{L^{-1}} \|_{k+1, \ka}    \leq         9 K_0  L^3    \|E\|_{k, \ka}  \\
\ee

\subsection{normalization}

The     previous estimate  has a      growth factor    $\cO(1)L^3$.   We   can cancel some or all of this 
if  we  remove  relevant terms from  $E$.  Such a  functional  will be  called normalized.   We   give  a definition  appropriate for     our model.
The following treatment roughly follows   \cite{BDH98}.

The functional  $E  \in \cK_k$     is said to be \textit{normalized}   if  
\footnote{$1$  means the function  $x  \to 1$  and  $x_{\mu}$   means the  projection   $x \to  x_{\mu}$}
\begin{equation}  \label{normalization}
\begin{split}
E(X,0)  =& 0 \\
E_2(X, 0;  1,1 ) = &0\\
E_2(X, 0;  1,x_{\mu} ) = &0\\
\end{split}
\end{equation}
Here  the derivatives can be evaluated  by
   \begin{equation}
   \begin{split}
     E_n(X, \phi;  f_1,  \dots ,f_n)   
     =  &  \frac{ \pa^n}  { \pa t_1 \cdots \pa t_n} E_n(X, \phi  + t_1 f_1  +  \cdots  +  t_nf_n) |_{t_i =0}\\
   \end{split} 
\end{equation}
Note  that  all odd derivatives at zero vanish.   This is  due to our assumption that  $E(X, \phi)$ is even in $\phi$.

It is convenient to  make a distinction  between  \textit{small polymers} $X$
which have  $d_M(X)  <  L$   and  \textit{large polymers}  which have  $d_M(X) \geq   L$.  
(A similar   distinction was first exploited in   \cite{BrYa90}).  We    generally    only require  normalization    for small polymers.  The set  of all small polymers is denoted $\cS$   and the large 
polymers are denoted $\bar  \cS$

Now  define   
\begin{equation}
\cK_k^{norm}  =  \{  E  \in \cK_k:     E(X, \phi)    \textrm{ is  normalized for small  } X   \}
\end{equation}
This is a closed subspace of  $\cK_k$.  We  also need  the real  closed  subspace
$\textrm{Re} (\cK_k^{norm})  =  \cK_k^{norm}  \cap  \textrm{Re}  (\cK_k )$.

Then  we have the following improvement of  (\ref{shotgun}).

\begin{lem}   
\label{scalinglem}  Let   $E \in  \cK^{norm}_k  $.   Then  for   $L$ sufficiently large and  
  $\la_k$  sufficiently small   (depending on  $L,M$)
\begin{equation}   \label{crude}  
  \|  ( \cB E)_{L^{-1}} \|_{k+1, \ka}  \leq         \cO( 1) L^{-\ep}      \|E\|_{k, \ka} 
\end{equation}
\end{lem}
\bigskip

\pr         
Let   $1_{\cS}, 1_{\bar \cS}$   be the characteristic functions of  small polymers and large polymers.  We  write
$ ( \cB E)_{L^{-1}} =  ( \cB1_{\bar \cS} E)_{L^{-1}}  +   ( \cB 1_{\cS} E)_{L^{-1}}$.

For the large set term  we  follow the proof of the previous lemma.   
In  (\ref{slum}) we can arrange to have  $\ka_0 +1$ instead of  $\ka_0$ (with a change in $\ka'$).  Then since    $ d_M(X)  \geq   L$ 
   \begin{equation}
  \sum_{X \in \bar \cS:  \bar  X  =     Y  }   e^{ -  ( \ka_0+1)  d_M(X)  }  
\leq   e^{- L}     \sum_{\bar  X  =     Y  }   e^{ -    \ka_0  d_M(X)  }  
\leq   e^{-L}  K_0L^3 |Y|_{LM}
\end{equation}
   Since    $  e^{-L} 9 K_0L^3    \leq  L^{-1}$ for $L$ sufficiently large,       we  get   instead of  (\ref{shotgun})
\begin{equation}   
  \|  ( \cB1_{\bar \cS} E)_{L^{-1}} \|_{k+1, \ka}  \leq        L^{-1}    \|E\|_{k, \ka} 
\end{equation}

 Now  consider the contribution of small  polymers which  is 
  \begin{equation}
  ( \cB1_{\cS} E)_{L^{-1}} (Z,  \phi)   =  \sum_{X \in \cS : \bar  X  =     LZ }   E(X, \phi_L)
  \end{equation}  
  We  will show that   for  $X \in \cS$ and $\phi \in  \cR_{k+1}$   we  have  
  \begin{equation}  \label{snort17}
  \|E(X,   \phi_L)\|   \leq   \cO(1) L^{-3-\ep}   \| E(X) \|_k
  \end{equation}
  Then  
  \be   \label{samsung3}
   \| (\cB1_{\cS} E)_{L^{-1}} (Z)\|_{k+1}   \leq    \cO(1) L^{-3-\ep}   \sum_{ \bar  X  =     LZ }    \| E(X) \|_k
  \ee
  This is the same as  (\ref{study0}),  except that  $Y=LZ$  and there is the extra factor   $L^{-3-\ep}$.
  Following the argument  (\ref{study0}) -(\ref{study00})    the factor  $L^3$ there is canceled.   Using also  
  $d_{LM}(LZ)= d_M(Z)$ and  $\ka< \ka'$ we have that  (\ref{samsung3}) is bounded by   $ \cO(1) L^{-\ep}      \| E(X) \|_{k, \ka}e^{-\ka d_M(Z)}$.
and therefore  instead  of  (\ref{shotgun})
\begin{equation}   
  \|  ( \cB1_{ \cS} E)_{L^{-1}} \|_{k+1, \ka}  \leq      \cO(1) L^{-\ep}         \|E\|_{k, \ka} 
\end{equation}

 To establish     (\ref{snort17})
   we   make a Taylor expansion of  $E(X,  \phi_L)$   in the field.  
At   $\phi_L=0$   we   get  zero by the normalization condition.   Also   odd  derivatives   vanish
since    the  functional is even in  $\phi_L$.   Therefore  
\begin{equation}  \label{stunned}
 E(X,  \phi_L )
=   \frac12  E_{2}(X,0;  \phi_L ,  \phi_L) +
   \frac{1}{2\pi i}   \int_{|t |= \frac12    L^{3/4+3\ep} }    \frac{  E(X,  t\phi_L ) } {t^4(t-1)}   dt \
\end{equation}

Note that   since     $\phi_L   \in    L^{-3/4- 3\ep} \cR_k$     on the circle  $ |t |= \frac12    L^{3/4+3\ep} $   we  have     $ t\phi_L   \in   \frac12  \cR_k$.      It  follows that  the second   term   in   (\ref{stunned})    is  bounded  by   $\cO(1)  L^{-3- 12 \ep}  \|E(X) \|_k$
which suffices.

For the first    term in   (\ref{stunned})   we  we   pick a point  $x_0  \in   X$  and      insert  into   $E_2(X,0;   \phi_L,   \phi_L)$
the expansion   
\begin{equation}
\phi_L(x)  =  \phi_L(x_0)   +(x-x_0) \cdot \pa \phi_L(x_0)   + \De_{\phi_L} ( x, x_0)
\end{equation}
where  
\begin{equation}
   \De_{\phi_L} ( x, x_0)
=    \int_{x_0}^x  (  \pa \phi_L(y) - \pa \phi_L(x_0) )\cdot dy 
\end{equation}
Then 
\begin{equation}  \label{spread}
\begin{split}
E_2(X,0;   \phi_L,   \phi_L)  
 =   &  E_2(X,0; \phi_L(x_0) , \phi_L(x_0) )  \\
 + & 2   E_2(X,0; \phi_L(x_0) ,(x-x_0) \cdot \pa \phi_L(x_0)  )     \\
+  &  E_2(X,0; (x-x_0) \cdot \pa \phi_L(x_0)  , (x-x_0) \cdot \pa \phi_L(x_0)  )   \\
+  &2  E_2(X,0; (x-x_0) \cdot \pa \phi_L(x_0) ,  \De_{\phi_L}  )   \\
+  &   E_2(X,0;   \De_{\phi_L} ,   \De_{\phi_L}   )  \\
+ & 2   E_2(X,0; \phi_L(x_0) , \De_{\phi_L} )   \\
\end{split}
\end{equation}
\bigskip
The first and second terms vanish due to our normalization conditions.

The remaining terms will be estimated by  Cauchy inequalities.   In  general
if     $f_1  \in a_1\cR_k ,  f_2  \in a_2\cR_k$   we  can write   
\begin{equation}
\begin{split}
E_{2} (X,  0; f_1,f_2)  
=  &   \frac{ \pa} {\pa  t_1}  \frac{ \pa} {\pa  t_2}\Big[E (X, t_1 f_1 + t_2 f_2) \Big ]_{t_1=t_2 =0}\\
 = & \frac{1}{(2\pi i)^2}  \int_{|t_1| =(2a_1)^{-1}, |t_2|  =(2a_2)^{-1}  }    \frac{ dt_1}{t_1^2}
  \frac{ dt_1}{t_1^2}E (X, t_1 f_1 + t_2 f_2)  \\ 
  \end{split}
\end{equation}
This   gives the estimate
\begin{equation}  \label{awkward}
|E_{2} (X,  0; f_1,f_2)  |   \leq  \cO(1)  a_1a_2   \| E (X)\|_k   
\end{equation}

Now  we  claim  that  if   $\phi \in   \cR_{k+1}$  and   $x, x_0  \in  X$  then   
 \begin{equation}   \label {stupor}
\begin{split}
   \phi_L(x_0) &  \in     L^{-3/4-3\ep} \cR_k  \\
 (x-x_0) \cdot \pa \phi_L(x_0)  &  \in   L^{-7/4 - 2 \ep}\cR_k  \\
  \De_{\phi_L}  &   \in    L^{-7/4  - \al   - \ep}  \cR_k  \\
\end{split}
\end{equation}
Then     the third  term  in  (\ref{spread})    has  a  factor     $ L^{-7/2-  4\ep}$, 
  the fourth  term  has a factor   $L^{ -7/2 - \al - 3 \ep}$ ,
the fifth term  has  a  factor  $ L^{-7/2  - 2\al   - 2\ep}$ ,  
 and the  sixth  term  has  a  factor   $  L^{ - 5/2  - \al   -  \ep}$.   This  easily 
gives 
\begin{equation}
|E_{2} (X,  0;  \phi_L, \phi_L)  |    \leq    \cO(1)L^{-3-\ep} \|E (X)  \|_k  
\end{equation}
and completes the proof of (\ref{snort17}).

For  the first inclusion in  (\ref{stupor})  we  already have by  (\ref{bonnie}),
$| \phi_L(x_0)|   \leq    L^{-3/4- 3\ep} \la_k^{-1/4- 3\ep}$.
The  derivatives  vanish so this establishes  $ \phi_L(x_0)   \in     L^{-3/4- \ep} \cR_k$.

 For  the second   inclusion in  (\ref{stupor}) 
 note that   if $X$ is small then  $|X|_M \leq   9(d_M(X) +1)  \leq    9(L+1)$ so the largest distance between
 points in $X$  is   $9M(L+1)$.  Then    by  (\ref{bonnie})  for  $\la_k$  sufficiently small
 \begin{equation}
 \begin{split}
&| (x-x_0) \cdot \pa \phi_L(x_0) | 
 \leq   9M(L+1) L^{-7/4- 2 \ep}  \la_{k}^{-1/4-2 \ep} \\
   \leq &  (9M(L+1)\la_k^{\ep})   L^{-7/4-2\ep}  \la_{k}^{-1/4- 3\ep}  
 \leq   L^{-7/4-2\ep}  \la_{k}^{-1/4- 3\ep}  \\
 \end{split}
\end{equation}
Furthermore   the  derivative   is   the  constant    
\begin{equation}  \label{bonnie3}
\left|\pa  \Big[ (x-x_0) \cdot \pa \phi_L(x_0) \Big]  \right|
= |  \pa    \phi_L(x_0)| \leq     L^{-7/4- 2 \ep}  \la_k^{-1/4 - 2\ep} 
\end{equation}
The  difference  of derivatives is  zero,   so  $ (x-x_0) \cdot \pa \phi_L(x_0)   \in    L^{-7/4 - 2 \ep}\cR_k $.

 For  the third       inclusion in  (\ref{stupor})  we  estimate   by  (\ref{bonnie})
\begin{equation}  
\begin{split}
| \De_{\phi_L}  (x)|   =    & | \int_{x_0}^x  (  \pa \phi_L(y) - \pa \phi_L(x_0) ) \cdot  dy | \\
\leq  &   L^{-7/4- \al-\ep } \la_k^{-1/4-\ep }  \int_{x_0}^x  d(y, x_0)^{\al} |dy| \\
\leq      &     L^{-7/4- \al-\ep } \la_k^{-1/4-\ep }( 9M(L+1))^{1+ \al} \\
\leq      &   L^{-7/4- \al } \la_k^{-1/4-3\ep }  \\
\end{split}
\end{equation}
and also    
\begin{equation}
\begin{split}
|\pa \De_{\phi_L}  (x)|
= &   |   \pa \phi_L(x) -  \pa \phi_L(x_0) | \\
\leq   &     L^{-7/4  - \al -  \ep}   \la_{k}^{-1/4-\ep}   d(x ,x_0) ^{\al}  \\
\leq   &     L^{-7/4  - \al -  \ep}   \la_{k}^{-1/4-\ep}( 9M(L+1))^{ \al} \\
\leq      &  L^{-7/4  - \al  - \ep}     \la_{k}^{-1/4-2\ep}\\
\end{split}
\end{equation}
Similarly 
   \begin{equation}  \label{bonnie4}
   \begin{split}
|\pa \De_{\phi_L}  (x)  - \pa \De_{\phi_L}  (y)|
=   & |  \pa \phi_L(x) -    \pa \phi_L(y) | 
 \leq       L^{-7/4  - \al -  \ep}  \la_{k}^{-1/4 - \ep}    d( x,y )^{\al}  \\
\end{split} 
\end{equation}
The  last  three bounds  imply  $\De_{\phi_L}    \in    L^{-7/4  - \al - \ep }  \cR_k $.   This completes  the 
proof  of  (\ref{stupor})  and  the theorem.
\bigskip

We   also  explain how to arrange  the normalization   for small polymers.
Given  $E \in \cK_k$  we  define    $\cR E \in    \cK_k$   as  follows. 
  If   $X$ is small  ($X  \in \cS$) then  $\cR E(X)$  is 
defined by 
\begin{equation}  \label{renorm}
\begin{split}
& E( X, \phi)  
=   \al_0(E,X)  \Vol( X )
+  \al_{2}(E,X)   \int_X  \phi^2        + \sum_{\mu}  \al_{2, \mu}(E,X)       
   \int_X  \phi \  \pa_{\mu} \phi   \    +   \cR E( X, \phi)    \\
\end{split}
\end{equation} 
where
\begin{equation} 
\begin{split}
\al_0(E, X)   =& \frac{1}{ \Vol (X)}  E(X,0)  \hs   \al_{2}(E,X) =      \frac{1}{2 \ \Vol(X)}   
 E_2 ( X, 0; 1,1 )\\
   \al_{2, \mu}(E,X)    =&\frac{1}{\Vol(X)} \left( E_2 ( X, 0; 1,x_{\mu} - x^0_{ \mu}) -   \frac{1 }{ \Vol(X)}  E_2 ( X, 0; 1,1)\int_X  x_{\mu} - x^0_{ \mu}\right)\\
\end{split}
\end{equation}
The last   is independent of  the base point  $x^0$,  which we take to be  in $X$. 
Then  it is straightforward to check that    $\cR  E$  is normalized for small polymers.    If   $X$ is large then  $\cR E(X)   =  E(X)$.

\begin{lem}    \label{rlem}    For  $E  \in \cK_k$  and  $\la_k$ sufficiently small
\begin{equation}
\|    \cR E (X)  \|_k    \leq  \cO(1)    \|   E(X)   \|_k      
\end{equation}
\end{lem}
\bigskip

\pr  It  suffices to check for small polymers $X$.    We  check that  every other  term  in (\ref{renorm})  satisfies  such a bound.
 This  is  immediate for   $E(X,\phi)$  and $\al_0(E,X) \Vol(X)= E(X,0)$.  
 
    For the next term     
 note that      since   $1  \in    \la_k^{1/4 +3\ep }  \cR_k$   we  have  by    (\ref{awkward})
\begin{equation}  \label{saint}
| E_{2} ( X, 0; 1,1 ) |  \leq  \cO(1)  \la_k^{1/2 +6\ep } \|  E (X)  \|_{k}
 \end{equation}
Also for     $\phi  \in \cR_k$       we  have  $| \int_X  \phi^2     |  \leq    \Vol(X)    \la_k^{-1/2 -6\ep } $.  Therefore   
 \begin{equation}
\Big |   \al_2(E,X)  \int_X  \phi^2 \Big |         
 \leq   \cO(1)    \|  E(X ) \|_k   
 \end{equation}

 For the next  term  note that    
\begin{equation}
\begin{split}
|x_{\mu} - x^0_{\mu}|   \leq   &\  9M(L+1)  \leq \   ( 9M(L+1)\la_k^{\ep} ) \la_k^{1/4  + 2\ep} \la_k^{-1/4  -3 \ep} 
\leq   [  \la_k^{1/4  + 2\ep} ]\la_k^{-1/4  -3 \ep}     \\
|  \pa_{\nu} ( x_{\mu} - x^0_{ \mu})  |  \leq    &\  1   \leq  [ \la_k^{1/4+   2 \ep} ] \la_k^{-1/4- 2 \ep}  \\
\end{split}
\end{equation}
Therefore     $x_{\mu} - x^0_{ \mu}  \in  \la_k^{1/4 +2 \ep}  \cR_k$   and  so   
\begin{equation}
| E_{2} ( X, 0; 1,x _{\mu}- x^0_{\mu}) | \leq  
\cO(1)  \la_k^{1/2 +5 \ep} \|  E (X  ) \|_k
  \end{equation}  
Also   $|\int_X  x_{\mu} - x^0_{ \mu}|  \leq  \Vol(X)   \la_k^{-\ep}$   and    for     $\phi \in \cR_k$ we have   
$| \int_X  \phi \  \pa_{\mu} \phi |  \leq   \Vol(X)    \la_k^{-1/2 -5\ep }$.  These combine to give
\begin{equation}
\Big| \al_{2, \mu}(E,X)  \int_X  \phi \  \pa_{\mu} \phi  \Big |  \leq  \cO(1)   \|  E(X ) \|_k   
\end{equation}
  This  completes the proof
\bigskip
 
 Inserting (\ref{renorm})  into   $E =  \sum_X E(X)$  and   defining   $ \cR E =  \sum_X \cR E(X)$
   we  find   we have extracted  energy and mass terms:
 \begin{equation}  \label{renorm2}
  E =     -   \vep(E)  \Vol(  \bbT_{\sM+ \sN-k} ) -   \frac12  \mu(E)   \|  \phi \|^2
+    \cR E
\end{equation} 
Here 
\begin{equation}
\begin{split}
\vep(E)  =  &    -   \sum_{X \supset \square,  X \in \cS}   \al_0(E,X) \\
\frac12  \mu(E)  =& - \sum_{X \supset \square,  X \in \cS}    \al_2(E,X)  \\
\end{split}
\end{equation}
are independent of   $\square$  by translation invariance.
We have also  used  
\begin{equation}
  \sum_{X \supset \square,  X \in \cS}     \al_{2, \mu}(E,X)    =0
\end{equation}
which  follows by  choosing $x^0$  in the center of  $\square$   and  using  $  \al_{2, \mu}(E,r_{\mu}X)  =  -    \al_{2, \mu}(E,X)$
where $r_{\mu}$ is reflection  thru the plane  $x_{\mu} = x^0_{\mu}$.

  \begin{lem}   \label{study}
 \begin{equation}  
\begin{split}
|\vep(E)  |  \leq  &   \cO(1) \| E  \|_{k, \ka}      \\
 | \mu(E)|  \leq     &\  \cO(1)   \la_k^{1/2 +6\ep }  \| E  \|_{k, \ka}       \\
\end{split}
\end{equation}
\end{lem}
  \bigskip
  
\pr    For  the first  bound  we  have
   $ \vep(E)   \leq         \sum_{X \supset \square}  \| E( X ) \|_k    \leq    K_0  \|E \|_{k, \ka}$    as  in  (\ref{summing}).
 The second bound  uses  (\ref{saint})  and follows in the same way.

 \section{The  RG  transformation with small fields}

\subsection{the theorem} 

Now   we study   the  RG transformation   with the   potential,  but modified with  a small field assumption.
The starting point is  still    the density   
\begin{equation}
\rho_0  (\Phi_0)  =  \exp   \Big(  -S_0(\Phi_0 )   -  V_0(\Phi_0)  \Big)  
\end{equation}
where  $\Phi_0:  \bbT^0_{\sM + \sN}  \to   \bbR$   and   
\begin{equation}
\begin{split}
S_0 (  \Phi_0  )  = &  \frac12  \|  \pa \Phi_0  \|^2  +  \frac 12   \bar \mu_0  \| \Phi_0 \|^2 \\
V_0  ( \Phi_0  )    = &   \vep_0     \Vol (\bbT^0_{\sM+ \sN} )     +    \frac12     \mu_0   \| \Phi_0 \|^2 
  +  \frac{1}{4}   \la_0       \sum_x  \Phi_0 (x)^4   \\
\end{split}
\end{equation}
But now instead of  (\ref{kth})   we  add  some characteristic functions and    define   $\rho_k$  recursively as follows.    For  $\Phi_k:  \bbT^0_{\sM+ \sN -k}  \to   \bbR$   and    $\Phi_{k+1}:  \bbT^1_{\sM+ \sN -k}  \to   \bbR$
let  
\begin{equation}
\begin{split}  \label{newkth}
& \tilde  \rho_{k+1} ( \Phi_{k+1}) \\
 = & \cN^{-1}_{aL ,  \bbT^1_{\sM + \sN-k}}  \int \exp \left(-  \frac{a}{2L^2} 
\|\Phi_{k+1}-Q\Phi_k\|^2 \right) \chi^w_k\Big(C_k^{-1/2}( \Phi_k -  \Psi_k) \Big)  \chi_k(\Phi_k)    \rho_{k} ( \Phi_k)\ \ d\Phi_k \\
\end{split}
\end{equation}
and   as  before for      $\Phi_{k+1}:  \bbT^0_{\sM+ \sN -k-1}  \to   \bbR$
\begin{equation}
 \rho_{k+1} ( \Phi_{k+1})  =  \tilde  \rho_{k} ( \Phi_{k+1,L}) L^{ -   | \bbT^1_{\sM +\sN -k} |/2}   
\end{equation}
Here the characterstic  functions   are
\begin{equation}
\begin{split}
\chi^w_k(W)  =   &  \chi  (   |W|  \leq   p_{0,k})  \\
\chi_k(\Phi_k)  =&   \chi(  \Phi_k  \in \cS_k ) \\
\end{split}
\end{equation}       
With  the  free minimizer  $\Psi_k =   \Psi_k(\Phi_{k+1}, \phi^0_{k+1})$ defined in  (\ref{kingmaker})   the  function   $  \chi^w_k\Big(C_k^{-1/2}( \Phi_k -  \Psi_k) \Big) $  enforces
 that  the fluctuation  field   $ \Phi_k -  \Psi_k$    be  small.
The size  is  determined by  
\begin{equation}
p_{0,k}  =   p_0(\la_k)   =  (-   \log   \la_k  )^{p_0}  
 \end{equation}
This has    the same form  as  $p_k$  but with  a smaller  integer    exponent    $p_0 < p$.   
The  function   $\chi_k(\Phi_k)$   is the small field restrtiction.    Actually  we  only consider  $\Phi_{k+1} \in S_{k+1}$
in which case this restriction is unnecessary as we will see.

Adding the characteristic functions gives us the leading term  in an expansion of  the full integral  into various blocks of field values.  
This is developed in paper  II.

\bigskip

We   are going to assert       that   after   $k$   steps
 the modified  density  can  be  written in    the  following    local   form.  For   $\Phi_k  \in S_k$  
 \begin{equation}  \label{basic}
\rho_k(  \Phi_k )  
=  Z_k    \exp \Big ( - S_k( \Phi_k, \phi_k )   - V_k(\phi_k) +  E_k (\phi_k)  \Big) 
\end{equation}
where  $\phi_k:   \bbT^{-k}_{\sM + \sN -k}   \to \bbR$   is    $\phi_k = a_kG_k Q_k^T \Phi_k$   and  
 \begin{equation}    \label{basic20}
\begin{split}
   S_k(\Phi_k, \phi_k)    = &  \frac12  a_k   \|\Phi_k-Q_k  \phi_k\|^2     +\frac12   \|  \pa \phi_k  \|^2 
      +\frac12  \bar \mu_k  \|  \phi_k  \|^2\\
V_k(\phi_k)    = &\vep_k  \Vol( \bbT_{\sM+\sN-k} )    
 +\frac12   \mu_k \|  \phi_k  \|^2 
 +  \frac{1}{4} \la_k \int   \phi_k^4 (x)  dx   \\
\end{split}
\end{equation}
  We  further assert that the  functional  $E_k(\phi)   $  is   defined and analytic   in the larger set    $\phi \in \cR_k$   
  and   can  be  written  
  \begin{equation}  \label{basic3}
E_k(\phi)   =  \sum_{X}  E_k(X,  \phi)     
\end{equation} 
where    $E_k(X,  \phi)  \in  \textrm{Re} (   \cK^{norm}_k ) $ and so   is normalized for small polymers.

\begin{thm}   \label{lanky} 
Let   $L,M$  be sufficiently large,  let $\la_k$  be  sufficiently small   (depending on $L,M$). 
Suppose    $\rho_k(\Phi_k)$  has the representation   (\ref{basic})- (\ref{basic3})     for  $\Phi_k \in \cS_k$ and
   \begin{equation}  
|\mu_k|     \leq   \la_k^{1/2}       \hs   \|E_k\|_{k, \ka}   \leq   1
 \end{equation}
 Then  $\rho_{k+1}(\Phi_{k+1}) $  has  a representation of the same form   for  $\Phi_{k+1}  \in \cS_{k+1}$.
    The bounds   are not the same   but we do have     
 \begin{equation}  \label{recursive}
\begin{split}
\vep_{k+1}   =&  L^3 \vep_k  + \cL_1E_k   +  \vep_k^*(\la_k, \mu_k,  E_k) \\
\mu_{k+1}   =&   L^2 \mu_k  +  \cL_2E_k  + \mu_k^*(\la_k, \mu_k,  E_k)  \\
\la_{k+1}   =&  L\la_k   \\ 
E_{k+1}   =&    \cL_3 E_k  +  E^*_k(\la_k,    \mu_k,  E_k)  \\
 \end{split}
\end{equation}
where   the $\cL_i$  are linear   operators    which   satisfy  
\begin{equation}
\begin{split}
| \cL_1 E_k |   \leq &\  \cO(1) L^{-\ep} \|E_k \|_{k, \ka}\\
| \cL_2 E_k |   \leq &\  \cO(1) L^{-\ep}   \la_k^{1/2  + 6 \ep} \|E_k \|_{k, \ka}\\
\| \cL_3 E_k \|_{k+1, \ka}   \leq &\  \cO(1) L^{-\ep} \|E_k \|_{k, \ka}\\
\end{split}
\end{equation}
and where  
\begin{equation}
|\vep_k^*| \leq    \cO(1)L^3 \la_k^{1/4 - 10 \ep}  \hs  |\mu_k^*|  \leq   \cO(1) L^3\la_k^{3/4 - 4 \ep}   \hs    \|E^*_k \|_{k+1, \ka}     \leq  \cO(1)L^3 \la_k^{1/4 - 10 \ep}
\end{equation}
\end{thm}
\bigskip

\res   The unstarred terms  represent   scalings and rearrangements  of  the  existing terms,   but 
not the effects of the  fluctuation integral.  The  starred terms   are the effect of the fluctuation integral.
To put it another way the unstarred terms are zeroeth order perturbation theory,  and the starred terms 
are all higher order contributions.
  The starred terms   are   not  necessarily  smaller  than the unstarred terms,  although they do have
better bounds.

This flow  shows strong growth, but  it    is tolerable due to the ultraviolet origin of the problem:
 we  start with very small coupling constants.  Our  concern  will be  that the growth is not too rapid.
We  want to finish at  a good place. 
  
The   main     idea
 is that we have removed mass and  energy terms  from $E_k$ by normalizing,   and included them in corrections
to      $\vep_k,  \mu_k$. These  are  the   fastest growing  terms   and  in this form  they will be susceptible to analysis.

\subsection{start of the proof} 

We  study   $\tilde   \rho_{k+1}  (\Phi_{k+1}) $  for    $\Phi_{k+1}  \in \cS_{k+1}^0$
defined as  all  $\Phi_{k+1}:  \bbT^{1}_{\sM + \sN-k} \to \bbR$  
such that   
\begin{equation}  \label{charm}
\begin{split}
 | \Phi_{k+1}  -  Q_{k}    \phi^0_{k+1}|  \leq & L^{-1/2} p_{k+1} \\
 |\pa   \phi^0_{k+1}|  \leq &L^{-3/2} p_{k+1}    \\   
 |  \phi^0_{k+1}|  \leq & L^{-1/2} \la_{k+1}^{-1/4} p_{k+1}  \\
\end{split}
\end{equation} 
where  $\phi_{k+1}^0=  \phi^0_{k+1}(\Phi_{k+1})$.
This is the appropriate choice  since   we  eventually  want  $ \tilde  \rho_{k} ( \Phi_{k+1,L})$
for   $\Phi_{k+1}:  \bbT^{0}_{\sM + \sN-k-1} \to \bbR$   and  $\Phi_{k+1} \in  \cS_{k+1}$.  These
conditions  imply     $\Phi_{k+1,L}  \in  \cS^0_{k+1}$ as  can be demonstrated using  
$\phi^0_{k+1}(\Phi_{k+1,L})   = [  \phi_{k+1}(\Phi_{k+1})]_L$.

Returning  to  $\Phi_{k+1}  \in \cS^0_{k+1}$   we  note  that this condition implies
 $ \phi^0_{k+1}(\Phi_{k+1}) \in  L^{-\frac34 - \ep}\cR_{k}$.    (Proof:    Then   $\Phi_{k+1, L^{-1}}  \in  \cS_{k+1}$ which
implies    $\phi_{k+1}(\Phi_{k+1, L^{-1}})  \in  \cR_{k+1} $ by lemma \ref{strong}. 
 But     $\phi_{k+1}(\Phi_{k+1, L^{-1}}) = [\phi^0_{k+1}(\Phi_{k+1})]_{L^{-1}}$   so     
$\phi^0_{k+1}(\Phi_{k+1})  \in     L^{-\frac34 - \ep}\cR_{k}$   by  lemma  \ref{scaling}.)
\bigskip

Now   for  $\Phi_{k+1} \in   \cS^0_{k+1}$  we  have 
\begin{equation}  \label{stanley2}
\begin{split}
\tilde   \rho_{k+1}  (\Phi_{k+1})  
= & Z_k\ \cN^{-1}_{aL ,  \bbT^1_{\sM + \sN-k}} \int \exp \left(-  \frac{a}{2L^2} 
\|\Phi_{k+1}-Q\Phi_k\|^2  -   S_k ( \Phi_k, \phi_k) - V_k(\phi_k)   +  E_k (\phi_k) \right) \\
& \chi^w_k\Big(C_k^{-1/2}( \Phi_k -  \Psi_k)\Big)\chi_k(\Phi_k)  \ d\Phi_k \\
\end{split}
\end{equation}
We  expand  in  $\Phi_k$  
  around the miniminzer   $\Psi_k$   for   the first  two  terms  
in the exponential  by  writing  $\Phi_k  =  \Psi_k  + Z$.
As  in  (\ref{eighty}) this  generates
  $ S^0_{k+1}(\Phi_{k+1}, \phi_{k+1}^0)  +        
    \frac12 \Big<Z,  (\De_{k} + aL^{-2} Q^TQ  )  Z  \Big>$ .
We  also have  $\phi_k ( \Psi_k + Z)   =  \phi^0_{k+1}  + \cZ_k$  where  $\cZ_k =   a_kG_k Q^TZ$ as in
(\ref{undo}).
Changing the integration variable from  $\Phi_k$ to $Z$
yields
\begin{equation}  \label{stanley3}
\begin{split}
&\tilde   \rho_{k+1}  (\Phi_{k+1})  
= Z_k\  \cN^{-1}_{aL ,  \bbT^1_{\sN + \sM-k}} \exp  \Big( - S^0_{k+1}(\Phi_{k+1}, \phi_{k+1}^0)  \Big)\\
& \int \exp     \Big( E^+_k( \phi^0_{k+1}  +  \cZ_k)  -  \frac12 <Z,  (\De_{k} +  \frac{a}{L^2} Q^TQ  )Z >       \Big)   
\chi_k(    \Psi_k  + Z)    \chi^w_k\Big(C_k^{-1/2}Z)\Big) \ dZ \\
\end{split}
\end{equation}
Here    we have introduced
\begin{equation}
E_k^+( \phi)   =     E_k ( \phi) -  V_k( \phi) 
\end{equation}
If  we  define
\begin{equation}
V_k(\square,  \phi)    =      \vep_k  \Vol( \square )    
 +\frac12   \mu_k \|  \phi \|_{\square}^2 
 +  \frac{1}{4} \la_k \int_{\square}   \phi_k^4 (x)  dx   
\end{equation}
then   $ V_k( \phi)   = \sum_{\square}  V_k(\square, \phi) $.   If we also   define    $V_k(X, \phi) =0$  for  $|X|_M \geq  2$ then 
  $ V_k( \phi)   = \sum_X  V_k(X, \phi) $.   Together  with  (\ref{basic3})   this gives    a   local expansion
  \begin{equation}
E_k^+ (\phi)  =  \sum_{X} \  E_k^+(X, \phi)
\end{equation}

  Recall also  that      $
C_k  =    \left(  \De_k +  aL^{-2} Q^TQ\right)^{-1}$
and let  $\mu_{C_k}$  be the Gaussian measure with covariance  $C_k$.
Then (\ref{stanley3})   can be   written     
\begin{equation}  \label{stanley4}
\begin{split}
&\tilde  \rho_{k+1}  (\Phi_{k+1})  
= Z_k   \cN^{-1}_{aL ,  \bbT^1_{\sM + \sN-k}}  (2 \pi)^{| \bbT^0_{\sM +\sN-k}|/2}  (\det  C_k)^{1/2} 
 \exp  \Big( - S^0_{k+1}(\Phi_{k+1}, \phi_{k+1}^0) \Big)   \\
&\int \exp   \Big( E^+_k( \phi^0_{k+1}  +  \cZ_k)     \Big)    
\chi_k(    \Psi_k  + Z)    \chi^w_k(C_k^{-1/2}Z) \  d\mu_{C_k}(Z)  \\
\end{split}
\end{equation}
If  we  multiply by  $L^{- | \bbT^1_{\sM+\sN -k}|/2}$ we  can identify the constant in front  as  $Z_{k+1}$  by  (\ref{Ziterate}).
  We  make one further adjustment  in the  integral  
by  changing from   Gaussian  $Z: \bbT^0_{\sM + \sN -k}  \to  \bbR$   with covariance   $C_k$     to   $Z=  C_k^{\frac12}  W$   where  Gaussian  $W: \bbT^0_{\sM + \sN -k}  \to  \bbR$ has  identity   covariance.
Thus we  have  
\begin{equation}  \label{stanley5}
\begin{split}
&\tilde   \rho_{k+1}  (\Phi_{k+1})  L^{- | \bbT^1_{\sM+\sN -k}|/2}
= Z_{k+1} 
 \exp  \Big( - S^0_{k+1}(\Phi_{k+1}, \phi_{k+1}^0)  \Big)   \\
& \int \exp     \Big( E^+_k( \phi^0_{k+1}  +  \cW_k)     \Big)   \chi_k(   \Psi_k  +  C_k^{\frac12}  W)   \   \chi^w_k( W)   d\mu_{I}(W)  \\
\end{split}
\end{equation}
where   $\cW_k: \bbT^{-k}_{\sM  + \sN  -k }  \to  \bbR$ is given by
\footnote{If  $k=0$ this is  $\cW_0 = ( C_0 )^{\frac12}  W  =  ( -\De  + a L^{-2} Q^TQ)^{-\frac12} W$}
\begin{equation}
\cW_k     = \phi_k(C_k^{\frac12}W ) =        a_k G_k  Q_k^T C_k^{\frac12}W
\end{equation}

We  will need a more explicit  representation  of  $C_k^{\frac12}$.
  For   $\la >0$  
\begin{equation}
\la^{-1/2}  =  \frac{1}{\pi}  \int_0^{\infty}  \frac{   dr}{   \sqrt{r}} (\la+ r)^{-1}  
\end{equation}
Hence   we have the operator identity.
\begin{equation}  \label{one}
\begin{split}
C_k^{\frac12}  =&     \frac{1}{\pi}  \int_0^{\infty}  \frac{dr}{\sqrt{r}}  C_{k,r} \\
C_{k,r}  =  &  \Big(   \De_k +  \frac{a}{L^2}Q^TQ  +r\Big)^{-1}  \\
\end{split}
\end{equation}
In  appendix   \ref{C} we  establish  
\begin{equation}  \label{two}
 C_{k,r} =   A_{k,r}  +   a_k^2  A_{k,r} Q_k  G_{k,r} Q_k^T A_{k,r}
\end{equation}
where 
\begin{equation}  \label{three}
\begin{split}
A_{k,r}   =&   \frac{1}{a_k+r}  (I - Q^TQ)   +   \frac{1}{ a_k + aL^{-2}  +r}   Q^T Q  \\
G_{k,r}  = &  \Big( -\De  + \bar \mu_k  +  a_k Q_k^TQ_k    -   a_k^2Q_k^T   A_{k,r} Q_k\Big)^{-1}
  \\
 \end{split}
 \end{equation}
 An  alternative     expression for   $G_{k,r}$
 is
 \begin{equation}   \label{alt}
 G_{k,r}  =   \Big( -\De  + \bar \mu_k  +   \frac{a_k r}{a_k + r}Q_k^TQ_k    +\frac{a_k^2 aL^{-2}} {(a_k +r)(a_k + a L^{-2} +r)} Q_{k+1}^TQ_{k+1}
\Big)^{-1}
 \end{equation}
 This  shows that we are inverting a positive operator.
 Note   that   $G_{k,r}$ interpolates between   $G_{k,0}  =  G_{k+1}^0$  (use  (\ref{ak}))  and   $G_{k, \infty}  = G_k$.

\subsection{a  simplification}

The next  lemma shows that  we  can  drop  
$\chi_k(     \Psi_k  +  C_k^{\frac12}  W) $  from  the   expression  (\ref{stanley5}).

\begin{lem}  For  
 $\Phi_{k+1} \in \cS^0_{k+1}$    and  $|W|  \leq  p_{0,k}$  we have
 $   \Psi_k  +  C_k^{\frac12}  W  \in \cS_k$  and hence
\begin{equation}
  \chi_k(   \Psi_k  +  C_k^{\frac12}  W)=1  
\end{equation}
\end{lem}
\bigskip
\pr  
  We  must show 
\begin{equation}  \label{blank}
\begin{split}
 |     \Psi_k  +  C_k^{\frac12}  W -  Q_{k}   \phi_k (     \Psi_k  +  C_k^{\frac12}  W)|  \leq &\  p_k \\
 |\pa   \phi_{k}(     \Psi_k  +  C_k^{\frac12}  W)|  \leq &\  p_k    \\   
 |  \phi_{k}(     \Psi_k  +  C_k^{\frac12}  W))|  \leq &  \la_k^{-1/4}\ p_k  \\
\end{split}
\end{equation} 
We   give  separate bounds on the  terms  involving   $\Psi_k$    and  $W$.

For  the $\Psi_k$ terms  we  identify        $  \phi_k (   \Psi_k )   = \phi^0_{k+1}$   and show  
\begin{equation}  \label{blank2}
\begin{split}
 |     \Psi_k  -  Q_{k}   \phi^0_{k+1} |  \leq &\  \frac12  p_k \\
 |\pa   \phi^0_{k+1}|  \leq &\  \frac12 p_k    \\   
 |  \phi^0_{k+1}|  \leq &\  \frac12  \la_k^{-1/4} p_k  \\
\end{split}
\end{equation} 
These follow from   (\ref{charm}).
The  last follows from     $ |  \phi^0_{k+1}|  \leq   L^{-1/2}   \la_{k+1}^{-1/4} p_{k+1}  \leq    L^{-3/4} \la_{k}^{-1/4} p_{k}$. 
The    second follows  by  $     |\pa  \phi^0_{k+1}|    \leq   L^{-3/2}  p_{k+1}     \leq    L^{-3/2}  p_k$.
The  first  follows by     
 \begin{equation}
 |  \Psi_{k}
 -  Q_{k}   \phi^0_{k+1} | 
\leq     | Q^T \Big(\Phi_{k+1} -  Q_{k+1}   \phi^0_{k+1}\Big)| 
\leq    \| \Phi_{k+1} -  Q_{k+1}   \phi^0_{k+1}\|_{\infty} 
\leq      L^{-1/2} p_{k+1}    \leq  L^{-1/2}p_k
\end{equation}
Here we have used the explicit expression  (\ref{kingmaker})   for  $\Psi_k$.

For  the  $W$   terms  we  need
\begin{equation}  \label{needed}
\begin{split}
 |    C_k^{\frac12}  W -  Q_k  \cW_k|  \leq &\ \frac12   p_k \\
|\pa  \cW_k|  \leq &\  \frac 12p_k    \\   
 |   \cW_k|  \leq &\  \frac 12   \la_k^{-1/4} p_k  \\
\end{split}
\end{equation} 
In  the  next  lemma we show   that    $|C_k^{\frac12}  W |, |\cW_k|,  | \pa \cW_k|$   are all
bounded by  a constant times  $p_{0,k}$
Then   if  $\la_k$ is sufficiently small  we  have   $p_{0,k}  / p_k   =    ( - \log \la_k)^{p_0 - p}$       as small as
we  like  since  $p_{0}< p$.  Hence  these functions are bounded  by say  $\frac14  p_k$ which
suffices  to prove  (\ref{needed}). 

\bigskip

\begin{lem}  \label{11}
If   $| W |  \leq   p_{0,k}$  then 
  \begin{equation}   \label{wbounds}
|C_k^{\frac12}  W|,  |\cW_k|,  | \pa  \cW_k|,    | \de_{\al} \pa  \cW_k|     \leq  C    p_{0,k}
\end{equation}
  Furthermore  $\cW_k   \in   \la_k^{1/4} \cR_k$
\end{lem}
\bigskip

\pr   We  use  the representation  (\ref{one}), (\ref{two}), (\ref{three})  of  $C_k^{\frac12}$.   These  express   $C_k^{\frac12}$ in 
terms  of  $D_{k,r}  =Q_k  G_{k,r} Q_k^T$.   In  appendix  \ref{E}  we  establish a random walk expansion  for $G_{k,r} $  which 
leads to $L^2$   bounds.    For  the kernel  $D_{k,r}(y,y') =  <Q_k^T\de_y,  G_{k,r} Q_k^T\de_{y'}>  $   
these   say
\be     |D_{k,r}(y,y')|   \leq   C  e^{-\ga_0 d(y,y')} \|Q_k^T\de_y\|_2\| Q_k^T\de_{y'}\|_2       \leq   C  e^{-\ga_0 d(y,y')}
\ee
This  gives the  $L^{\infty}$  bound    $|  D_{k,r}  W|   \leq   C   \|W \|_{\infty}$ .  
We  also have       $|A_{k,r}W|   \leq   \cO(1)  (1+r)^{-1}  \|W\|_{\infty}$.    Hence       $C_{k,r}  =      A_{k,r}  +   a_k^2  A_{k,r}  D_{k,r} A_{k,r}$
satisfies   
$|C_{k,r}W|   \leq   C (1+r)^{-1}  \|W\|_{\infty} $  and    so   $|C_k^{\frac12} W|  \leq    C   \|W\|_{\infty} \leq   C\  p_{0,k} $ as announced.

The other bounds follow  by  (\ref{gk2}).  For example
\begin{equation}
 |\cW_k|   =  |a_k   G_k Q_k^TC_k^{\frac12}  W|  \leq   C \| C_k^{\frac12}  W \|_{\infty}   \leq   
   C\  p_{0,k}  \leq  p_k  \leq  \la_k^{-\ep}   \leq   \la_k^{-3 \ep} 
 \end{equation}  
The last bound is the one needed for    $\cW_k  \in   \la_k^{1/4} \cR_k$.    The bounds on  $\pa \cW_k$ 
and  $\de_{\al}  \pa  \cW_k$   are similar.

\subsection{fluctuation integral}
With the characteristic  function  gone we now have 
\begin{equation}  \label{stanley6}
\tilde  \rho_{k+1}  (\Phi_{k+1})   L^{- | \bbT^1_{\sM+\sN -k}|/2}
=  Z_{k+1}\     \exp  \Big( -S^0_{k+1}(\Phi_{k+1}, \phi^0_{k+1}) \Big)  \Xi_k(  \phi^0_{k+1})  
\end{equation}
where  \begin{equation}
\Xi_k(  \phi)   =  \int \exp     \Big( E^+_k( \phi  +  \cW_k)     \Big)     \   \chi^w_k( W)   d\mu_{I}(W)
\end{equation}
This is the fluctuation integral.
We   are  going to study  it  for  $\phi \in  \frac12 \cR_k$.      It is well defined with this  restriction
since   $E^+$ is defined on   $ \cR_k$  and   $\cW_k   \in   \la_k^{1/4}   \cR_k   \subset    \frac 12 \cR_k$.     Note  also   that the point of interest   $\phi^0_{k+1}  \in  L^{-3/4-3\ep}  \cR_k$   is  included in $ \frac 12 \cR_k$.

We    make   a couple of adjustments in   $\Xi_k$.  
First   change to  the    probability measure  
\begin{equation}
  d \mu^*_k (W)  =  \cN_{\chi,k} ^{-1}\     \chi^w_k( W)   d\mu_{I}(W) 
  \end{equation}
  Here the normalizing factor is   
  \begin{equation} 
  \begin{split} 
  \cN_{\chi,k}  =&   \int    \chi^w_k( W)   d\mu_{I}(W)  \\
  =&   \prod_{x \in   \bbT^0_{\sM + \sN -k}}  \int    \chi^w_k( W(x))   d\mu_{I}(W(x))\\
  =   & \prod_{x \in   \bbT^0_{\sM + \sN -k}} \exp( - \vep_k^0)
  =     \exp\Big( - \vep_k^0  \Vol( \bbT^0_{\sM + \sN -k})\Big)\\
  \end{split} 
  \end{equation}
 where    $\vep_k^0 >0$  is defined by  
  \begin{equation}
   \vep_k^0   =-  \log \Big(    \int    \chi^w_k( W(x))   d\mu_{I}(W(x))  \Big)
  \end{equation}
It  is straightforward to show  
\begin{equation}
| \int    \chi^w_k( W(x))   d\mu_{I}(W(x) )  -1 |   \leq  \cO(    e^{ - p^2_{0,k}/2})
\end{equation}
and    hence      $  \vep_k^0   \leq  \cO(    e^{ - p^2_{0,k}/2})$ as  well.   It is very small
\bigskip

 Secondly  define     $ \de    E_k^+ ( \phi,  \cW_k)$
 by   
 \begin{equation}
   E_k^+ (  \phi +  \cW_k)=    E_k^+ ( \phi )  
 +\de  E_k^+ (\phi,   \cW_k ) 
\end{equation}
There is also a local decomposition inherited from  $E_k^+$.
The term   $ E_k^+ ( \phi  )$   is  pulled out of the integral.  It is   not necessarily small and would make  subsequent  estimates   awkward.

 Now    we  have   
  \begin{equation}   \label{sunny}
  \begin{split}
\Xi_{k} (\phi)
=  &  \exp  \Big( - \vep_k^0 \Vol( \bbT^0_{\sM + \sN -k})  +  E_k^+ ( \phi )\Big)   \Xi'_{k}(\phi)\\
  \Xi'_{k}(\phi) = & \int \exp     \Big(   \de E_k^+ ( \phi ,  \cW_k )   \Big)    d\mu^*_k(W)   \\
  \end{split}
   \end{equation}

 To   analyze  $ \Xi'_{k}(\phi)$  
 we  start with a  general   bound   on  the local  pieces $ \de E_k^+ (X,  \phi ,  \cW_k)   $   which are small.   
\begin{lem}    \label{12}
 For      $\phi  \in  \frac12  \cR_k$   and  $|W|  \leq   p_{0,k}$.
\begin{equation}  \label{unsung}
 | \de E_k^+ (X, \phi ,  \cW_k ) |   \leq   \cO(1)    \la_k^{1/4-10\ep}  e^{- \ka  d_M(X)}
\end{equation}
\end{lem}
\bigskip

\pr   We  have  $  \de E_k^+  =    \de E_k -      \de V_k     $  and we first consider   $\de V_k$.  We  have
\begin{equation}
\begin{split}
\de V_k(\square)  =&    \   \frac14  \la_k    \int _{\square}  \Big[ ( \phi + \cW_k)^4   -  \phi ^4\Big]  
+      \frac12  \mu_k    \int _{\square}   \Big[ ( \phi +  \cW_k)^2   -  \phi ^2  \Big]   \\
\end{split}
\end{equation}
Now    $|\phi|   \leq \frac12    \la_k^{-1/4 - 3\ep}$   and  $|\cW_k|  \leq   p_k$  so 
the   first term has a contribution 
\begin{equation}
\left| 2  \la_k    \int _{\square}  \phi^3  \cW_k   \right|
\leq    2   M^3 \cdot    \la_k^{1/4 - 9\ep}p_k    \leq  \la_k^{1/4  - 10 \ep}
\end{equation}
The other contributions to the first term are smaller.
Similarly   the second term has a contribution
\begin{equation}
\Big| \mu_k  \int_{\square} \phi \cW_k \Big|  \leq  \la_k^{1/2}   M^3  \cdot  \la_k^{-1/4 - 3\ep} p_k    \leq    \la_k^{1/4-4\ep} 
\end{equation}
The   other term is smaller.  
Overall then   $|\de V_k(\square)| \leq   \cO(1) \la_k^{1/4  - 10 \ep}$.

 By  lemma  \ref{11}  we have  $\cW_k  \in  \la_k^{1/4}  \cR_k $.     So  if      $|t|  \leq   \la_k^{-1/4}/4$ then   $t\cW_k  \in   1/4 \cR_k$   and      $ \phi  +  t\cW_k  \subset   3/4   \cR_k$.    
 The  function  $t  \to   E_k ( \phi  +t\cW_k)  $  is analytic in this domain and so  
 \begin{equation}
 \de    E_k ( X,\phi,\cW_k)  =   \frac{1}{2\pi i} \int_{ |t|  =   \la_k^{-1/4}/4}     \frac{ dt}{ t(t-1)}   E_k (X, \phi  +t\cW_k)  
 \end{equation}
 Since 
 \begin{equation}  
 |  E_k (X, \phi  +t\cW_k)   |   \leq   \|E_k\|_{k ,\ka}   e^{-\ka d_M(X) }    \leq   e^{- \ka d_M(X) } 
 \end{equation}
 this gives the bound   
  \begin{equation}  \label{take1}
  | \de    E_k (X, \phi,  \cW_k) |  \leq  \cO(1)  \la_k^{1/4}  e^{- \ka d_M(X) } 
  \end{equation}
  which is sufficient.
  \bigskip

  \re
  It is convenient at this point to reblock  to  $LM$ cubes  defining   $(\de    E^+_k)' (Y)=(\cB   \de    E^+_k)(Y) $.      Then  we  have    by   (\ref{unsung})   and   lemma \ref{citizen}    for  $Y \in \cD^0_{k+1}$
  \begin{equation} \label{take2}
  |(\de    E^+_k )'(Y, \phi,  \cW_k) |  \leq  \cO(1) L^3 \la_k^{1/4-10 \ep}  e^{- \ka' d_{LM}(Y) } 
 \end{equation} 
 and  now   $\de E^+_k  =  \sum_Y  (\de E^+_k)'(Y)$.

 \subsection{localization}

In preparation for the cluster expansion    we  localize    the dependence  of $ ( \de    E^+_k )'(Y, \phi,  \cW_k ) $  in  $W$.
Consider the random walk expansion   $G_k   =  \sum_{\om} G_{k,\om} $  of  section  \ref{randomwalk}.   However to match 
the  $LM$ polymers    in  $ ( \de    E^+_k )'(Y ) $  we  take  an expansion based on $LM$ cubes rather than  $M$ cubes.
As   explained  in    section  \ref{decoupling}   we  introduce  a   variable   $s=  \{ s_{  \square}  \}$ with  $ 0 \leq s_{  \square}  \leq 1$ for every  $LM$- cube $ \square$.   In  the  random   walk   expansion  $G_k   =  \sum_{\om} G_{k,\om} $   we  weaken the coupling
through   $\square$    by  introducing    
\begin{equation}  \label{again}
G_{k}(s) =    \sum_{\om} s_{\om}  G_{k,\om}
\end{equation}
We   can   also   give  a weakened   form for $C_k^{1/2}$   as  follows.   We  use the representation
(\ref{one}),  (\ref{two}),   (\ref{three}) for  $C_k^{1/2}$ in terms  of  $G_{k,r}, A_{k,r}$.    Now    $G_{k,r}$ also    has   a  random walk  expansion   $G_{k,r}   =  \sum_{\om} G_{k,r,\om} $  as  explained in  appendix  \ref{E},  but now also taken based on $LM$ cubes.
Hence  it    has  a weakened  version   
 \begin{equation}
G_{k,r}(s) =    \sum_{\om} s_{\om}  G_{k,r,\om}
\end{equation}
The  weakened form  for  $C_k^{1/2}$  is now
\begin{equation}
\begin{split}
C_k^{1/2}(s)  = &   \frac{1}{\pi}  \int_0^{\infty}  \frac{dr}{\sqrt{r}}  C_{k,r}(s) \\
C_{k,r}(s) =  &  A_{k,r}  +   a_k^2  A_{k,r} Q_k  G_{k,r}(s) Q_k^T A_{k,r} \\
\end{split}
\end{equation}
Combining these we  get  a weakened   form  for  $\cW_k  =    a_k G_k  Q_k^TC_k^{1/2}W$
which is  
\begin{equation}
\cW_k(s)  \equiv    a_k G_k(s)  Q_k^TC^{1/2}_k(s)W
\end{equation}

    The  term    $(\de E_k^+ )'(Y, \phi, \cW_k)  $   is local  in  $\phi,  \cW_k$,  but   not    in  $W$  because $\cW_k$ at  any point depends on $W$ at every  point.
  We  remedy
 this with the following localization expansion.
 Break     the coupling   outside of  $Y$  by    interpolating with       $\de  E_k^+( Y,   \phi,\cW_k(s) ) $.  
 Use the identity    
 \begin{equation}
 f(s_{\square} = 1)  =  f(s_{\square} = 0)  +  \int_0^1   ds_{\square} \frac{\pa f}{\pa  s_{\square} }
 \end{equation}
 successively in each variable  in   $s_{Y^c}  =  \{s_{\square} \}_{\square  \in  Y^c}$  and obtain
\begin{equation}  \label{summer}
\begin{split}
(\de  E_k^+)'( Y)   
= & \sum_{Z \supset Y}        \de  E_k^+( Y,Z) \\
(\de  E_k^+)( Y,  Z; \phi,    W)   
= &  \int   ds_{Z-Y} 
 \frac { \pa  }{ \pa s_{Z-Y}}   \left[  (\de E_k^+)'(Y,    \phi,  \cW_k(s))   \right]_{s_{Z^c} = 0, s_Y=1}\\
 \end{split}
 \end{equation}
Now  we   write 
\begin{equation}   
\de E_k^+   =  \sum_Y  (\de  E_k^+)'( Y)    =   \sum_Y      \sum_{Z \supset Y}        \de  E_k^+( Y,Z) 
=  \sum_{Z}    (\de E^{+}_k )^{\loc}(Z) 
\end{equation}
where  
\begin{equation}    \label{newloc}
    (\de E^{+}_k )^{\loc}(Z) =  \sum_{Y  \subset  Z}        \de  E_k^+( Y,Z)   
\end{equation}
Here  $Y$ is connected  but  $Z$  may not  be.  However we have:

\begin{lem}   In the expansion    $\de E_k^+ =  \sum_{Z}    (\de E^{+}_k )^{\loc}(Z) $  we  can restrict to connected  $Z$, i.e.  $Z \in \cD^0_{k+1}$.  Furthermore   $ (\de E^{+}_k )^{\loc}(Z) =   (\de E^{+}_k )^{\loc} (Z,  \phi,  W)$ only  depends on $ \phi,  W$  on  $Z$.
\end{lem}
\bigskip

\pr
Consider  the  random walk expansion   (\ref{again}) for   $G_k(s)|_{s_{Z^c} =  0}$  which occurs   in $\cW_k(s)|_{s_{Z^c} =  0}$. 
 If  $\square  \subset  Z^c$ then  
$s_{\square}  = 0$  and  so 
then   $s_{\om} =0$  for any  path  $\om$  such that   $X_{\om}  \supset  \square$.   Thus in  $G_k(s)|_{s_{Z^c} =  0}$ 
paths such that  $X_{\om}$  intersect   $Z^c$  do not occur,    and we must have  $X_{\om}  \subset  Z$.
But    $X_{ \om }$  is connected   so  only  paths  such   that  $X_{\om}$   is   in a single  connected component of $Z$
contribute.    This  means  that   $G_k(s)|_{s_{Z^c} =  0}$  preserves the subspaces of     functions  on   the  various   connected components  of  Z.
The   same  is  true  of  $C^{1/2}_k(s)|_{s_{Z^c} =  0}$ and  
$\cM_k(s) )|_{s_{Z^c} =  0}   \equiv    a_k G_k(s)  Q_k^TC^{1/2}_k(s)|_{s_{Z^c} =  0}$.   But  we  are interested in
$\cW_k(s) )|_{s_{Z^c} =  0} =  \cM_k(s))W|_{s_{Z^c} =  0}$    on   $Y$
   which means  that  only   $ \cM_k(s))|_{s_{Z^c} =  0}$   restricted to   functions on   the component of 
$Z$  containing $Y$   contributes.     Therefore   derivatives   in   $  \pa  / \pa s_{Z-Y}$ for cubes  in other connected components of  $Z$   give  zero.   Hence  in (\ref{summer})    we  can restrict  the sum over  $Z \supset  Y$   to    connected  $Z$  which proves the first statement.
Furthermore we  see that   $\de  E_k^+( Y,  Z; \phi,    W)   $  and  hence   $   (\de E^{+}_k )^{\loc} (Z,  \phi,  W)$   only  depends on 
  $\phi, W$  in $Z$.  This completes the proof.

   \begin{lem}   \label{snort}
 For    $ \phi  \in  \frac12  \cR_{k}$   and       $|W|   \leq  p_{0,k}$
  \begin{equation}   \label{stinger}
 |  (\de E^{+}_k )^{loc}(Z,  \phi,  W)|   \leq     \cO(1)L^3  \la_k^{1/4-10\ep}   e^{ -L(\ka - 2\ka_0-2)  d_{LM}(Z) } 
 \end{equation}
 \end{lem}
\bigskip

\pr   For  $\square \subset  Z-Y$     we   consider  $s_{\square}$  complex and  satisfying  $|s_{\square}|  \leq  M^{1/2}$.    As   explained
in   section  \ref{decoupling}   the operator $G_k(s)$   satisfies bounds of the same form   as  $G_k$.     In the same 
way   $C_k^{1/2}(s)$   satisfies   bounds of the same  form   as   $C_k^{1/2}$.  Hence  $\cW_k(s)$  satisfies 
bounds of the same form as   $\cW_k$.     Therefore    $( \de E_k^+)'(Y,    \phi,  \cW_k(s) )   $    is analytic   
   in  $|s_{\square}|  \leq  M^{1/2}$   and  satisfies there  
\begin{equation}  \label{unsung2}
 | (\de E_k^+ )'(Y, \phi ,  \cW_k(s) ) |   \leq   \cO(1) L^3   \la_k^{1/4-10\ep}  e^{- \ka'  d_{LM}(Y)}
\end{equation}
  just  as  in  (\ref{take2}).  If we  let  $\ka_1  =  \frac12 \log M$   we can write  the condition as  $|s_{\square}|  \leq  e^{ \ka_1}$.
   Now  if  we   restrict  to   $|s_{\square}|  \leq   1$   we  get  Cauchy  bounds on the  derivatives:
   \begin{equation}
\Big| \frac { \pa  }{ \pa s_{Z-Y}}   \left[ ( \de E_k^+)'(Y,    \phi,  \cW_k(s))   \right]   \Big|
   \leq   \cO(1)L^3 \la_k^{1/4-10\ep}    e^{ - (\ka_1-1) |Z-Y|_{LM}  }   e^{- \ka'  d_{LM}(Y)}
\end{equation}
     We   can assume $\ka_1-1   \geq  \ka'$.  
Using this,  integrating over  $s_{Z-Y}$    and  summing   over    $Y \subset  Z$  yields  
   \begin{equation}  \label{tuna}
|   (\de E^{+}_k )^{loc} (Z,  \phi,  W)  |
\leq      \cO(1)L^3 \la_k^{1/4-10\ep}     \sum_{Y \subset  Z }     e^{ - \ka' |Z-Y|_{LM}    - \ka'  d_{LM}(Y)  }     
\end{equation}
We   show  below  that      $      |Z-Y|_{LM}       +  d_{LM}(Y)  \geq  d_{LM}(Z) $.
Then we can extract a  factor    $  e^{- (\ka'  - \ka_0)   d_{LM}(Z)}$ 
and obtain   
  \begin{equation}  \label{tuna2}
|   (\de E^{+}_k )^{loc} (Z,  \phi,  W)  |
\leq      \cO(1) L^3\la_k^{1/4-10\ep}   e^{- (\ka'  - \ka_0)   d_{LM}(Z)}   \sum_{Y \subset  Z }  e^{  - \ka_0  d_{LM}(Y)}        
\end{equation}
But  the sum is bounded by  $\cO(1)|Z|_{LM}$ by (\ref{sudsy}) in the appendix.   Furthermore   by  (\ref{ninety})
$|Z|_{LM}  \leq  \cO(1)( d_{LM}(Z)  +1) \leq   \cO(1)e^{d_{LM}(Z)}$.  Since  $\ka' - \ka_0 -1  \geq     L(\ka - 2\ka_0-2)$ this gives the result.

\begin{lem}  For  $X,Y \in \cD_k$ and  $X  \subset Y$: 
\begin{equation}  \label{salsa}
Md_{M}(Y)    \leq   M  |Y-X|_{M}       + M d_M(X)
\end{equation} 
\end{lem}
\bigskip

\pr 
Let  $\tau$ be    a minimal  tree  on  the $M$-cubes in $X$  of  length   $M d_M(X)$. 
Let  $(Y-X)_i$,   be the connected components   of  $Y-X$. 
Every component  $(Y-X)_i$  has  a cube  $\square_i$  adjacent to  a cube in  $\square_i'   \subset   X$ across
a  2-dimensional face.     Let   $x'_i$   be  the point in  $\square_i'$ which is a vertex of  $\tau$.   Now  extend the
tree  $\tau$  by  taking a  line from   $x'_i$  to  the translated point   $x_i$   in  $\square_i$.      Then extend it to
all  of   $(Y-X)_i$   by  taking  lines  across two dimensional faces  joining  translates of  $x_i$.  For  each  $i$ 
this adds a length  $M|(Y-X)_i|_M$.   Thus we have  constructed a tree joining  all the blocks  of  $Y$  of length
$M d_M(X)  +  \sum_i  M|(Y-X)_i|_M   =  M d_M(X)  +   M|Y-X|_M $.   This must be greater than the length  of 
a minimal tree   $Md_{M}(Y)    $.

\subsection{cluster expansion}

 The     fluctuation integral  is     now  
 \begin{equation}
 \Xi'_k(\phi)=  
  \int \exp     \Big(    \sum_{Y  \in \cD^0_{k+1}}     (\de E^{+}_k )^{\loc}(Y,  \phi,  W)  \Big)     d\mu^*_k(W)   
\end{equation}
The cluster  expansion gives this a local structure.  The result is:

\begin{lem}  \label{cluster0}(cluster expansion)  Let  $\la_k$ be sufficiently small.
For   $\phi  \in   \frac12   \cR_k$  
\begin{equation}  \label{sunshine}
  \Xi'_k(\phi)       = \exp  \Big(  \sum_{Y \in \cD^0_{k+1} }      \ E^\#_k(Y,\phi)  \Big)
\end{equation}
 where
\begin{equation}  \label{osprey}
| E^\#_{k}(Y, \phi    ) |  \leq   \cO(1)  L^3 \la_k^{1/4-10\ep} 
  e^{ - L(  \ka     -5 \kappa_0 -5 )  d_{LM}(Y)}  
\end{equation}
\end{lem}
\bigskip

 For the standard    proof   see appendix  \ref{B}.   Here it is applied with $LM$ cubes.   The  bound  (\ref{osprey}) follows from the  bound  
(\ref{stinger}).     The latter   is   small  enough to  fall  within the range of validity of the cluster expansion   if  
 $ \cO(1)L^3\la_k^{1/4-10\ep}  \leq  c_0$.

  Inserting this result  into   (\ref{sunny})  and defining   $E^\#(\phi)  =  \sum_Y  E^\#(Y, \phi)$     we have
   \begin{equation}
 \Xi_{k}(\phi)=   \exp \Big(  - \vep_k^0  \Vol(  \bbT^0_{\sM + \sN -k})  +  E_k^+ ( \phi )  +    \ E^\#_k(  \phi)  \Big)
\end{equation}
Insert this into  (\ref{stanley6})  and  obtain
\begin{equation}  \label{stanley7}
\begin{split}
&\tilde  \rho_{k+1}  (\Phi_{k+1})   L^{ -  | \bbT^1_{\sM +\sN -k} |/2}  \\
=&  Z_{k+1}\    \exp 
 \Big( -S^0_{k+1}(\Phi_{k+1},   \phi^0_{k+1})  - \vep_k^0  \Vol(  \bbT^0_{\sM + \sN -k})  +  E_k^+ ( \phi^0_{k+1} ) 
  +    \ E^\#_k(  \phi^0_{k+1} \Big)  
\end{split}
\end{equation}

\subsection{scaling}

From the last  expression we form  $ \rho_{k+1} ( \Phi_{k+1})  =  \tilde  \rho_{k} ( \Phi_{k+1,L}) L^{ -  | \bbT^1_{\sM +\sN -k} |/2} $.

     We  have  seen in lemma \ref{tiny}  that   $\phi^0_{k+1}(\Phi_{k+1, L})  =\phi_{k+1,L}$ and that    that  
$S^0_{k+1}(\Phi_{k+1},   \phi^0_{k+1})$  scales to  $S_{k+1}(\Phi_{k+1},   \phi_{k+1})$.  We   also  have   
$ \vep^0_{k} \Vol(  \bbT_{\sM + \sN -k}) = L^3  \vep^0_{k}  \Vol(  \bbT_{\sM + \sN -k-1})$.

In  $E_k^+  =  E_k   -  V_k$   we have 
\begin{equation}
V_k(\phi_{k+1,L})  = L^3 \vep_{k} \Vol(\bbT_{M+N-k-1} )  
  +  \frac 12  L^2 \mu_{k}   \|  \phi_{k+1}   \|^2    +  \frac14  L \la_k  \int  \phi_{k+1}^4 
\end{equation}
For     $E_k$   we  reblock before scaling,   and   have for  $\phi \in \cR_{k+1}$   that   $ E_k( \phi_{L} )  =  (\cB E_k)_{L^{-1}}(  \phi)$.
Since   $E_k$  is normalized for small polymers    lemma \ref{scalinglem} says
\begin{equation}  \label{bb1}
\| (\cB E_k)_{L^{-1}}\|_{k+1, \ka}  \leq  \cO(1) L^{-\ep}\|  E_k\|_{k, \ka} 
\end{equation}

The function $ E^\#_k $ is already reblocked.   We   have
\be   E^\#_k( \phi_L) =  \sum_{Y \in \cD^0_{k+1}}  E^\#_k(Y, \phi_L) =  \sum_{X \in \cD_{k+1}}  E^\#_k(LX, \phi_L)
=    \sum_{X \in \cD_{k+1}}  E^\#_{k, L^{-1}}(X, \phi)  \equiv   E^\#_{k, L^{-1}}( \phi) 
\ee
For  $\phi  \in \cR_{k+1}$ we have   $\phi_L  \in  \frac12  \cR_k$    and so
by     (\ref{osprey})  
\be   | E^\#_{k, L^{-1}}(X, \phi) |  \leq    
 \cO(1)  L^3 \la_k^{1/4-10\ep} 
  e^{ - L(  \kappa    -5 \kappa_0 -5 )  d_{LM}(X)}  
\ee
We  need    $L( \kappa    -5 \kappa_0 -5)   \geq  \ka$   or equivalently     $\ka  \geq   5L(L-1)^{-1}(\ka_0 +1)$.   Since 
$L \geq  2$  it suffices that    $\ka  \geq  10(\ka_0 +1)$   which we assume.  
Then
\be    \label{bb2}
 \| E^\#_{k, L^{-1}} \|_{k+1, \ka}    \leq    
 \cO(1)  L^3 \la_k^{1/4-10\ep} 
\ee

 Altogether  then 
\begin{equation}  \label{stanley8}
\begin{split}
  \rho_{k+1}  (\Phi_{k+1})  
=&  Z_{k+1}\   \exp 
 \Big( -S_{k+1}(\Phi_{k+1},   \phi_{k+1})  - L^3 (\vep_k + \vep_k^0 )\Vol(  \bbT_{\sM + \sN -k-1})   \\ &    - \frac12  L^2 \mu_{k}   \|  \phi_{k+1}   \|^2   
  -  \frac14      \la_{k+1}  \int  \phi_{k+1}^4  +  (\cB E_k)_{L^{-1} }( \phi_{k+1} ) 
  +     E^\#_{k,L^{-1}}(  \phi_{k+1}) \Big)  
\end{split}
\end{equation}

\subsection{completion of the proof}

Neither   $ (\cB E_k)_{L^{-1} }$    nor     $E^\#_{k,L^{-1} }$  are normalized   for small polymers,   and we
need this feature to complete the induction.
\footnote{  If  we  had   normalized  $E_k(X)$   for all polymers, not just small polymers,  then  $(\cB E_k)_{L^{-1}}(X)$  would still
be  normalized   and  $\cL_1, \cL_2$   would not  appear  below.   This  strategy is possible,  but presents 
other  difficulties.}
We  remove  energy and mass terms to normalize them.

By     (\ref{renorm2}) 
 \begin{equation}  
 ( \cB E_k)_{L^{-1}}(\phi_{k+1}) =     -   \cL_1E_k   \Vol(  \bbT_{N+M-k-1} ) 
 -   \frac12   \cL_2E_k   \|  \phi^2_{k+1} \| 
+  (\cL_3 E_k) (\phi_{k+1})
\end{equation} 
where    
 \begin{equation}
\begin{split}
\cL_1E_k  =&   \vep( ( \cB E_k)_{L^{-1}})   
    \\
 \cL_2E_k  =&   \mu(( \cB E_k)_{L^{-1}})          \\
 \cL_3 E_k  =  &     \cR  (( \cB E_k)_{L^{-1}} )      \\ 
\end{split}
\end{equation}
From the bound    (\ref{bb1})
and       lemma \ref{rlem}  and   lemma   \ref{study} we have that
$| \cL_1E_k  |  $   and    $\|\cL_3E_k \|_{k+1, \ka}$   are  bounded by   $ \cO(L^{-\ep}) \|E_k\|_{k , \ka}  $
and that   $| \cL_2E_k  |  $  is bounded by   $ \cO(L^{-\ep})   \la_k^{1/2 +6\ep }  \|E_k\|_{k, \ka}$.
These are the required bounds

We   also   apply   (\ref{renorm2})   to   $E^\#_{k,L^{-1} }$   but now tack on the extra term  $\vep_k^0$    
We   have
 \begin{equation}  
E^\#_{k,L^{-1} }(\phi_{k+1})   -  L^3  \vep_k^0     \Vol(  \bbT_{N+M-k-1} ) 
=  -   \vep_k^*  \Vol(  \bbT_{N+M-k-1} ) 
 -   \frac12  \mu_k^* \|  \phi^2_{k+1} \| 
+  E_k^*  (\phi_{k+1})
\end{equation} 
where
\begin{equation}
\begin{split}
\vep_k^*  =    &   L^3  \vep_k^0      +       \vep( E^\#_{k,L^{-1} }) \\
\mu_k^*  =  &    \mu( E^\#_{k,L^{-1} })   \\
E_k^*   =  &  \cR( E^\#_{k,L^{-1} }  ) \\
\end{split}
\end{equation}
From the bound   (\ref{bb2})   and       lemma \ref{rlem}  and   lemma   \ref{study},  
$|\vep_k^*| $   and     $\|E_k^* \|_{k+1, \ka}$   are bounded by  $\cO(1) L^3  \la_k^{1/4- 10 \ep}$
and    $|  \mu_k^*|$  is   bounded  by $ \cO(1) L^3  \la_k^{3/4- 4 \ep}$.  These  are the required bounds.

Insert   these  expansions  into   (\ref{stanley8})   and obtain the final form 
\begin{equation}  \label{stanley9}
\begin{split}
  \rho_{k+1}  (\Phi_{k+1})  
=&  Z_{k+1}\   \exp 
 \Big( -S_{k+1}(\Phi_{k+1},   \phi_{k+1})  - \vep_{k+1} \Vol(  \bbT_{\sM + \sN -k-1})   \\ & 
    -  \frac12 \mu_{k+1}   \|  \phi_{k+1}   \|^2   
  -  \frac14     \la_{k+1}  \int  \phi_{k+1}^4  +   E_{k+1}( \phi_{k+1})   \Big)
   \end{split}
\end{equation}
where   $ \vep_{k+1},  
\mu_{k+1},  E_{k+1} $     are given by    (\ref{recursive}).  This completes the proof of theorem \ref{lanky}.

\subsection{derivatives}

The  previous proof was  carried out under the assumption  that  $\la_k$ is  small  and    $\mu_k  \in  \bbR,  E_k  \in  \textrm{Re}  (\cK_k^{norm})$  satisfy 
$|\mu_k |     \leq  \la_k^{1/2} $  and     $   \|  E_k  \|_{k, \ka}  \leq  1$.
In  this   domain  $\mu^*_k = \mu^*_k(\la_k,  \mu_k, E_k)$    and $ E_k^*  =  E_k^*   (\la_k,  \mu_k, E_k)$
satisfy  the bounds
\begin{equation}  \label{newt}
\begin{split}
|\mu_k^*|  \leq &   \cO(1)L^3\la_k^{3/4  - 4 \ep}  \\
\|E_k^* \|_{k, \ka}  \leq  &  \cO(1)L^3 \la_k^{1/4-  10  \ep}    \\
\end{split}
\end{equation}
However the proof  works as well  for     $\mu_k  \in  \bbC,  E_k  \in \cK_k^{norm}$ 
with exactly the same  bounds,   and  
 one can show  that       $\mu_k^*,   E_k^*$
are  analytic functions  of  $\mu_k, E_k$  on  this domain. This  means we can use     Cauchy bounds 
to  get estimates on partial  derivatives   in  a slightly smaller  region.

\begin{lem}     \label{smooth}
In the region   $|\mu_k|  \leq   \frac12    \la_k^{1/2}   $  and     $   \|  E_k  \|_k  \leq  \frac 12$     we  have 
\begin{equation}
\left|\frac{  \pa  \mu_k^*}{  \pa \mu_k}\right|    \leq   \cO(1)L^3 \la_k^{1/4-4 \ep} 
\hs   \left\|\frac{\pa  \mu_k^*}{  \pa E_k}\right\|    \leq   \cO(1)L^3 \la_k^{3/4- 4 \ep} 
\end{equation}
\begin{equation}
\left|\frac{\pa   E_k^*}{  \pa \mu_k}\right|    \leq   \cO(1)L^3 \la_k^{-1/4- 10 \ep} 
\hs   \left\|\frac{\pa  E_k^*}{  \pa E_k}\right\|    \leq   \cO(1) L^3\la_k^{1/4- 10 \ep} 
\end{equation}
\end{lem}
\bigskip

\pr
We  have 
\begin{equation}
\frac{\pa  \mu_k^*}{  \pa \mu_k}  =   \frac{d}{ dt} \Big[   \mu_k^*(\la_k, \mu_k  + t,E_k)\Big]_{t=0}
=  \frac{1}{2 \pi i}   \int_{|t|= \frac 12\la_k^{1/2}}\frac{1}{t^2}  \mu_k^*(\la_k, \mu_k + t,E_k)  dt
\end{equation}
whence
\begin{equation}
\left|\frac{\pa  \mu_k^*}{  \pa \mu_k} \right|
\leq  \cO(1)    \la_k^{-1/2}   ( L^3 \la_k^{3/4 - 4\ep})
\leq   \cO( 1)L^3 \la_k^{1/4-  4 \ep}  
\end{equation}
We also have  for  $\|  \dot{E}  \|_{k, \ka}  \leq  1$
\begin{equation}
<\frac{\pa  \mu_k^*}{  \pa E_k}, \dot{E}>  
\equiv      \frac{d}{ dt} [   \mu_k^*(\la_k,   \mu_k  ,E_k+t\dot{E}) ]_{t=0}   
=  \frac{1}{2 \pi i}   \int_{|t|=\frac12}\frac{1}{t^2}  \mu_k^*(\la_k, \mu_k,E_k+t    \dot{E})  dt
\end{equation}
whence 
\begin{equation}
|<\frac{\pa   \mu_k^*}{  \pa E_k}, \dot{E}>  |   \leq         \cO(1)L^3\la_k^{3/4- 4 \ep} 
\end{equation}
Then   
\begin{equation}
\|\frac{\pa   \mu_k^*}{  \pa E_k}   \| =   \sup_{  \|\dot{E}\|_{k, \ka}   \leq  1}
|<\frac{\pa    \mu_k^*}{  \pa E_k}, \dot{E}>  |     \leq  \cO(1)L^3\la_k^{3/4- 4 \ep}
\end{equation}
The  estimates  on the derivatives of $E^*_k$   are  similar.

\section{the flow}

We    seek  well-behaved    solutions of   the    RG   equations   (\ref{recursive}).  
\footnote{This  section is not particularly due to Balaban.  We   study  the RG flow by a discrete dynamical systems approach.
Somewhat  similar methods can be  found in   \cite{BDH98},  \cite{Bry07}.   However
those papers  are concerned with  infrared problems,  not ultraviolet  problems of the type considered here.}
 We  continue to treat   $\la_k$  as a parameter, not a dynamical variable.    
 Thus  the equations  of  interest are
 \begin{equation}  \label{recursive3}
\begin{split}
\vep_{k+1}  = &  L^3  \vep_k   +  \cL_1 E_k   + \vep_k^*  \\
\mu_{k+1}   =&   L^2 \mu_k  +  \cL_2E_k  + \mu_k^*  \\
E_{k+1}   =&    \cL_3 E_k  +  E^*_k \\
 \end{split}
\end{equation}
Keep in mind that  the quantities   $\vep_k,  \mu_k, \la_k,  E_k$  determine a density $\rho_k$    on the lattice
$\bbT^0_{\sM+\sN -k}$  as given by     (\ref{basic}) (\ref{basic20}).

The  transformation is defined
as   long as   $\la_k$ is sufficiently small   and       $|\mu_k|   \leq   \la_k^{1/2}   $  and   $  \|  E_k  \|_k  \leq  1$.  
 We  make no restriction on 
the size of  the bare  coupling   $\la$   but the initial  values  $\la_0 = \la_0^{\sN} =
L^{-\sN} \la$      will be  be small  enough    
for  $\sN$  sufficiently large, and we  assume the other conditions are satisfied  initially.   We   iterate  it as long as the conditions are satisfied.
Our  goal is to show  that for   any    $\sN$   we can choose  the  initial   point     so  that the solution exists   for  $k=0,1,  \dots,  K$
with    $K = \sN - \De$   and    $\De \geq   0 $ independent of  $\sN$.
Then  at  $k=K$   we  are on the lattice  $\bbT^0_{\sM+\sN- K} =  \bbT^0_{\sM+\De}$
and  can    make estimates  on  $  \rho_K$   uniformly in $\sN$  (for small fields).

To accomplish  this  tuning    we  do  not  at  first  specify  the initial values  for  $\vep_k,    \mu_k$  but instead 
specify  final values   for   these  quantities  which for simplicity we  take to be zero.    Thus  we look for solutions  $\vep_k,  \mu_k,  E_k$    for    $k = 0,1,2,  \dots  , K$
satisfying  
\begin{equation} \label{bc}
\vep_K  = 0       \hs      \mu_K  = 0     \hs     E_0  =0    
  \end{equation}
  This is  non-perturbative  renormalization -  the initial  values for  $\vep,  \mu$  will depend on $K$ and hence  $N$.  Note that the total   mass at  level  $K$  is then   $\bar  \mu_K  +   \mu_K  =  \bar  \mu_K  $.
  
At this  point  we  temporarily  drop $\vep_k$   as  a variable since it does not 
afffect the others.  Then  we    rewrite  the flow equation   as   
 \begin{equation}  \label{recursive4}
\begin{split}
\mu_k   =&   L^{-2}( \mu_{k+1} -  \cL_2E_k - \mu_k^*)  \\
E_k   =&    \cL_3 E_{k-1}  +  E^*_{k-1} ) \\
 \end{split}
\end{equation}
 The  first  equation is     for     $k=0,1,2,  \dots,  K-1$   with    value at  $K$  given by   $  \mu_K  =  0 $.   
The  last  equation    for  $k=1,2,  \dots, K$   with  $E_0  = 0$.
These equations  have the same solutions as  (\ref{recursive3}), but  are   contractive and hence more tractable.  
We   analyze them as a fixed point problem.

Let    $\xi_k  =  (\mu_k, E_k)$  be  an  element of the real  Banach space 
  $\bbR \times  \textrm{Re} ( \cK_k^{norm} )   $    and consider sequences      
\begin{equation}
\underline{ \xi }  =  ( \xi_0,  \dots  ,   \xi_K)
 \end{equation}
 Pick    a fixed  $\beta$  satisfying
\begin{equation}
0<  \beta   <  \frac 14 -  10 \ep
\end{equation}
and  let    $\sB$   be the  Banach  space of all   such   sequences
 with norm   
  \begin{equation}
\|  \underline{\xi } \|  =  \sup_{0 \leq  k   \leq  K}       \{   \la_k^{-\frac12 - \beta}  | \mu_k|, 
 \la_k^{-  \beta}     \| E_k \|_{k, \ka}   \}  
\end{equation}
This anticipates the kind of growth we can establish  for solutions.
Let  $\sB_0$   be the   subset  of all sequences satisfying the  boundary 
 conditions.
 Thus   
 \begin{equation}
 \sB_0  =  \{\underline{ \xi}   \in  \sB:    \mu_K  =  0,        E_0  =0\} 
 \end{equation}
  This is a complete metric space  with distance  
  $\| \underline {  \xi } - \underline {\xi' }  \|$.
 Finally   let  
 \begin{equation}
 \sB_1  = \sB_0 \cap    \{\underline{ \xi}   \in  \sB:    \| \underline{\xi} \| <1  \}   
 \end{equation}

 Next  define an    operator  $\un{\xi' } =T \un{\xi } $    by      
  \begin{equation}  \label{recursive5}
\begin{split}
\mu'_k   =&   L^{-2}( \mu_{k+1} -  \cL_2E_k - \mu_k^*)   \\
E'_k   =&      \cL_3 E_{k-1}  +  E^*_{k-1} )  \\
 \end{split}
\end{equation}
Then   $\underline{  \xi }$ is a solution  of  (\ref{bc}), (\ref{recursive4}) iff  it is a fixed point for  $T$ on 
$\sB_0$.     We    look for  such   fixed points  in $\sB_1$.

We  proceed under the assumption that  
 \begin{equation}
 \la_K =   \la^{\sN}_K  =  L^{-(\sN-K)}  \la  = L^{-\De} \la 
 \end{equation}
is sufficiently small.    
This  can  be  arranged either    by taking $\la$ small
   (in which case we can take  $\De = 0$  and  $K=\sN$),  or  more generally    by  taking $\De$ large.
If       $\la_K =  L^{-\De}\la$  is sufficiently small then    $T$   is defined on     $\sB_1$.   
 This follows since  we have   $\la_k \leq  \la_K$  small  and   
 \begin{equation}
| \mu_k| \la_k^{-1/2}   \leq      \la_k^{\beta}      \hs     \| E_k  \|_{k, \ka}   \leq   \la_k^{\beta}   
 \end{equation}
 which is well within the allowed region    $| \mu_k| \la_k^{-1/2}   \leq  1,    \| E_k  \|_{k, \ka}  \leq  1$.

\begin{lem}   Let  $\la_K  =   L^{- \De}\la   $  be sufficiently small.  Then   for all  $\sN  \geq  \De$    and   $K =  \sN-   \De$
\begin{enumerate}
\item   The  transformation   $T$     maps   the set
  $\sB_1$  to itself.  
\item   There is a unique fixed point   $T\underline{ \xi}  = \underline{ \xi }$  in this  set. 
\end{enumerate}
\end{lem}
\bigskip

\pr  (1.) We   use the bounds of  theorem   \ref{lanky}  for   $\cL_2,   \cL_3$
(replacing  $\cO(1)L^{-\ep}$ by   $1$ )   and  for  $\mu_k^*,  E_k^*$.
To   show the the map sends  $\sB_1$ to itself
we  estimate
\begin{equation}  \label{jelly}
\begin{split}
\la_k^{- \frac12  -\beta} |\mu'_k| 
 \leq   &
  \la_{k}^{- \frac12 -\beta} L^{-2}
\Big(   | \mu_{k+1}|  + \la_k^{1/2 + 6 \ep}  \| E_k\|_{k, \ka} +          \cO(1)  L^3  \la_k^{3/4  - 4 \ep}   \Big)\\
  \leq   &
   L^{\beta-  \frac32}\Big[\la_{k+1}^{- \frac12 -\beta}  | \mu_{k+1}|\Big]   +
    L^{-2} \la_k^{1/4 + 6 \ep}\Big[ \la_k^{-\beta} \| E_k\|_{k, \ka} \Big]  +
    \cO(1) L  \la_{k}^{ \frac14 -\beta- 4 \ep}\\
    \leq  & \frac12 (  \|  \underline  \xi  \|  +   1   )  \leq   1  \\
\end{split}
\end{equation}
Here  we use that   $L^{\beta - 3/2}  \leq   1/4$  for  $L$ large,   that   $\la_{k+1}^{- \frac12 -\beta}  | \mu_{k+1}| \leq   \|  \underline{\xi}  \|$, 
that  $L^{-2} \la_k^{6 \ep}   \leq  1/4$,   
  that   $ \la_k^{-\beta} \| E_k\|_{k, \ka}  \leq   \|  \underline{\xi}  \|$  and   that  
  $  \cO(1) L  \la_{k}^{ \frac14 -\beta- 4 \ep}  \leq  1/2$  for  $ \la_k$ small  (depending on  $L$).  
 We   also   
have  for  $L$  sufficiently large
\begin{equation}
\begin{split}
\la_k^{-\beta} \|E'_k\|_{k, \ka}  
   \leq   &  \la_{k}^{-\beta}   \Big( \|  E_{k-1} \|_{k-1,\ka}   +      \cO(1)L^3 \la_{k-1}^{1/4-  10  \ep}          \Big)\\
   \leq    &    L^{  - \beta}  \Big[  \la_{k-1}^{-\beta} \| E_{k-1} \|_{k-1, \ka} \Big]  +  \cO(1) L^{3- \beta}
     \la_{k-1}^{1/4- \beta -10 \ep}\\
       \leq  & \frac12 (  \|  \underline  \xi  \|  +   1   )  \leq   1  \\
     \end{split}
\end{equation}
 Combining this  with  (\ref{jelly}) yields  $ \|  T (\underline{ \xi  } ) \|   \leq   1$  as  required.
\bigskip

\noindent
(2.)  By the standard fixed point theorem in a  complete  metric space it suffices to show   that   the mapping is a contraction.
 We show  that   under our assumptions
\begin{equation}
\| T( \underline{ \xi_1}) -   T( \underline{ \xi_2}) \|  \leq  \frac12
\|  \underline{ \xi_1}- \underline{ \xi_2} \| 
\end{equation}

First  for the $\mu$  terms   we have   (suppressing the dependence of  $\mu_k^*$  on  $\la_k$)
\begin{equation}
\mu'_{1,k}-\mu'_{2,k}   =   L^{-2}\Big( ( \mu_{1,k+1}-  \mu_{2, k+1} )  -  
\cL_2   (   E_{1,k}  -   E_{2,k}  )
  - (\mu_k^*(  \mu_{1,k}, E_{1,k}) -  \mu_k^*(  \mu_{2,k}, E_{2,k}))  \Big)
\end{equation}
Then     \begin{equation}
\begin{split}
&  \la_k^{- \frac 12  - \beta}|\mu'_{1,k}-\mu'_{2,k}| \leq  
 L^{-2}\la_k^{- \frac 12  - \beta}  | \mu_{1,k+1}-  \mu_{2, k+1} |  \\
 + &  L^{-2}  \la_k^{- \beta   + 6 \ep}  \|    E_{1,k}  -   E_{2,k}    \|_{k, \ka}
+  L^{-2}  \la_k^{- \frac 12  - \beta} \Big|\mu_k^*(  \mu_{1,k}, E_{1,k}) -  \mu_k^*(  \mu_{2,k}, E_{2,k})\Big|  \\
  \end{split}
\end{equation}
The  first   term  is 
\begin{equation}
 L^{\beta - \frac32}\Big[\la_{k+1}^{- \frac 12  - \beta}   | \mu_{1,k+1}-  \mu_{2, k+1} |\Big]
\leq     L^{\beta - \frac32}     \|  \underline{ \xi_1}- \underline{ \xi_2} \| 
\end{equation}
The second term  is   
\begin{equation}
 L^{-2}  \la_k^{ 6 \ep} \Big[ \la_k^{-\beta} \|    E_{1,k}  -   E_{2,k}    \|_{k, \ka}  \Big]
 \leq     L^{-2}  \la_k^{ 6 \ep}   \|  \underline{ \xi_1}- \underline{ \xi_2} \| 
\end{equation}
For the last  term  we  write   with  $\mu(t)   =t\mu_{1,k} +  (1-t) \mu_{2,k}$
and
 $E(t)   =tE_{1,k} +  (1-t) E_{2,k}$  and
\begin{equation}
\begin{split}
&\mu^*_k(\mu_{1,k}, E_{1,k}) -\mu^*_k(\mu_{2,k}, E_{2,k}) \\
=  &  \mu^*_k(\mu_{1,k}, E_{1,k}) - \mu^*_k(\mu_{2,k}, E_{1,k}) +
\mu^*_k(\mu_{2,k}, E_{1,k}) -\mu^*_k(\mu_{2,k}, E_{2,k}) \\
=    &    \int_0^1  \frac{\pa  \mu^*_k}{ \pa  \mu_k}(\mu(t), E_{1,k})(\mu_{1,k}- \mu_{2,k} ) dt
 + \int_0^1 \Big< \frac{  \pa \mu^*_k}{    \pa  E_k}(\mu_{2,k}, E(t)),   E_{1,k}  -   E_{2,k}   \Big>        \\
\end{split}
\end{equation}
We  use the bounds    $ |   \pa \mu^*_k/  \pa  \mu_k| \leq   \cO(1)L^3\la_k^{1/4- 4\ep}$  
 and  $\|  \pa \mu^*_k  /  \pa  E_k  \|_k \leq  \cO(1)L^3\la_k^{3/4- 4 \ep}$   from  lemma   \ref{smooth}. 
       Thus we have   
 \begin{equation}
\begin{split}
 &   \la_k^{- \frac 12  - \beta} \Big|\mu_k^*(  \mu_{1,k}, E_{1,k}) -  \mu_k^*(  \mu_{2,k}, E_{2,k})\Big|  \\
 \leq  &  \la_k^{- \frac 12  - \beta}   \Big( \cO(1)L^3 \la_k^{1/4- 4 \ep}| \mu_{1,k}- \mu_{2,k} |    +    
 \cO(1)L^3 \la_k^{3/4- 4 \ep}\| E_{1,k}- E_{2,k} \|_{k, \ka} \Big)  \\
  \leq  &  \cO(1)L^3  \la_k^{1/4- 4 \ep}\Big[     \la_k^{- \frac 12  - \beta}  | \mu_{1,k}- \mu_{2,k} |  \Big]  +    
  \cO(1)L^3  \la_k^{1/4- 4 \ep}\Big[     \la_k^{  - \beta} \| E_{1,k}- E_{2,k} \|_{k, \ka} \Big] \\
   \leq    &  \cO(1)L^3 \la_k^{1/4- 4 \ep}  \|  \underline{ \xi_1}- \underline{ \xi_2} \| \\
 \end{split}
\end{equation}
Altogether then  for $L$ large and $\la_k$  small   
 $ \la_k^{- \frac 12  - \beta}|\mu'_{1,k}-\mu'_{2,k}| \leq  1/2 \|  \underline{ \xi_1}- \underline{ \xi_2} \| $ as  required.
  \bigskip

Now consider the $E$ terms.  We have  
\begin{equation}
E'_{1,k}-E'_{2,k}   =   \cL_3 (E_{1,k-1}  -  E_{2,k-1})    +   ( E^*_{1,k-1}(\mu_{1, k-1}, E_{1, k-1})  - E^*_{2,k-1}(\mu_{2, k-1}, E_{2, k-1}))          
\end{equation}
Then
\begin{equation}
\la_k^{  - \beta}\|E'_{1,k}-E'_{2,k}\|_{k, \ka} 
\leq   L^{- \beta }    \la_{k-1}^{  - \beta}
\Big(  \|  E_{1,k-1}  -      E_{2,k-1}\|_{k-1, \ka}     +    \| E^*_{1,k-1}  -   E^*_{2,k-1}\|_{k-1, \ka}          \Big ) 
\end{equation}
The first  term  is  bounded by   $ L^{- \beta }  \|  \underline{ \xi_1}- \underline{ \xi_2}\| $.
For the   second    term  let
  $\mu(t)   =t\mu_{1,k-1} +  (1-t) \mu_{2,k-1}$
and
 $E(t)   =tE_{1,k-1} +  (1-t) E_{2,k-1}$ and write
  \begin{equation}
\begin{split}
& E_{k-1}^*(\mu_{1,k-1}, E_{1,k-1}) - E_{k-1} ^*(\mu_{2,k-1}, E_{2,k-1}) \\
=    &    \int_0^1  \frac{ \pa  E_{k-1}^*}{ \pa \mu_ {k-1}}(\mu(t), E_{1,k-1})(\mu_{1,k-1}- \mu_{2,k-1} ) dt
 - \int_0^1  \left< \frac{ \pa  E_ {k-1} ^*}{ \pa E_ {k-1}}(\mu_{2,k-1}, E(t)),   E_{1,k-1}  -   E_{2,k-1} \right>    \\
\end{split}
\end{equation}
We  use the bounds    $|  \pa E_ {k} ^*/ \pa \mu_k|   \leq \cO(1)L^3 \la_ {k} ^{-1/4- 10 \ep}$  
 and  $\|  \pa  E_k^*/ \pa E_ {k} \|_ {k}  \leq \cO(1) L^3\la_ {k-1} ^{1/4- 10 \ep }$   from  (\ref{smooth}). Then we have   
 \begin{equation}
\begin{split}
&   L^{- \beta  }  \la_{k-1}^{  - \beta}     \|E^*_{1,k-1}  -   E^*_{2,k-1}\|_{k-1, \ka}      \\
\leq &    L^{- \beta  }   \la_{k-1}^{  - \beta}     \Big( \cO(1)L^3  \la_ {k-1} ^{-1/4- 10 \ep } |\mu_{1,k-1}- \mu_{2,k-1} |
+  \cO(1)L^3  \la_ {k-1} ^{1/4- 10 \ep } \| E_{1,k-1}  -   E_{2,k-1} \|_{k-1, \ka}    \Big)   \\
\leq   &   \cO(1)     L^{3- \beta  }  \la_{k-1}^{1/4- 10 \ep } \Big[  \la_{k-1}^{- \frac12  - \beta} |\mu_{1,k-1}- \mu_{2,k-1} |  \Big]
 +  \cO(1)   L^{3- \beta  }  \la_{k-1}^{1/4- 10 \ep }\Big[ \la_{k-1}^{ - \beta}  \| E_{1,k-1}  -   E_{2,k-1} \|_{k-1, \ka}     \Big]   \\
 \leq     &  \cO(1)  L^{3- \beta  }  \la_k^{1/4- 10 \ep }  \|  \underline{ \xi_1}- \underline{ \xi_2} \| \\
\end{split}
\end{equation}
Altogether then for  $L$  large and $\la_k$ small we have   $\la_k^{  - \beta}\|E'_{1,k}-E'_{2,k}\|_{k, \ka} 
 \leq   \frac12\| \underline{ \xi_1}- \underline{ \xi_2} \|$
which  completes the proof.
 \bigskip

 Now  we  can state:

 \begin{thm}  \label{gsf}  Let   $\la_K   =  L^{-\De}  \la$  be  sufficiently  small.     Then  for    $N \geq  \De$  there is a unique sequence
  $\vep_k,  \mu_k,  E_k$    for    $k = 0,1,2,  \dots  , K=   \sN-\De$   
satisfying of  the dynamical equation   (\ref{recursive3}),   the boundary conditions  (\ref{bc}),     
and     
\begin{equation}  \label{somewhat}
  | \mu_k|  \leq     \la_k^{\frac12 + \beta}   \hs
      \| E_k \|_{k, \ka}   \leq    \la_k^{\beta} 
\end{equation}
Furthermore  
 \begin{equation}  \label{eg}
|\vep_{k}|   \leq     \cO(1) \la_k^{\beta} 
 \end{equation}
\end{thm}
\bigskip

\pr   This  solution is the fixed point from the previous  lemma and  the bounds  (\ref{somewhat})  are
a consequence.   

To complete the proof    we check  the  estimate
on the vacuum energy.
Once  $\mu_k,  E_k$  are  fixed   $\vep_k$   is  determined  by  
$\vep_{k+1}   =  L^3 \vep_k   +  \cL_1  (E_k)   +     \vep_k^*( \mu_k,  \la_k,  E_k) $ 
or by 
\begin{equation} 
\vep_k    =  L^{-3}(\vep_{k+1}  -    \cL_1  (E_k)    -  \vep_k^*) 
\end{equation}
starting with  $\vep_K  = 0$.
We  have    $|\cL_1  (E_k)|   \leq  \| E_k   \|_{k , \ka}  \leq  \la_k^{\beta}$  
and    $|  L^{-3}\vep_k^*|  \leq    \cO(1)  \la_k^{\frac14 - 10\ep}  \leq   \cO(1)  \la_k^{\beta}$.   Therefore   for some constant  $b= \cO(1)$ 
\begin{equation} 
|\vep_k|    \leq     L^{-3}|\vep_{k+1}|   +   b  \la_k^{\beta}   
\end{equation}
  At   $k = K-1$  it says  $   |\vep_{K-1}|    \leq   b   \la_{K-1}^{\beta  }  $.   This  gives an inequality for
  $|\vep_{K-2}|$  and we  repeat this process.  We  claim  that 
  in  general
\begin{equation}  \label{out}
  |\vep_{K -n}|       \leq   b  \Big (   \sum_{j=0}^{n-1}     L^{(\beta- 3)j}   \Big)  \la_{K-n}^{\beta}
  \end{equation} 
  Suppose it is true   for  $K-n$.  Then  
  \begin{equation}
  \begin{split}
  |\vep_{K -n-1}|     \leq &  L^{-3}   b  \Big (   \sum_{j=0}^{n-1}     L^{(\beta- 3)j}   \Big)  \la_{K-n}^{\beta}    
  +     b   \la_{K-n-1} ^{\beta}  \\
    \leq &     b  \Big (   \sum_{j=0}^{n}     L^{(\beta- 3)j}   \Big)  \la_{K-n-1}^{\beta}      \\
  \end{split}
  \end{equation}
Here we  used     $\la_{K-n}^{\beta}   =  L^{\beta}  \la_{K-n-1}^{\beta}$.    Thus  (\ref{out}) is true for  $K-n-1$,   hence
(\ref{out})  is established,   and  this  implies the result   (\ref{eg}) since the series converges.  
\bigskip

\res    \begin{enumerate}
\item    Our   method  is efficient,   but   the  estimates  are  not very sharp.  
 For example we  get    $\vep_k  =   \cO(    \la_k^{ \beta}  )$   and   $ \mu_k      =    \cO(    \la_k^{\frac 12 + \beta}  )$  for  $\beta< \frac 14$,  
 whereas pertubation theory suggests that  both  are   $\cO(    \la_k  )$.

\item    In  the second paper  we   analyze  the renormalization group transformations without the
small field assumptions   in this paper.  This  is accomplished   with  by splitting the 
fluctuation integrals  into  large and small field  region  at  each  step.       The result  is an
expansion with terms  labeled by     decreasing  sequence of  small   field  regions.   The leading term in this expansion  is     the   case  where  each  small field region is the whole  torus  - the case   considered in this paper.     The  bounds of this paper will also be useful in estimating the other terms in the expansion,  leading to a proof of  theorem  \ref{major}.
 \end{enumerate}

\newpage

\appendix

\section{estimates}      \label{basicsection}  

Let  $X$  be an   $M$-polymer  as defined in section  \ref{overview},  although not necessarily  in dimension $d=3$.  So $X$ is a connected union of  $M$-blocks $\square$ centered on lattice points.
  Let  $|X|_M$  be the number of  $M$-blocks in  $X$  and   let      $M d_M (X)$    be  the   length of a minimal   tree connecting    the  blocks   in    X. \bigskip

\begin{lem}    \label{totalbasic} {  \  }
\begin{enumerate}
\item  
There  are  constants   $a,b$  such that    for any   $\square$
\begin{equation}   \label{basic1}  
\sum_{X : \  X  \supset  \square}
\exp  (  -  a  |X|_M  )   \leq  b
\end{equation}
\item  There  are  constants   $\ka_0,K_0$  such that    for any   $\square$
\begin{equation}   \label{basic2}  
\sum_{X : \  X  \supset  \square}
\exp  (  -   \ka_0   d_M (X)   )   \leq  K_0
\end{equation}
\end{enumerate}
\end{lem}
\bigskip

\re  
The sums are independent of  $M$ so it suffices to prove it for  $M=1$.    In this case we drop the
subscript $M$.  The constants  depend only on the dimension.

A version of (\ref{basic2}) holds even if one drops the condition that  $X$ is connected.  This is presented in paper II.
\bigskip

\pr   \cite{GlJa87}  
\begin{enumerate}
\item
 We  have  
\begin{equation}
\sum_{X : \  X  \supset  \square}
\exp  (  -  a  |X|  )   =
\sum_{n \geq 1}  e^{-an}  |\{X  \supset  \square:  |X|  = n\}| 
\end{equation}
Thus we have to estimate   the number of polymers $X$  with  $|X| =n$ containing $\square$.
For each such   $X$  
consider  the connected graph with lines joining the centers of  adjacent cubes. Delete lines until you have  a tree.  The tree will connect all the cubes  in  $X$   and have  
  $n-1$  lines  of unit length.   The tree  can be traversed  with a path starting at $\square$    that  goes over each line  twice   and has  length   $2(n-1)$.  Distinct polymers give distinct  paths  so  the number of polymers is bounded  by   the number of  paths of length $2(n-1)$.
  But the latter can be estimated
by   $(2^d)^{2(n-1)}$.   Thus
\begin{equation}
\sum_{X : \  X  \supset  \square}
\exp  (  -  a  |X|  )   \leq  
\sum_{n  \geq  1}   e^{-an} ( 2^d )^{2(n-1)}   = 2^{-2d}  \sum_{n \geq1}  
\exp  (  (-a  + 2d \log 2)n  )   \leq   b
\end{equation}
for suitable  $b$ provided    $a>  2d \log 2$. 
\item   We  use the inequality  
$d(X)  \geq  3^{-d} |X|  -1$   quoted in   (\ref{ninety}).  Therefore
\begin{equation}
\sum_{X : \  X  \supset  \square}
\exp  (  -   \ka_0   d (X)   )   \leq  e^{\ka_0}     \sum_{X : \  X  \supset  \square}
\exp  (  - 3^{-d}  \ka_0   |X|   ) 
\leq   K_0   
\end{equation}  
provided  $\ka_0  \geq   3^d a $   and  $K_0  \geq  e^{\ka_0} b$.
\end{enumerate}

\begin{cor}   \label{cranberry}
\begin{equation}  \label{sudsy}
\begin{split}
\sum_{X:   X  \cap   Y  \neq   \emptyset}  e^{- a|X|_M}  \leq &   b    |Y|_M      \\
\sum_{X:   X  \cap   Y  \neq   \emptyset}   e^{- \ka_0  d_M(X)}  \leq &  K_0    |Y|_M       \\
\end{split}
\end{equation}
\end{cor}
\bigskip

\pr   Again it suffices to take  $M=1$.
The first      follows by  
\begin{equation}
\sum_{X: X \cap   Y    \neq   \emptyset}   e^{- a |X|   }  \leq    
\sum_{\square \subset Y }   \sum_{ X  \supset  \square}  e^{-a |X|   } \leq b   |Y| 
\end{equation}
The   second  is similar.

\section{cluster expansion}  \label{B}

We   give  a treatment of the standard    cluster  expansion adapted to our    circumstances.    
General references  are   \cite{Cam82},  \cite{Sei82},   \cite{Bry86},   \cite{GlJa87}.  We  present an ultralocal version favored by Balaban.

Consider  fields   $\Phi$    and $M-$polymers  $X$        on  a    $d$-dimensional  unit  toroidal  lattice.
We   are  given localized functionals       $H(X, \Phi)$  depending    on $\Phi$  only   in   $X$
and  integrals of the form     
\begin{equation}
\Xi =  \int \exp     \left(    \sum_X    H( X,   \Phi  )  \right)    d\mu(\Phi)   
\end{equation}
where     $d \mu(\Phi)=  \prod_x  d \mu(\Phi(x))   $  is  an  ultralocal   probability measure.    These do not occur naturally  in quantum field theory,   but
can be arranged as we have seen in the text.   Our goal is  to give a local structure to this integral.   This is particularly important if there
are other spectator fields which for which we want to localize the dependence.

\begin{thm}  \label{cluster}      (cluster expansion)  There is a constant $c_0$  depending only 
on the dimension such that if      $H(X,\Phi)$  satisfies 
\begin{equation}
|H(X,\Phi)  |   \leq    H_0 e^{ - \kappa   d_M( X)  }
\end{equation}
 on the support of  $\mu$
with  $\kappa  >  3 \kappa_0  +  3$   and      $H_0  \leq c_0$ 
then    
\begin{equation}
\Xi   = \exp  \Big(  \sum_Y   H^\#(Y)  \Big)
\end{equation}
where $H^\#(Y)$  only depends  on   $H(X)$  for  $X \subset  Y$  and 
\begin{equation}  \label{sunshine0} 
|H^\#(Y)  |   \leq  \cO(1)  H_0  e^{ - (\ka -  3 \ka_0-3)   d_M( Y)  }
\end{equation}
The   constant   $\cO(1)$  depends only on the dimension.
\end{thm}
\bigskip

\pr   
 \textbf{step  1:}   Start with a Mayer  expansion which  yields
\begin{equation}
\begin{split}
\exp    \left(    \sum_X   H( X,     \Phi  )  \right) =&     \prod_X  \Big( (e^{H(X, \Phi)} -1)  + 1  \Big)   \\
=&   \sum_{ \{X_i\}  } \prod_i   ( e^{H(X_i, \Phi)} -1)\\
=  &  \sum_{   \{Y_j\}  } \prod_j    K(Y_j,    \Phi  ) \\
\end{split}
\end{equation}
Here  the  product over  $X$  is  written  as a sum over collections of distinct  polymers  $\{ X_i \}$.    Then
 terms in this sum  are grouped  together   into   collections of disjoint  polymers     $\{ Y_j \}$  (possibly empty),  defining  
 for connected  $Y$
 \begin{equation}
 K(Y, \Phi  ) 
=   \sum_{ \{ X_i\}:     \cup X_i   =  Y}  \prod_i   (e^{H(X_i, \Phi)} -1)
  \end{equation}
In this sum we  require that  the $ \{ X_i\}$  cannot be divided into  two disjoint sets.
Instead of   unordered  $\{ X_i\}$  we  can  write this as a sum over ordered sets  $(X_1,  \dots  X_n)$  by
 \begin{equation}
 K(Y, \Phi  ) 
=    \sum_{n=1}^{\infty} \frac{1}{n!}  \sum_{(X_1,\cdots  X_n) : \cup_i   X_i   =    Y}    \prod_i   (e^{H(X_i, \Phi)} -1)
  \end{equation}
still with the same conditions  on the $X_i$.

If    $H_0  \leq  \log2$  then    on the support of  $\mu$
\begin{equation}
| e^{   H( X, \Phi )}  -1 |  \leq   2H(X, \Phi)    \leq   2H_0 e^{ - \kappa   d_M( X)  }
\end{equation}
and so 
\begin{equation}
 |K(Y, \Phi)| 
\leq     \sum_{n=1}^{\infty} \frac{1}{n!}  \sum_{(X_1,\cdots  X_n) : \cup_i   X_i   =    Y}   \prod_{i=1}^n  2H_0 e^{ - \kappa   d_M( X_i)  }
\end{equation}

Next  we  claim  that    if    $\cup_{i=1}^n X_i   = Y$  as  above,   then 
\begin{equation}  \label{clams}
Md_M(Y)   \leq  \sum_{i=1}^n  M  d_M( X_i)  +  M (n-1)
\end{equation}
Indeed  let  $\tau_i$  be a minimal  tree  on  $X_i$ of length  $Md_M(X_i)$
Also  consider the connected graph whose edges are  pairs  $\{X_i, X_j\}$  such that 
$X_i \cap X_j \neq  \emptyset$.    Take  a tree  which is a subgraph  with  $(n-1)$  edges.
Each  pair   $\{X_i, X_j\}$  in this tree will have  a block  $\square$ in common.  For each pair  add a  line in $\square$
  joining the point  in $\tau_i$  to the point in  $\tau_j$.   This line  has length at most  $M$.  The tree  graph consisting of the $\tau_i$ 
  and the  $(n-1)$  extra  lines  now joins  all the blocks   in   $Y$  and has  length less than  $ \sum_{i=1}^n    Md_M( X_i)  +  M (n-1)$.  The  minimal tree must have shorter length which is the claim.

 We    use  this  to  extract a factor    $\exp  \Big(- (\ka - \ka_0) ( d_M(Y)  -  (n-1))  \Big)$.   Dropping all conditions on
 the  $X_i$  except  $X_i \subset  Y$  we  have      
\begin{equation}
\begin{split}   
| K(Y, \Phi  )  |   
\leq   &  e^{- (\ka - \ka_0) d_M(Y) }  \sum_{n=1}^{\infty} \frac{1}{n!} e^{(\ka- \ka_0)n}  \sum_{(X_1,\cdots  X_n)   \subset    Y^n}  \prod_{i=1}^n   2 H_0   e^{ - \ka_0   d_M( X_i)} \\
\leq   &e^{- (\ka - \ka_0) d_M(Y) }  \sum_{n=1}^{\infty} 
           \frac{1}{n!}  \Big(e^{\ka- \ka_0} \sum_{X \subset Y} 2 H_0   e^{ - \kappa_0   d_M( X)}  \Big)^n \\
\leq   & e^{- (\ka - \ka_0) d_M(Y) } \sum_{n=1}^{\infty}  \frac{1}{n!}  \Big( 2 H_0K_0 e^{\ka- \ka_0} |Y|_M \Big)^n \\
 \leq  &e^{- (\ka - \ka_0) d_M(Y) }  2 H_0K_0e^{\ka- \ka_0} |Y|_M  \exp  \Big( 2 H_0 K_0 e^{\ka- \ka_0} |Y|_M  \Big)  \\
\end{split}
\end{equation}
Now     $|Y|_M   \leq   3^d  ( 1+ d_M(Y) )   \leq  \ka_0 ( 1+ d_M(Y))$.  
Furthermore we  assume  $c_0$  is small enough  so that  $2 c_0K_0 \ka_0 e^{\ka- \ka_0}  \leq  1$.
(So  $c_0$  does depend  on  $\ka$.)
   Then  the exponent    is bounded by   $\cO(1)e^{d_M(Y)}$  and downstairs
$ 2 H_0K_0e^{\ka- \ka_0} |Y|_M$   is bounded by  $\cO(1) H_0e^{d_M(Y)}$.  
Altogether then 
 on the support of  $\mu$
\begin{equation}
| K(Y, \Phi  )  |   \leq   \cO(1)     H_0 e^{ - (\ka - \ka_0 -2)  d_M( Y)  }
\end{equation}

\bigskip
\noindent
\textbf{step  2:}
Now because the $Y_j$  are disjoint and because     fields at different sites are independent  random variables
 \begin{equation}
   \int   \Big(    \sum_{   \{Y_j\}  } \prod_j    K(Y_j,\Phi  )  \Big)     d\mu(\Phi)   
= \sum_{   \{Y_j\}  } \prod_j    K^\#(Y_j   ) 
\end{equation}
where  
\begin{equation}
 K^\#( Y )=   \int    K(Y,\Phi  )    d\mu(\Phi)   
\end{equation}
satisfies  the same  bound
\begin{equation}  \label{owl}
| K^\#(Y) | \leq      \cO(1) H_0   e^{  - (\ka - \ka_0 -2)  d_M(Y)   } 
\end{equation}
\bigskip

\noindent
\textbf{step  3:}
Next  we  claim that  
\begin{equation}
 \sum_{   \{Y_i\}  } \prod_i    K^\#(Y_i  )   =  \exp \Big(     \sum_Y   H^\#(Y)     \Big)
 \end{equation}
 where   
 \begin{equation}    \label{hstar}
   H^\#(Y)  
 =    \sum_{n=1}^{\infty}
 \frac{1}{n!}  \sum_{(Y_1, \dots,  Y_n): \cup_i Y_i  =Y}  \rho^T(Y_1, \dots,  Y_n)  \prod_i K^\#( Y_i )
 \end{equation}
and  $\rho^T(Y_1, \dots,  Y_n) $   vanishes  if the $Y_j$  can be divided into disjoint sets.  
At  first   we  demonstrate the  identity  as  formal  series.   Afterwards we  demonstrate convergence.

Start  by   writing 
\begin{equation}
 \sum_{   \{Y_i\}  } \prod_i    K^\#(Y_i  ) = 1+    \sum_{n=1}^{\infty}   \frac{1}{n!}   \sum_{(Y_1, \dots,  Y_n):  Y_i \cap Y_j  = \emptyset}
\prod_{i=1}^n   K^\#(Y_i)
\end{equation}
where the sum is now over ordered  $n$-tuples of  polymers.
Next   let    $\zeta(X,Y)  =1$  if   $X \cap Y= \emptyset$  and  $\zeta(X,Y)  =0$  if     if   $X \cap Y  \neq  \emptyset$ .
Then  this can be  written  
\begin{equation}   \label{sunrise}
  1+    \sum_{n=1}^{\infty}   \frac{1}{n!}   \sum_{(Y_1, \dots,  Y_n): }
\prod_{i=1}^n   K^\#(Y_i) \prod_{\{i,j\}  \subset (1,\dots,  n)} \zeta( Y_i, Y_j) 
\end{equation}
Next   write 
\begin{equation}
\prod_{\{i,j\} }\zeta( Y_i, Y_j) 
=   \prod_{\{i,j\} }\big[ 1 +(\zeta( Y_i, Y_j)-1)\big]  
 = \sum_G    \prod_{ \{ i,j\}   \in   G }  (  \zeta( Y_i,Y_j)  -1  )           
\end{equation}
Here  in the second step we  expand out the product and identify the sum
with a sum over  collections of    pairs  $\{ i,j\}$  from  $(1, \dots n)$,   that   is  with  graphs  $G$ on 
$(1, \dots,  n)$.    Each  graph determines  a  partition  $ \{  I_1, \dots,  I_K  \} $    of   $(1,  \dots, n)$   and  we  group together
terms  which give the same  partition.      
Then  we  have    
\begin{equation}  \label{sunset}
\prod_{\{i,j\}} \zeta( Y_i, Y_j)   = \sum_{K=1}^n  \sum_{   \{  I_1, \dots,  I_K  \}  \in  \pi_{n,K}  } \prod_{k=1}^K     \rho^T(Y_{I_k})  
\end{equation}
where   $\pi_{n,K}$  is the partitions of  $(1, \dots,  n)$  into  $K$ subsets.  We  have  defined  $\rho^T(Y) =1$  and for  $n \geq 2$
 \begin{equation}
 \rho^T(Y_1, \dots  Y_n)    = \sum_G     \prod_{ \{ i,j\} \in  G }  (  \zeta( Y_i,Y_j)  -1  ) 
 \end{equation}
where   the sum is now over    connected   graphs      $ G $   on   $(1, \dots,   n)$.    We do have       $ \rho^T(Y_1, \dots  Y_n)=0 $
if the  $Y_j$   can be divided into disjoint  sets.

Inserting  (\ref{sunset})  into  (\ref{sunrise})   we  have
\begin{equation}   \label{midday}
\begin{split} 
 \sum_{   \{Y_i\}  } \prod_i    K^\#(Y_i  )=&1+  \sum_{n=1}^{\infty}   \frac{1}{n!}   \sum_{(Y_1, \dots,  Y_n): }
\prod_{i=1}^n   K^\#(Y_i)  \sum_{K=1}^n    \sum_{  \{  I_1, \dots,  I_K  \}  \in  \pi_{n,K}     } \prod_{k=1}^K     \rho^T(Y_{I_{k}})   \\
=&1+  \sum_{n=1}^{\infty}   \frac{1}{n!}   \sum_{K=1}^n   \sum_{   \{  I_1, \dots,  I_K  \}  \in  \pi_{n,K}      }
  \prod_{k=1}^K    \sum_{  Y_{I_k}  }   \rho^T(Y_{I_k})     \prod_{i\in I_k}     K^\#(Y_i)   \\
= & 1+   \sum_{n=1}^{\infty}   \frac{1}{n!}   \sum_{K=1}^n   \sum_{   \{  I_1, \dots,  I_K  \}  \in  \pi_{n,K}     }  \prod_{k=1}^K   f(|  I_k |)
\end{split}
\end{equation}
 In the last  step we  defined  for  $N \geq  1$
\begin{equation}
f(N)  =    \sum_{(Y_1, \dots  Y_N  )  }\rho^T(Y_1, \dots,  Y_N)  
\prod_{i=1}^N  
  K^\#(Y_i) 
  \end{equation}

Replace  the sum over  partitions  $ \{  I_1, \dots,  I_K  \}  $  by  a sum over ordered  partitions    $( I_1, \dots,  I_K )  $.  The  
summand only  depends on  the number of elements  $N_k =  |I_k|$  in  each set.    For each   $(N_1, \dots, N_K)$  with  $N_k  \geq  1$    the  number
of  partitions with these  numbers  is   $n!/ N_1! \dots  N_K!$.   Thus  we  have  
\begin{equation}
\begin{split}
  \sum_{   \{  I_1, \dots,  I_K  \}  \in  \pi_{n,K}     }  \prod_{k=1}^K   f(|  I_k |)
=   &  \frac{1}{K!}     \sum_{  (  I_1, \dots,  I_K)       }  \prod_{k=1}^K  f(|  I_k |)\\
 =   &
 \frac{1}{K!}     \sum_{  (  N_1, \dots,  N_K):  \sum_k N_k  = n     } \frac{n!}{N_1! \dots  N_K!} \prod_{k=1}^K  f(N_k) \\
 \end{split}
\end{equation}

Insert this into  (\ref{midday})   and  change the order of summations  
\begin{equation}   \label{midday2}
\begin{split} 
 \sum_{   \{Y_i\}  } \prod_i    K^\#(Y_i  )= &   1 +    \sum_{n=1}^{\infty}   \frac{1}{n!}   \sum_{K=1}^n   \frac{1}{K!}     \sum_{  (  N_1, \dots,  N_K):  \sum_k N_k  = n     } \frac{n!}{N_1! \dots  N_K!} \prod_{k=1}^K  f(N_k) \\
= &   1 +   \sum_{K=1}^{\infty}    \frac{1}{K!}      \sum_{n  \geq  K }^{\infty}     \sum_{  (  N_1, \dots,  N_K):  \sum_k N_k   =n    } \frac{1}{N_1! \dots  N_K!} \prod_{k=1}^K  f(N_k) \\
= &   1 +   \sum_{K=1}^{\infty}    \frac{1}{K!}        \sum_{  (  N_1, \dots,  N_K):   N_k \geq  1  } \frac{1}{N_1! \dots  N_K!} \prod_{k=1}^K  f(N_k) \\
= &   1 +   \sum_{K=1}^{\infty}    \frac{1}{K!}    \Big(  \sum_{N =1}^{\infty}    \frac{1}{N!}  f(N)  \Big)^K
 =  \exp \Big(     \sum_{N=1}^{\infty} \frac{1}{N!}  f(  N)  \Big)
\end{split}
\end{equation}
The  result  now follows from  
\begin{equation}
  \sum_{N=1}^{\infty}    \frac{1}{N!}  f(  N)  =  \sum_Y   H^\#(Y)
\end{equation}
\bigskip

\noindent
\textbf{ step 4: }  We  now  demonstrate  that under our assumptions  the series  (\ref{hstar})   defining  $H^\#$  converges.

Each indivisible     $n$-tuple $(Y_1, \dots,  Y_n)$  
determines a connected   graph  $g$ on  $(1, \dots, n)$:   a pair  $\{i,j\} \in g$  if  $Y_i \cap Y_j  \neq  \emptyset$.
We  write  $(Y_1, \dots,  Y_n)  \to   g$
The expression    $ \rho^T(Y_1, \dots  Y_n) $  only depends on the  graph  (it is a certain sum over subgraphs) and  one can show
that  
\begin{equation}
|\rho^T(Y_1, \dots  Y_n)|  \leq     \textrm{  number of tree graphs contained in } g  
\end{equation}
Now  fix      $n$     and  $Y$    and   let us restrict  to  sums  over  $(Y_1, \dots,  Y_n)$  such that     $\cup_i  Y_i =Y$.
\begin{equation}  \label{sundry}
\begin{split}
 \sum_{(Y_1, \dots,  Y_n)}  \rho^T(Y_1, \dots  Y_n)  \prod_i K^\#( Y_i )
\leq   & \sum_g  \sum_{  (Y_1, \dots , Y_n)   \to g}   \rho^T(Y_1, \dots  Y_n)  \prod_i K^\#( Y_i )\\
\leq   & \sum_g       \sum_{   (Y_1, \dots , Y_n) \to   g}   \sum_{\tau  \subset g}   \prod_i K^\#( Y_i )\\
=  &\sum_{\tau}   \sum_{g  \supset  \tau}        \sum_{ (Y_1, \dots , Y_n)  \to   g}    \prod_i K^\#( Y_i )\\
=  &    \sum_{\tau}         \sum_{  (Y_1, \dots , Y_n)  \to  g:   g  \supset   \tau  }    \prod_i K^\#( Y_i )\\
\end{split}
\end{equation}

If  $\ka_2  =  \ka -  \ka_0 -1$  we have by  (\ref{owl})   the bound
\begin{equation}
\left| \sum_{ (Y_1, \dots,  Y_n) }  \rho^T(Y_1, \dots  Y_n)  \prod_i K^\#( Y_i ) \right|
\leq ( \cO(1) H_0)^n   \sum_{\tau}            \sum_{  (Y_1, \dots , Y_n)  \to  g:   g  \supset   \tau  }  \prod_i  e^{-\ka_2 d_M(Y_i) }
\end{equation}
Next use the inequality (\ref{clams})   to  bound this by  
\begin{equation}
 e^{-(\ka_2-2 \ka_0) d_M(Y) } \Big(  \cO(1) H_0 \Big)^n   \sum_{\tau}           \sum_{  (Y_1, \dots , Y_n)  \to  g:   g  \supset   \tau  }  \prod_i  e^{-2\ka_0 d_M(Y_i) }
\end{equation}

After relabeling  a  tree graph $\tau$   can be thought of as a map  $\tau$ from  $(1, \dots n)$ to itself  such that   $\tau(j) < j$.
The  restrictions in the  sum  over  $(Y_1, \dots,  Y_n)$   are then   that  $Y_j \cap  Y_{\tau(j)} \neq  \emptyset$.
For   the sum  over  $Y_n$ we    we  have  by  (\ref{sudsy}) 
\begin{equation}
\sum_{Y_n \cap  Y_{\tau(n)}  \neq   \emptyset  }  e^{ -2 \ka_0   d_M( Y_n)  }
\leq   K_0 | Y_{\tau(n)}|_M
\end{equation}
Continue summing  over   $Y_{n-1},  Y_{n-2},  \dots  $.    
By the time  we  get  to the $j^{th}$   vertex we will have accumulated   a factor  
$|Y_j|^{ \tau^{-1}(j)}= |Y_j|^{ d_j-1} $  where  $d_j $      is the incidence number at  $j$ for   $\tau$.
Then we  estimate
\begin{equation}
\begin{split}
\sum_{Y_j \cap  Y_{\tau(j)}  \neq   \emptyset  }  e^{-  2 \ka_0 d_M(Y_j) }
 |Y_j|_M^{d_j-1}  
  \leq  &(d_j-1)!  \sum_{Y_j \cap  Y_{\tau(j)}  \neq   \emptyset  }   e^{-  2 \ka_0 d_M(Y_j) }e^{   |Y_j|_M  }   \\
 \leq   & (d_j-1)! K_0 e^{\ka_0} |Y_{\tau(j)} | \\
\end{split}
\end{equation}
Here  we  used  again   $ |Y_j|_M   \leq  \ka_0  ( 1  + d_M(Y_j) )$
The last step   is
\begin{equation}
\begin{split}
\sum_{Y_1  \subset   Y   }  e^{-  2\ka_0 d_M(Y_1) }
& |Y_1|_M^{d_1-1}  
  \leq  (d_1-1)!  \sum_{Y_1  \subset   Y }   e^{-  2 \ka_0 d_M(Y_j) }e^{   |Y_1|_M  }   \\
 \leq   & (d_1-1)! K_0 e^{\ka_0} |Y |_M 
 \leq    (d_1-1)! K_0\ka_0e^{\ka_0} e^{d_M(Y)}  \\
\end{split}
\end{equation}
Combining the above  yields
\begin{equation}   \label{spit1}
\left| \sum_{ (Y_1, \dots,  Y_n) }  \rho^T(Y_1, \dots  Y_n)  \prod_i K^\#( Y_i ) \right|
\leq   e^{-(\ka_2-2 \ka_0-1) d_M(Y) }    ( \cO(1)H_0)^n \sum_{\tau} \prod_{j=1}^{n} (d_j-1)! 
\end{equation}

By    Cayley's theorem    the number of trees    with incidence numbers
$d_j$   is   $(n-2)! / \prod_{j=1}^{n} (d_j-1)! $  so we  have   
\begin{equation}  \label{spit2}
\sum_{\tau} \prod_{j=1}^{n} (d_j-1)! 
=   \sum_{d_1, \dots, d_n    } \left( \prod_{j=1}^{n} (d_j-1)! \right) \sum_{\tau \textrm{ with }   d_j  }1
\leq    \sum_{d_1, \dots, d_n    }   (n-2)!    \leq    (n-2)! 4^{n-1} 
\end{equation}
In the last  step  we  used  that  a tree graph  has  $n-1$  lines  so  $\sum_{j=1}^n (d_j -1)  =  2(n-1) -n  =  n-2$
and   
\begin{equation}
  \sum_{d_1, \dots, d_{n}:  \sum_j (d_j-1)  =n-2    } 1
 \leq  2^{n-2}   \sum_{(d_1, \dots, d_{n})   }   2^{   - \sum_j  ( d_j-1)  }   
\leq    2^{n-2} 2^n  \leq   4^{n-1}
\end{equation}

Now  use  (\ref{spit2}) in (\ref{spit1}),  divide by $n!$   and sum over $n$  to get a bound on $H^\#(Y)$.  We have  
\begin{equation}
|   H^\#(Y)  |
  \leq     e^{-(\ka_2 - 2\ka_0-1) d_M(Y) }   \sum_{n=1}^{\infty} (\cO(1)H_0)^n   
  \leq    \cO(1)  H_0   e^{-(\ka_2 - 2\ka_0-1) d_M(Y) } 
 \end{equation}
 provided  $H_0  \leq  c_0$  and  $c_0$ is sufficiently small.
Since  $\ka_2 - 2\ka_0-1  =  \ka - 3\ka_0-3$  this completes the proof.

\section{an identity} \label{C}

We   seek an expression for   
$
  C_{k,r}  =\Big( \De_k +  \frac{a}{L^2} Q^TQ+ r\Big)^{-1}
$

\begin{lem}   
\begin{equation}  \label{lion}
 C_{k,r} =   
A_{k,r}  +   a_k^2  A_{k,r} Q_k  G_{k,r} Q_k^T A_{k,r}
\end{equation}
where 
\begin{equation}
\begin{split}
A_{k,r}   =&   \frac{1}{a_k+r}  (I - Q^TQ)   +   \frac{1}{ a_k + aL^{-2}  +r}   Q^T Q  \\
G_{k,r}  = &  \Big( -\De  + \bar \mu_k  +  a_k Q_k^TQ_k      -     a_k^2Q_k^T   A_{k,r} Q_k\Big)^{-1}
  \\
 \end{split}
 \end{equation}
\end{lem}
\bigskip

\pr   Start with 
\begin{equation}
\exp  ( \frac12  <f,C_{k,r} f>  )
=  \const \int   d\Phi   \exp \left(  <\Phi, f>  - \frac{a}{2L^2}  \|Q\Phi \|^2 - \frac{r}{2}\|\Phi\|^2 - \frac12  <\Phi,  \De_k \Phi>  \right)  
\end{equation}
and   from section \ref{freeflow}
\begin{equation}
\exp \left(  - \frac12  <\Phi,  \De_k \Phi>  \right)  
=   \const   \int   \exp \left(   -\frac{a_k}{2} \| \Phi    -   Q_k  \phi  \|^2  - \frac12  < \phi,  ( -\De  + \bar \mu_k  )  \phi  >   \right)   \ d \phi
\end{equation}
Insert the second into the first   and  do the integral  over  $\Phi$  which is 
\begin{equation}
\begin{split}
&  \int   d\Phi   \exp \left(  <\Phi, f>  - \frac{a}{2L^2}  \|Q\Phi \|^2 - \frac{r}{2}\|\Phi\|^2   -\frac{a_k}{2} \| \Phi    -   Q_k  \phi  \|^2   \right) \\
=&     \int   d\Phi   \exp \left(  <\Phi, f +a_k Q_k \phi>  
-\frac12  < \Phi,  \Big(   a_k + r  +   aL^{-2}Q_k^TQ_k  \Big)  \Phi  >    -  \frac{a_k}{2}  \|  Q_k  \phi  \|^2   \right) \\
=&  \const    \exp   \Big(  \frac12  \Big<(f  + a_kQ_k),   A_{k,r}  (f  + a_kQ_k)  \Big> -  \frac{a_k}{2}  \|  Q_k  \phi  \|^2  \Big) \\
\end{split} 
\end{equation}
Here  we  used 
$
\Big(   a_k + r  +   aL^{-2}Q_k^TQ_k  \Big) ^{-1}   
 =A_{k,r}
$
which follows since $Q_k^TQ_k$ is a projection. 
Hence 
\begin{equation}
\begin{split}
&\exp  ( \frac12  <f,C_{k,r} f>  )\\
=&   \const
\int   \exp   \Big(  \frac12  \Big<(f  + a_kQ_k\phi),   A_{k,r}  (f  + a_kQ_k\phi)\Big>   
- \frac12  < \phi,  ( -\De  + \bar \mu_k  + a_k Q_k^TQ_k)  \phi  >  \Big)   \ d \phi
 \\
= &     
 \const \exp   \Big(  \frac12  \left<f ,   A_{k,r}  f    \right>   \Big)
\int    \exp   \Big(    \left< \phi, a_k Q_k^T A_{k,r}  f \right>    - \frac12  < \phi,  G_{k,r}^{-1}  \phi  >   \Big)
  \ d \phi
 \\
=& \const \exp \Big(    \frac12   \left<f ,   A_{k,r}  f    \right>  
+   \frac{a_k^2}{2}  <   f,  A_{k,r}Q_k  G_{k,r}  Q_k^T A_{k,r}  f>   \Big)  \\
\end{split}
  \end{equation}

\section{Estimate on   $G_k( \tilde  \square  )$  }  \label{D}

As in the text let  $\square$  be a $M= L^m$-cube in a partition of  $\bbT^{-k}_{\sM +\sN -k}$,  
let   $\tilde \square$  be a $3M$ enlargement of $\square$,   and let  
 $G_k(\tilde \square)  = 
  [ -\De  +  \bar \mu_k   +  a_k Q_k^T  Q_k ]_{\tilde \square}^{-1}  $  be the Green's functions on $\tilde \square$ with Neumann 
  boundary conditions.  
  We   sketch a proof  of  the bounds  (\ref{sycamore}), (\ref{maple}) which say  for   $x,x' \in \De_y,   \supp f  \subset  \De_{y'}$
  \begin{equation}  \label{sycamore2}
\begin{split}
|( G_k(\tilde   \square)f )(x) |  \leq & C  e^{  -   \ga_0  d(y,y') } \|f\|_{\infty}\\
|( \pa G_k(\tilde   \square)f)(x)  |  \leq &  C      e^{  -   \ga_0  d(y,y') } \|f\|_{\infty}\\
|(\de_{\al}  \pa G_k(\tilde   \square)f)(x)  |  \leq &  C      e^{  -   \ga_0  d(y,y') } \|f\|_{\infty}\\
\end{split}
 \end{equation}
The proof follows   \cite{Bal83b}  with improvements  suggested by    \cite{Bal96b}.

 We  consider the more general case  
 \begin{equation}
 G_k(\Om)  = 
  [ -\De  +  \bar \mu_k   +  a_k Q_k^T  Q_k ]_{\Om}^{-1}  
 \end{equation}
where $\Om  \subset  \bbT^{-k}_{\sM +\sN -k}$  is a union  of  $M$  cubes ,   and  we impose
Neumann boundary conditions.  Eventually  we  want  $\Om$  to be rectangular,  but for the first results
it is any union of $M$ cubes.   Another  restriction is that  $\Om$   should be    small  enough so that it can be identified
with a  subset  of  $(L^{-k} \bbZ)^3$  with the  same distances.  Then we can   study  $G_k(\Om)$ as an operator 
on functions  on  $ \Om \subset   (L^{-k} \bbZ)^3$.      We  want    pointwise bounds,  but   start with $L^2(\Om)$  bounds.   

\begin{lem}   \label{twentynine} The following holds    for  a  constant  $c_0 = \one$.
\begin{enumerate}
\item  For  a     unit   cube  $\De$,   as operators on  $L^2(\De)$
 \begin{equation}
  \Big[ -\De  +  \bar \mu_k   +  a_k Q_k^T  Q_k  \Big]_{\De}  \geq   c_0     (- \De  + I) 
\end{equation}  
\item   For    $\Om $  a union  of  $M$  cubes,  as operators   on  $L^2(\Om)$
 \begin{equation}
  \Big [ -\De  +  \bar \mu_k   +  a_k Q_k^T  Q_k  \Big]_{\Om}  \geq   c_0  ( -\De  + I) 
\end{equation} 
\end{enumerate} 
\end{lem}
\bigskip

\re  The idea is  that  the  averaging  operator   $ a_k Q_k^T  Q_k $  supplies an effective mass.       The parameter  $\bar \mu_k=  L^{-2(\sN - k)}  \bar \mu$   is generally   tiny and cannot help with this uniform   bound.   However
if  $k$ is a bounded distance from   $\sN$  then    $\bar   \mu_k $ is not tiny.  Then   the   $ a_k Q_k^T  Q_k $ is unnecessary  and   the  $\bar  \mu_k$  is sufficient for  a lower bound.
\bigskip

\pr   If   $f  \in L^2(\De)$ is constant  we have   for  a  constant  $c_0 = \one$
\begin{equation}
<f,  a_k [Q_k^T  Q_k ]   f  >  =  a_k \|f\|^2   = \one \|f\|_2^2
\end{equation}
If     $f  \in L^2(\De)$ is orthogonal to constants,  then since the   lowest non-zero  eigenvalue of $-\De$   is  $\cO(1)$  we  have 
\begin{equation}
<f,  [  - \De  ]   f  >\  \geq\    \cO(1)  \|f\|^2
\end{equation}
These combine to  give    
\be    <f,   \Big[  - \frac12\De +  a_k Q_k^T  Q_k     \Big]   f  >\  \geq\    \cO(1)  \|f\|^2
\ee
which suffices to prove the the first  inequality .

For the  second   inequality    we have  for  $f  \in L^2(\Om)$
\begin{equation}
\begin{split}
\Big<f, \Big[ -\De  +  \bar \mu_k   +  a_k Q_k^T  Q_k \Big]_{\Om} f\Big> 
\geq  &   \sum_{\De \subset  \Om}   \Big< f_{\De}, \Big[ -\De  +  \bar \mu   +  a_k Q_k^T  Q_k \Big]_{\De} f_{\De} \Big> \\
\geq  &  c_0 \sum_{\De \subset  \Om}  \| f_{\De} \|^2_2  = c_0 \|f\|^2_2 \\
\end{split}
\end{equation}
Here in the first inequality we  take advantage of the Neumann  boundary conditions and   drop bonds connecting adjacent unit squares.
This completes the proof
\bigskip

Now we  consider   $G_k(\Om)  = 
  [ -\De  +  \bar \mu_k   +  a_k Q_k^T  Q_k ]_{\Om}^{-1}  $.
 The lemma implies that
  \begin{equation}
    \|G_k(\Om)f  \|_2,\     \|\pa   G_k(\Om)f  \|_2    \leq     \cO(1)   \|f \|_2
  \end{equation}
The next  result improves this.  

\begin{lem}   \label{thirty}
Let   $\supp f  \subset  \De_y,   \supp f' \subset  \De_{y'}  $   with  $ \De_y , \De_{y'}  \subset  \Om$. 
Then     with  $\de_0  =  \one $  we have 
\begin{equation}    \label{usher}
|<f,    G_k(\Om)  f'>|  
\leq  \cO(1)  e^{- \de_0  d(y, y')}  \|f \|_2 \|f'\|_2
\end{equation}
\end{lem}
\bigskip

\re   This result  also holds for Dirichlet or mixed boundary conditions.
\bigskip

\pr   (1.)   For  $q \in  \bbR^3$    let   $e_q$ be the exponential function  $e_q(x)  =  e^{q \cdot  x}  $.   For  $|q| \leq  1$ we    consider the operator
\be  
\sD_q   \equiv   e_{-q}\Big[ -\De  +  \bar \mu_k   +  a_k Q_k^T  Q_k\Big]_{\Om}e_q
\ee
We claim that there is a   constant  $c_1 = \one$ such that for       $f  \in L^2(\Om) $ 
\be   \label{city1}
  |<f, [\sD_q -  \sD_0] f> | \    \leq   \  c_1 |q|   <f,(-\De  + I) f> 
\ee

There  are  two terms to consider.   One is   $<f, [e^{-q}Q_k^T  Q_ke^q-Q_k^T  Q_k] f>$.  If we define for  $y \in  \Om \cap   \bbZ^3$
\be
( Q_{k,q}f)(y)  =   \int_{|x-y| <  \frac 12  }   e^{q \cdot  (x-y)} f(x)  dx
 \ee
then the identities       $ Q_ke_q  =  e_q Q_{k,q}$   and    $e_{-q} Q^T_k =   Q^T_{k,-q} e_{-q}$   hold.  Then this term  can be written
$ <f, [Q_{k,-q}^T  Q_{k,q} -  Q_{k}^T  Q_{k}]f>  $.
It is straightforward to establish      $\| Q_{k,q}  f   -  Q_k  f  \|_2    \leq     \one |q|  \|f  \|_2$  and   it  follows that  
\be    |<f, [Q_{k,-q}^T  Q_{k,q} -  Q_{k}^T  Q_{k}]f>|   \leq    \one |q| \|f \|_2^2 \ee
which  is sufficient.  

The other term  is   $<f,   [e^{-q}( -\De ) e^q  - (- \De)] f>$,  also written as   $<\pa  e^{-q}f, \pa e^q  f>  - <\pa f, \pa  f>    $.
If we  define  $\pa_q  =  e^{-q}\pa e^q$  then it is   $<\pa_{-q} f, \pa_q  f>  - <\pa f, \pa  f>    $.   It is straightforward to
show   
\be  \label{ugly}
   \| ( \pa_q  -  \pa )f \|_2   \leq    \one   |q|  \|f  \|_2
\ee
 Then    (\ref{city1}) is established  by     
\be 
\begin{split}
&  | <\pa_{-q} f, \pa_q  f>  - <\pa f, \pa  f>  |  \\  \leq &   |   <(\pa_{-q}-\pa) f, \pa_q  f> |   +   | <\pa f, (\pa_q  - \pa)  f>  |  \\
   \leq &    \one   |q|  \|f  \|_2( \| \pa f\|_2  + \one |q| \|f\|_2)   +  \|\pa  f\|_2 (  \one   |q|  \|f  \|_2)\\
       \leq  &     \one   |q|  ( \|\pa  f\|^2_2  + \|f\|_2^2)    =    \one  |q|  <f, (-\De+I) f>   \\
\end{split}
\ee      
\bigskip

\noindent
(2.)    Now we  write   
\be   <f, \sD_q  f>  =  <f,   \sD_0 f>  +   <f, [\sD_q -  \sD_0] f>
\ee
By the previous lemma   $ <f,   \sD_0 f> \  \geq  c_0    <f,(-\De  + I) f>  $.  Combining this  with  (\ref{city1})  we conclude
that  for  $|q|   \leq  \frac12 c_1^{-1} c_0 $     
\be   \label{city2}  | <f, \sD_q  f> | \    \geq  \  \frac12 c_0     <f,(-\De  + I) f> \  \geq  \ \frac12  c_0   \|f \|_2^2
\ee
Now  substitute      $f =  \sD_q^{-1} h$  and  get   $ | <h, \sD^{-1}_q  h> |     \geq     \frac12 c_0    \| \sD_q^{-1}h \|_2^2$
which implies   $   \| \sD_q^{-1}h \|_2  \leq   2c_0 ^{-1}  \| h \|  $.
    Since  $\sD_q^{-1}  = e_{-q}G_k(\Om) e_q$  this  reads
 \be
 \|   e_{-q}G_k(\Om) e_qh  \|_2  \leq       \one   \|h\|_2 
 \ee
Now let   $\de_0  =  \min \{  \frac12 c_1^{-1} c_0, 1 \} $.   
Then for  $|q| \leq  \de_0$  and   $\supp f  \subset  \De_y,   \supp f' \subset  \De_{y'}  $
\begin{equation}    \label{usher1.5}
|<f,    G_k(\Om)  f'>|   =  |<e_qf, [ e_{-q} G_k(\Om)  e_q ] e_{-q}  f '>|  \leq    \one   \| e_{q} f\|  \| e_{-q} f' \|  
\leq   \one  e^{q\cdot  ( y-y' ) }  
\end{equation}
Here we used that   $\| e_{q} f\|_2  \leq    \one e^{q\cdot y }    \|f \|_2 $. 
Take     $q  =   \de_0 [-(y-y')/ |y-y'|]$  and   get   the bound $ \one  e^{- \de_0| y-y' | } $. 
This  completes the proof.
\bigskip

We continue to assume     $\Om$  be a union of $M$ cubes in  $(L^{-k}\bbZ)^3 $,  and   consider the   operator 
\begin{equation}
\De_k(\Om)  =  a_k -  a_k^2  Q_k G_k (\Om) Q_k^T
\end{equation}
defined on  $\Om \cap  \bbZ^3$.  This  is a local version of the global operator
$\De_k=a_k -  a_k^2  Q_k G_k Q_k^T$  considered in the text.
We  study the inverse 
\begin{equation}
   C_k(\Om)  =  \Big[  \De_k(\Om)  +  \frac{a}{L^2} Q^TQ\Big]^{-1}
\end{equation}
By  a variation  of the identity  (\ref{lion})  at  $r=0$  and   with everything  restricted to $\Om$  we have   
\begin{equation}  \label{coin}
C_k( \Om)   =    A_k   + a_k^2 A_k Q_k G^0_{k+1} ( \Om)  Q_k^T A_k  
\end{equation}
where 
\begin{equation}
\begin{split}
   G^0_{k+1}(\Om)  = &  \Big[ -\De  +  \bar \mu_k   +   \frac{a_{k+1}}{L^2} Q_{k+1}^T  Q_{k+1} \Big]_{\Om}^{-1}\\
A_k ( \Om )  =  &   \Big [    a_k  +  \frac{a}{L^2} Q^TQ  \Big]_{\Om}^{-1}    \\  
\end{split}
\end{equation}
 For    $\supp f  \subset  \De_y,   \supp f' \subset  \De_{y'}  $  
 \begin{equation}    \label{usher2}
|<f,    G^0_{k+1}(\Om)  f'>|  
\leq  \cO(1)L^2  e^{- \de_0 L^{-2}  d(y, y')}  \|f \|_2 \|f'\|_2
\end{equation}
This follows   by scaling  up   (\ref{usher}) for  $G_{k+1}(L^{-1}\Om)$.
Since   $A_k,Q_k$ are  local operators,    it   follows that   the kernel   $ C_k(\Om; y,y') =  < \de_y,   C_k(\Om)  \de_{y'} >$   satisfies 
\begin{equation}   \label{Cbound}
|  C_k(\Om; y,y')|  \leq   \cO(1)L^2 e^{- \de_0 L^{-2}d(y,y')}
\end{equation}

Now
let  $\Om$  be  a   rectangular box    in  $(L^{-k} \bbZ)^3$  
which is a union of $M$  cubes.
Consider        the operator  $\cH_k(\Om )$ from functions on 
  $\Om  \cap     \bbZ^3$  to     functions  on   $\Om  $   defined by 
\begin{equation}
 \cH_k(\Om)    =   a_k  G_k(\Om)  Q_k^T
 \end{equation}      

\begin{lem}    
The  kernel   $\cH_k(\Om  ; x,y)  =   \Big(\cH_k(\Om  ) \de_y \Big)(x)$  satisfies for  $\de_1= \cO(1)$
\begin{equation}  \label{Hbound}
\begin{split}
|\cH_k(\Om   ; x,y)|  &  \leq    \cO(1) e^{- \de_1 d(x,y) }   \\
| \pa \cH_k(\Om  ; x,y)|  &  \leq    \cO(1) e^{- \de_1 d(x,y) }   \\
 |\de_{\al} \pa \cH_k(\Om ; x,x',y)|  &  \leq    \cO(1) e^{- \de_1 d(\{x,x'\},y) }   \\
\end{split}
\end{equation}
\end{lem}
\bigskip

\re
  $\cH_k(\Om) $  is easier to treat than    $G_k(\Om  ) $  since it has no short distance 
singularity.   \bigskip

\pr  \cite{Bal83b}.     First   establish the result    by Fourier  series on the whole  lattice.    Then     extend the result to  $\Om  $    by  multiple  reflections.     
\bigskip

\begin{lem}   \label{A3}  For  $\Om$ a rectangular union of $M$ cubes
define 
\begin{equation}
C'_k  ( \Om)   =   \cH_k(\Om)  C_k  (\Om)  \cH^T_k(\Om)
\end{equation}
Then    with    $\ga_0 \equiv \frac12    \de_0L^{-2}    < \de_1$
\begin{equation} 
|( C'_k(\Om)f )(x) |  \leq   C e^{-\ga_0d(x, \supp  f)} \|f\|_{\infty}
\end{equation}
with  the same bound   for   $\pa  C'_k(\Om)$ and $\de_{\al} \pa   C'_k(\Om)$.
\end{lem}
\bigskip

\pr     We  have 
\begin{equation}
  ( C'_k(\Om)f )(x)   =  \sum_{y, y'}  \cH_k(\Om)(x,y)  C_k  (\Om; y,y') ( \cH^T_k(\Om)f)(y')  \\
\end{equation}
 By   (\ref{Cbound})  and (\ref{Hbound})  
 \begin{equation}
 \begin{split}
|  ( C'_k(\Om)f )(x) |  \leq\  &  \cO(1)L^2  \sum_{y, y'}  e^{-\de_1d(x,y)}   e^{- \de_0L^{-1}d(y,y')}   e^{-\de_1 d(y',  \supp f)}  \|f\|_{\infty}
 \\
  \leq\   &  \cO(1)L^2e^{- \ga_0d(y,  \supp f)}  \sum_{y, y'}  e^{-\frac 12  \de_1d(x,y)}     e^{- \frac12   \de_0L^{-2}d(y,y')} \|f\|_{\infty}
\\
  \leq\  &  C e^{- \ga_0   d(y,  \supp f)} \|f\|_{\infty}
 \\
  \end{split}
\end{equation}

Now our main result is:

\begin{lem} Let  $\Om$  be  a rectangular union  of $M$  cubes.    Then  with  $\ga_0  =  \cO(L^{-2} )  $
\begin{equation}  \label{sycamore3}
\begin{split}
|( G_k(\Om)f )(x) |  \leq\   & C   e^{  -   \ga_0  d(x,  \supp f) } \|f\|_{\infty}\\
|( \pa G_k(\Om)f )(x) |  \leq \  & C   e^{  -   \ga_0  d(x,  \supp f) } \|f\|_{\infty}\\
|( \de_{\al} \pa G_k(\Om)f )(x,x') |  \leq \  &   C   e^{  -   \ga_0  d(\{x,x'\},  \supp f) } \|f\|_{\infty}\\
\end{split}
 \end{equation}
\end{lem}
\bigskip

\pr  Let     $f_{L}(x)  = f(x/L)$  in this proof only.  
The proof is based on the identity  (see \cite{Bal83b} or  \cite{Dim04})   
\begin{equation}
(G_k(\Om)f)(x)   =   \sum_{j=0}^{k-1} L^{-2(k-j)} 
\Big(    C'_j(L^{k-j} \Om)f _{L^{k-j} }\Big)     (L^{(k-j)}x ) 
\end{equation}
From lemma \ref{A3}   
\begin{equation}
\begin{split}
|\Big(     C'_j(L^{k-j} \Om) f_{L^{k-j}} \Big)(L^{k-j}x ) |  \leq  & C
  e^{  -   \ga_0 L^{-1} d(L^{k-j}x, \supp f_{L^{k-j}})} \| f\|_{\infty}   \\
= & C   e^{  -    \ga_0 L^{k-1-j} d(x, \supp f)} \| f\|_{\infty}   \\
\end{split}
\end{equation}
Therefore   
\begin{equation}
|(G_k( \Om)f)(x)|   \leq    C   \sum_{j=0}^{k-1} L^{-2(k-j)} 
 e^{  -    \ga_0 L^{k-1-j} d(x, \supp f)}\|f \|_{\infty}  
   \leq    C e^{  -   \ga_0 d(x, \supp f)}\|f \|_{\infty}  
\end{equation}
The other bounds are  similar.

\section{Random walk expansion    for  $G_{k,r}$}  \label{E}

We  want a random walk expansion  for   $G_{k,r} $   on   $\tk$  as  defined in  (\ref{alt}).   For this operator  $L^2$ bounds are 
sufficient.

To begin  we   get a local result  and consider  for    $\Om  \subset   (L^{-k} \bbZ)^3$,  a union of $M$ cubes,   the operator
\be  
G_{k,r}( \Om )  =   \Big [ - \De  + \bar \mu_k  +  \frac{a_k r}{a_k + r}Q_k^TQ_k 
   +\frac{a_k^2 aL^{-2}} {(a_k +r)(a_k + a L^{-2} +r)} Q_{k+1}^TQ_{k+1}\Big ]_{\Om}^{-1}
\ee

\begin{lem}   For unit cubes  $\De_y, \De_{y'}  \subset  \Om$  and  $r \geq  0$
 \begin{equation}    \label{usher2.5}
 \begin{split}
\| 1_{\De_y}  G_{k,r}(\Om) 1_{\De_{y'} } f\|_2     \leq  &  C  e^{- \de_0 L^{-2}  d(y, y')}  \|f \|_2 \\
\| 1_{\De_y} \pa  G_{k,r}(\Om) 1_{\De_{y'}} f\|_2     \leq  &  C  e^{- \de_0 L^{-2}  d(y, y')}  \|f \|_2\\
\end{split}
\end{equation}
\end{lem}
\bigskip

 \pr
 We  follow the proofs of   lemma \ref{twentynine},   lemma \ref{thirty}.   First 
 we  claim  that there is a constant,  again called  $c_0$, such that   as operators on $L^2(\Om)$
\begin{equation}
\Big[ - \De  + \bar \mu_k  +  \frac{a_k r}{a_k + r}Q_k^TQ_k 
   +\frac{a_k^2 aL^{-2}} {(a_k +r)(a_k + a L^{-2} +r)} Q_{k+1}^TQ_{k+1}\Big ]_{\Om}   \geq  c_0  ( - \De  +    L^{-2})  
 \end{equation}
 If    $r  \geq 1$  just drop the  second and   fourth terms   and get  the lower bound   $c_0  ( - \De  +  I) $ from  lemma  \ref{twentynine}.
 If  $0 \leq  r \leq 1$  drop  the second and third terms  and look for a lower bound on  $- \De   + a_{k+1} L^{-2}  Q_{k+1}^TQ_{k+1}$.   Argue just as
 in  lemma \ref{twentynine}  but now using  $L \De$ cubes  instead of unit cubes $\De$,   and  get the lower bound  $\one   ( - \De  +    L^{-2})$.

It  follows that   
 \be  \|G_{k,r}(\Om)f  \|_2,\     \|\pa   G_{k,r}(\Om)f  \|_2    \leq     \cO(1) L^2 \|f \|_2 
 \ee
 
 Continuing as in  lemma \ref{thirty},
   $\sD_q$  is  replaced    by  
 \be
\sD_{q,r}  =   e^{-q}  \Big [ -\De  +    \bar \mu_k   +  \frac{a_k r}{a_k + r}Q_k^TQ_k 
   +\frac{a_k^2 aL^{-2}} {(a_k +r)(a_k + a L^{-2} +r)} Q_{k+1}^TQ_{k+1}\Big]_{\Om}e^q
\ee
As  in     (\ref{city1}) there  is  a constant, again called  $c_1$,  such that  for  $|q|  <  L^{-1}$      
\be   \label{city2.5}
  |<f, [\sD_{q,r} -  \sD_{0,r}] f> | \    \leq   \  c_1 |q|   <f,(-\De  + I) f> 
\ee
Then    for     $|q|  \leq   \frac12 c_1^{-1} c_0   L^{-2}   =  \de_0 L^{-2}  $      we  have  $|q|< L^{-1}$  and
\be   \label{city2.6}  | <f, \sD_{q,r}  f> | \    \geq  \ \frac 12  c_0    <f,(-\De  + L^{-2}  ) f> \  \geq    \frac12 c_0  L^{-2}  \|f \|_2^2
\ee
As  in lemma \ref{thirty}   the last bound implies that  $\sD_{q,r}^{-1}  =  e^{-q} G_{k,r} e^q$  satisfies  
\be
\label{ugly2}
  \|  \sD_{q,r}^{-1} f \|_2  =   \| q^{-q} G_{k,r} e^q  f \|_2   \leq   C  \|f  \|_2
\ee
This leads to a bound of the form  (\ref{usher}).   But here we formulate it a little differently  and    write
\be   \label{city3}
\begin{split}
\| 1_{\De_y}  G_{k,r}(\Om) 1_{\De_{y'} } f\|_2 
  \leq  &   \one  e^{q\cdot y}    \|e^{-q}G_{k,r}(\Om)e^q 1_{\De_{y'}}e^{-q}  f\|_2 
  \leq     C  e^{q\cdot y}   \|  1_{\De_{y'}}e^{-q}  f\|_2 
  \leq    C  e^{q\cdot( y-y')}  \\
\end{split}
\end{equation}  
With  the choice   $q =   \de_0L^{-2} [-(y-y')/ |y-y'|]$  we  obtain the first  result  in  (\ref{usher2.5}).

For the second result   start with  
\be    \| \pa f  \|^2_2  =   <f,  (-\De)  f> \   \leq   \one  | <f,  \sD_{q,r} f>|
\ee
from  (\ref{city2.6}).  Let   $f  =  \sD_{q,r}^{-1} h$  and get   
$ \| \pa  \sD_{q,r}^{-1} h \|^2_2  \leq  \one  | <h,  \sD^{-1}_{q,r} h>|  \leq  C  \|h \|^2_2
$.
We also have   by   (\ref{ugly})  and    (\ref{ugly2})  that
$ \| (\pa_q - \pa)  \sD_{q,r}^{-1} h \|_2     \leq C   \|h \|_2
$.     The  last  two combine to give   $ \| \pa_q   \sD_{q,r}^{-1} h \|_2    \leq C   \|h \|_2  $  or  
\be \|   e^{-q}  (\pa   G_{k,r}  ) e^q h \|_2  \leq    C   \|h \|_2
\ee
Now  argue as in (\ref{city3})  to get the second  result in (\ref{usher2.5}).

\begin{lem}    
$G_{k,r}$  on  $\tk$    has  a random walk  expansion of the form  
$   G_{k,r}  =  \sum_{\om}    G_{k,r, \om} $ which converges  in  $L^2$ norm for  $M$ sufficiently large.   It yields the bounds
for  $r \geq  0$:
 \begin{equation}    \label{usher3}
 \begin{split}
\| 1_{\De_y}  G_{k,r} 1_{\De_{y'}}  f  \|_2     \leq  &  C  e^{- \ga_0  d(y, y')}  \|f \|_2 \\
\| 1_{\De_y} \pa  G_{k,r} 1_{\De_{y'}}  f\|_2     \leq  &  C  e^{- \ga_0 d(y, y')}  \|f \|_2\\
\end{split}
\end{equation}
\end{lem}

\pr      This follows by a random   walk  expansion   similar to lemma  \ref{jupiter}.    As in  (\ref{brand})   the random walk expansion has the form  
\begin{equation}  \label{brand2}
\begin{split}
G_{k,r} =
&
 \sum_{n=0}^{\infty}  \sum_{\om_0, \om_1,...,\om_n}
\Big( h_{\om_0} G_{k,r} (\tilde  \square_{\om_0})   h_{\om_0}\Big)
\Big(K_{r,\om_1}   G_{k,r} (\tilde  \square_{\om_1})   h_{\om_1}  \Big )
\cdots     \Big (K_{r,\om_n}   G_{k,r} (\tilde  \square_{\om_n})    h_{\om_n}\Big)
\\
\end{split}
\end{equation}
where  $\square$ is still  an $M$-cube,  and  $\tilde \square$ is an enlargement to a $3M$-cube.  The operator  $K_{r,z}$ is   
\be   K_{r,z}   =  -  \Big[\Big(-\De   +  \frac{a_k r}{a_k + r}Q_k^TQ_k 
   +\frac{a_k^2 aL^{-2}} {(a_k +r)(a_k + a L^{-2} +r)} Q_{k+1}^TQ_{k+1}\Big)  ,   h_z  \Big]
\ee
Estimates on this operator,   together with the bounds     (\ref{usher2}),  yield
\be    \|  1_{\De_y}    K_{r,z}   G_{k,r} (\tilde  \square_z) 1_{\De_{y'}} f \|_2    \leq CM^{-1}  e^{- \de_0 L^{-2}  d(y, y')} \|f\|_2
\ee
This is sufficient  to give  convergence of the expansion in the $L^2$ norm for  $M$ large,   and the decay   $ e^{- \ga_0  d(y, y')}$
  with   $\ga_0  =  \frac12   \de_0 L^{-2} $.

\end{document}